\documentclass{aa}  

\usepackage{graphicx}

\usepackage{txfonts}
\usepackage[citecolor=blue, linkcolor=blue]{hyperref}
\hypersetup{colorlinks=True}

\newcommand{\pyneb}{\texttt{PyNeb}}
\newcommand{\hcmuv}{\textsc{HCm-UV}}

\begin{document} 

\title{Ionized gas kinematics and chemical abundances of low-mass star-forming galaxies at $z\sim 3$}
   
\author{Llerena, M.\inst{1}\fnmsep\thanks{e-mail: mario.llerena@userena.cl}
\and Amor\'in, R. \inst{1,2}
\and Pentericci, L.\inst{3}
\and Calabrò, A.\inst{3}
\and Shapley, A. E.\inst{4}
\and Boutsia, K.\inst{5}
\and Pérez-Montero, E. \inst{6}
\and Vílchez, J.M. \inst{6}
\and Nakajima, K. \inst{7}
          }

   \institute{Departamento de Astronomía, Universidad de La Serena, Av. Juan Cisternas 1200 Norte, La Serena, Chile
         \and
             Instituto de Investigación Multidisciplinar en Ciencia y Tecnología, Universidad de La Serena, Raúl Bitrán 1305, La Serena, Chile
             \and
             INAF - Osservatorio Astronomico di Roma, Via di Frascati 33, 00078, Monte Porzio Catone, Italy
             \and 
            Department of Physics \& Astronomy, University of California, Los Angeles, 430 Portola Plaza, Los Angeles, CA 90095, USA
            \and 
            Las Campanas Observatory, Carnegie Observatories, Colina El Pino, Casilla 601, La Serena, Chile
             \and
            Instituto de Astrofísica de Andalucía, CSIC, Apartado de correos 3004, E-18080 Granada, Spain
             \and 
             National Astronomical Observatory of Japan, 2-21-1 Osawa, Mitaka, Tokyo 181-8588, Japan
             } 

   \date{Received ; accepted }

 
 \abstract
   {Feedback from massive stars plays a crucial role in regulating the growth of young star-forming galaxies (SFGs) and in shaping their interstellar medium (ISM). This feedback contributes to the removal and mixing of metals via galactic outflows and to the clearance of neutral gas, which facilitates the escape of ionizing photons.
   }   
   {
   Our goal is to study the impact of stellar feedback on the chemical abundances of the ISM in a sample of SFGs with strong emission lines at $z\sim3$.
   }
   { 
   We selected 35 low-mass SFGs ($7.9<\log$(M$_{\star}/{\rm M}_{\odot})<10.3$) from deep spectroscopic surveys based on their CIII]$\lambda$1908 emission. We used new follow-up near-infrared (NIR) observations to examine their rest-optical emission lines and identify ionized outflow signatures through broad emission line wings detected after Gaussian modeling of [OIII]$\lambda\lambda$4959,5007 profiles. We characterized the galaxies' gas-phase metallicity and carbon-to-oxygen (C/O) abundance using a T$_{\rm e}$-based method via the OIII]$\lambda$1666/[OIII]$\lambda$5007 ratio and photoionization models. 
   }
   {
   We find line ratios and rest-frame equivalent widths (EWs) characteristic of high-ionization conditions powered by massive stars. Our sample displays mean rest-frame EW([OIII]$\lambda$5007) of $\sim$560\r{A}, while about 15\% of them show EW([OIII]$\lambda\lambda$4959,5007)$>1000$\r{A} and EW(CIII])$>5$\r{A}, closely resembling those now seen in Epoch of Reionization (EoR) galaxies with JWST. We find high T$_{\rm e}$ values, which imply low gas-phase metallicities 12+log(O/H) $\sim$ 7.5 - 8.5 (mean of 17\% solar) and C/O abundances from 23\% to 128\% solar, with no apparent increasing trend with metallicity. 
   Our sample follows the mass-metallicity relation {at $z\sim$\,3, with some galaxies showing lower gas-phase metallicities thus resulting in significant deviations from the mass-metallicity-SFR relation}. 
   From our [OIII]$\lambda\lambda$4959,5007 line profile modeling, we find that 65\% of our sample shows an outflow component, which {is found both blue- or red-shifted} relative to the ionized gas systemic velocity, and have mean maximum velocities of $v_{\rm max} \sim280$ km s$^{-1}$. 
   We find {a weak correlation between $v_{\rm max}$ and the} star-formation rate surface density ($\Sigma_{\rm SFR}$) such as $v_{\rm max}=(2.41\pm0.03)\times\Sigma_{\rm SFR}^{(0.06\pm0.03)}$. 
   {Moreover, we} find that the mass-loading factor $\mu$ of our galaxy sample is typically lower than in more massive galaxies from literature but is higher than in typical local dwarf galaxies. In the stellar mass range covered by our sample, we find that $\mu$ increases with $\Sigma_{\rm SFR}$ thus suggesting that for a given stellar mass, denser starbursts in low-mass galaxies produce stronger outflows.   
   Our results complement the picture drawn by similar studies at lower redshift, suggesting that the removal of ionized gas in low-mass SFGs driven by stellar feedback is regulated by their stellar mass \textit{and} by {the strength and concentration of their star formation, i.e. $\Sigma_{\rm SFR}$}.
   }
   {
   }

\titlerunning{Ionized gas kinematics and chemical abundances in low-mass SFGs at $z\sim 3$}
\authorrunning{Llerena, M. et al.}

\maketitle
%

\section{Introduction}

Understanding the last phase transition of the Universe, known as the Epoch of Reionization (EoR), is one of the longstanding goals of extragalactic astronomy. The massive stars residing in the high-redshift($z$) star-forming  galaxies (SFGs) are suspected to be the dominant ionizing agents that drive reionization \citep[e.g.][]{Robertson2015,Finkelstein2019}. However, some studies suggest that low luminosity active galactic nuclei (AGN) may play a significant role in reionization \citep[e.g.][]{Madau2015,Dayal2020}. From recent high-resolution cosmological simulations of Lyman continuum (LyC, $<912$\r{A}) emitting sources in the EoR, $z>5$ simulated galaxies are studied with a detailed treatment of the multiphase interstellar medium (ISM) and stellar-feedback. For example, with the Feedback in Realistic Environments \citep[FIRE:][]{Hopkins2018} project, it has been found that a majority of LyC escape comes from the very young ($<10$ Myr), kpc-scale star-forming regions of a galaxy, with negligible contribution from an older ($>10$ Myr) stellar population. These ISM structures clear out the neutral gas column to allow the escape of ionizing photons at very low escape fractions (f$_{\rm esc}$) that can reach 10–20\% only for a small amount of time since the ionizing photon production from massive stars begins to decline after 3 Myr \citep[e.g.][]{Ma2015,Ma2020}. At the same time, most of the ionizing photons are consumed by surrounding neutral gas clouds \citep[e.g.][]{Ma2015,Kakiichi2021}. Also, the reionization phase is estimated to be a fast epoch where the Universe goes from 90\% neutral at $z\sim 8.22$ to 10\% neutral at $z\sim 6.25$ in $\sim$300 Myr by massive starburst galaxies \citep{Naidu2020}. Understanding the mechanisms that facilitate the escape remains crucial.

On the other hand, deep rest-frame ultraviolet (UV) spectra ($\lambda\sim1200-2000$\r{A}) of several high-$z$ galaxies \citep[$z\sim5-7$; e.g.][]{Stark2015,Mainali2018,Hutchison2019,Castellano2022,Tang2023} reveal prominent high-ionization nebular emission lines, such as He II$\lambda$1640, O III]$\lambda\lambda$1661,66, C III]$\lambda\lambda$1907,09 (C III] hereafter) and C IV$\lambda\lambda$1548,51 (C IV hereafter). To unveil their nature, the rest-frame UV is crucial, and analogs to these reionization sources at lower redshifts are key. The rest-UV spectra can foster our understanding of SFGs in terms of the stellar populations hosting massive stars and their impact on ISM physical conditions, chemical evolution, feedback processes, and reionization. This is important since James Webb Space Telescope (JWST) instruments such as NIRSpec \citep{Jakobsen2022} will cover blueward of $\sim$4500 \r{A} only in objects $z< 10$. As such, understanding the ISM properties from UV spectra will be essential for characterizing and interpreting the spectroscopic observations of high-$z$ systems. Large and deep surveys such as {Lyman Break Galaxy (LBG) survey of $z\sim3$ \citep{Steidel2003}}, VIMOS Ultra Deep Survey \citep[VUDS,][]{Lefevre2015,Tasca2017}, MUSE Hubble Ultra Deep Survey \citep{Bacon2017}, and VANDELS \citep{McLure2018,Pentericci2018,Garilli2021}, have targeted SFGs at mainly $z\sim 2-4$ to study the physical properties of SFGs with intense rest-UV emission lines. Significant improvements have also been made in the local universe with relatively large samples of metal-poor SFGs in The COS Legacy Spectroscopic SurveY \citep[CLASSY, ][]{Berg2022}. {From an observational point of view at $z\sim 3$, the escape fraction depends on galaxy properties such as equivalent width (EW) of Ly$\alpha$, stellar mass, and color excess by dust extinction \citep[e.g.][]{Steidel2018,Saxena2022weak,Begley2022,Pahl2022}. Moreover, the profile shape of Ly$\alpha$ is an essential predictor of the LyC escape fraction as this gives information about the covering fraction of neutral gas at the systemic velocity \citep{Izotov2020,Flury2022}. } Such indirect probes are essential since ionizing fluxes cannot be measured at $z>6$ due to the large opacity of the intergalactic medium (IGM) \citep{Inoue2014} and the measurement of the escape fraction of ionizing photons will heavily rely on indirect probes at high redshift \citep[e.g.][]{Xu2022,Naidu2022} or work on analogs at lower redshift \citep[e.g.][]{Schaerer2016,Flury2022I}.

One of the main suspects in aiding the escape of ionizing photons is galaxy scale outflows \citep[e.g.][and references therein]{Chisholm2017,Kim2020,Hogarth2020}. These outflows are expected to remove surrounding neutral gas, thus clearing a pathway for ionizing photons to efficiently escape. However, it is often challenging to reconcile the timescales of such strong galactic outflows with the timescale of production and escape of ionizing photons in a galaxy. Outflowing gas of SFGs at $z\sim2$ has been reported both from optical emission lines such as [OIII]$\lambda$5007 (hereafter [OIII]) or H$\alpha$ \citep[e.g.][]{Forster2019,Ubler2022} as well as UV absorption lines \citep[e.g.][]{Steidel2010,Jones2018,Calabro2022}. Optical rest-frame emission lines are able to trace denser outflowing gas, providing an instantaneous snapshot of the ongoing ejective feedback; therefore, in principle, they are less contaminated by tenuous gas around galaxies \citep{Concas2022}. These emission lines are typically modeled with one narrow and one broad Gaussian component. The latter is considered the outflow component. The properties of the outflows are probed to depend on the properties of the host galaxy. For example, it is observed that the velocity of outflows increases as a function of star formation rate (SFR) and galaxy stellar mass (M$_{\star}$) \citep[e.g.][]{Weiner2009}, which is also seen in simulations \citep[e.g.][]{Muratov2015}. 
In \cite{Freeman2019}, with the MOSFIRE Deep Evolution Field survey \citep[MOSDEF,][]{Kriek2015}, they found that the broad-to-narrow flux ratio increases with stellar mass. Understanding how the presence of highly ionized gas outflows affects the properties of host galaxies gives insights into how the stellar feedback from young stars plays a vital role in the escape of ionizing photons. 

Finally, analytic models explain that scaling relations such as the Mass-Metallicity relation \citep[MZR,][]{Tremonti2004} and the Fundamental Metallicity Relation \citep[FMR,][]{Mannucci2010,Curti2020,Sanders2021} arise due to an interplay between star formation, the infall of metal-poor IGM gas and the ejection of metal-rich ISM gas. The FMR is a signature of the smooth, long-lasting equilibrium between gas flows and secular evolution \citep[e.g.][]{Bouche2010,Lilly2013}. {Based on the FMR,} less massive galaxies (M$_{\star}<10^{11}$M$_{\odot}$) with smaller SFR produce less heavy elements that are more efficiently ejected due to their shallow potential wells; as a result, for a given M$_\star$, the gas metallicity decreases with SFR \citep{Dayal2013}. {On the other hand, the lack of any correlation between the ratio of a secondary element, like carbon (C) or nitrogen (N), to a primary one, like oxygen (O), with SFR \citep[e.g.][]{PM2009}  could indicate how stellar winds eject metals, leaving their proportions unaffected in the remaining gas.} The role of outflows in shaping the chemical properties of galaxies is another critical feature to understand in galaxy evolution. 

In this paper, we present an analysis of the physical properties of a sample of 35 SFGs at $z\sim 3$, searching for the effects of the stellar feedback on their chemical properties based on the kinematics of the ionized gas. In Sec. \ref{sec:data}, we present the sample selection and the optical and Near-infrared (NIR) spectroscopy used in this work. {In Sec. \ref{sec:properties}, we present the physical properties of the sample and the narrow+broad Gaussian modeling of the [OIII]$\lambda\lambda$4959,5007 profiles.} 
{Then, we present the results of our study in Sec. \ref{sec:results} describing the main source of ionization of our sample, their chemical abundances (basically, gas phase metallicity (O/H) and carbon-to-oxygen (C/O) abundance) and the properties of ionized gas kinematics based on the outflow component.}  
{Then, we present the discussion of our results in Sec. \ref{sec:discussion} based on how outflows and star formation histories (SFHs) may affect the chemical properties of their host galaxy.} 
Finally, in Sec. \ref{sec:conclusions}, we present our conclusions. 

Throughout this paper, we adopt a $\Lambda$-dominated, flat universe with $\Omega_\Lambda = 0.7$, $\Omega_M = 0.3$ and H$_0= 70$ km s$^{-1}$ Mpc$^{-1}$. All magnitudes are quoted in the AB system. Equivalent widths are quoted in the rest frame and are positive for emission lines. We adopt a \cite{Chabrier2003} initial mass function (IMF). We consider log(O/H)$_{\odot}=8.69$ and log(C/O)$_{\odot}=-0.26$ \citep{Asplund2009}.
\vspace{-2mm}
\section{Data and sample selection}\label{sec:data}
Our strategy is based on the combined analysis of the rest-UV+optical spectra of a diverse sample of SFGs at $z=2-4$. Our main selection criteria (the detection of CIII]) depends on the rest-UV emission lines that are obtained by optical spectrographs. Our parent sample is a combination of two large surveys carried out with the VIMOS spectrograph \citep{Lefevre2003}: the VUDS \citep{Lefevre2015,Tasca2017} and the VANDELS \citep{McLure2018,Pentericci2018} surveys. From this parent sample, we select galaxies with NIR spectroscopy targeting their rest-optical emission lines.   

Our final sample contains 17 VANDELS galaxies (hereafter, C3-VANDELS) and 18 VUDS galaxies (hereafter, C3-VUDS). We describe the selection of the final sample and the NIR observations in the next sections.

\subsection{{Sample selection} from rest-frame UV spectroscopy }
\subsubsection{VANDELS parent sample}
We use spectroscopic data from VANDELS \citep{McLure2018,Pentericci2018} – a deep VIMOS survey of the CANDELS fields – which is a completed ESO public spectroscopic survey carried out using the Very Large Telescope (VLT). VANDELS covers two well-studied extragalactic fields, the UKIDSS Ultra Deep Survey (UDS) and the Chandra Deep Field South (CDFS). The final VANDELS data release, DR4, contains spectra of $\sim$ 2100 galaxies in the redshift range $1.0 <z < 7.0$, with on-source integration times ranging from 20 to 80 hours, where $> 70$\% of the targets have at least 40 hours of integration time \citep{Garilli2021}. The spectral resolution of VANDELS spectra is R $\sim$ 600 in the wavelength range 480 $<\lambda_{obs}< $ 980nm. At the redshift range $2<z<4$ of our interest due to CIII] selection, 887 galaxies were observed with reliable redshift. From them, we select a parent sample of 280 SFGs that show S/N$>3$ in CIII] with EW(CIII])$>0$. Most of them (74\%) show EWs$<5$\r{A} while only $\sim 5$\% shows EWs$>10$\r{A}. To select this parent sample, we follow the methodology presented in \cite{Llerena2022} for the VANDELS DR3.  

\subsubsection{VUDS parent sample}
We also use observations from the VUDS survey \citep{Lefevre2015,Tasca2017}, a massive 640-hour ($\sim$80 nights) spectroscopic campaign reaching extreme depths (i'$<$25 mag) over three well-studied extragalactic fields: COSMOS, ECDFS, and VVDS-02h. Spectroscopic observations consisted of approximately 50400s of integration across the wavelength range 365 $<\lambda_{obs}< $ 935nm at a spectral resolution of R $=230$. At $2 < z < 3.8$, where the instrumental setup allows following the CIII] line reliably, 3899 SFGs were observed. Our parent sample is described in \cite{LeFevre2019} and is selected with $2 < z < 3.8$ and S/N $> 3$ in CIII]. They selected 1763 SFGs with EW(CIII])$>0$, most of them (75\%) show EWs$<5$\r{A} while only $\sim 8$\% shows EWs$>10$\r{A}.  

\begin{table}
\caption{{Coordinates and spectroscopic redshift of the final sample.}}
\label{tab:coordinates}
    \centering    
    \begin{scriptsize}
    \begin{tabular}{|c|c|c|c|c|}\hline\hline
    ID&RA\tablefootmark{(a)}&DEC\tablefootmark{(a)}&$z_{\rm{CIII]}}$\tablefootmark{(b)}&Note\tablefootmark{(c)}\\
    &deg&deg&&\\
\hline\multicolumn{5}{|l|}{C3-VANDELS sample}\\ \hline
UDS020394 & 34.54 & -5.16 & 3.308 & 1 \\
CDFS020954 & 53.02 & -27.73 & 3.496 & 1 \\
CDFS023527 & 53.08 & -27.71 & 3.110 & 1 \\
UDS021601 & 34.55 & -5.16 & 3.344 & 1 \\
CDFS022563 & 53.04 & -27.72 & 3.003 & 1 \\
UDS022487 & 34.48 & -5.15 & 3.064 & 1 \\
CDFS015347 & 53.06 & -27.78 & 3.516 & 1 \\
CDFS019276 & 53.02 & -27.75 & 3.400 & 1 \\
UDS020928 & 34.51 & -5.16 & 3.137 & 1 \\
CDFS019946 & 53.06 & -27.74 & 2.437 & 1 \\
UDS020437 & 34.52 & -5.16 & 3.207 & 1 \\
CDFS018182 & 53.01 & -27.76 & 2.317 & 1 \\
CDFS018882 & 53.03 & -27.75 & 3.403 & 1 \\
CDFS025828 & 53.05 & -27.69 & 3.350 & 1 \\
CDFS022799 & 53.06 & -27.72 & 2.544 & 1 \\
UDS021398 & 34.53 & -5.16 & 2.492 & 1 \\
UDS015872 & 34.52 & -5.19 & 2.301 & 1 \\
\hline\multicolumn{5}{|l|}{C3-VUDS sample\tablefootmark{(d)}}\\ \hline
5100998761 & 150.16 & 2.26 & 2.453 & 3 \\
5101444192 & 150.16 & 2.61 & 3.420 & 3 \\
510994594 & 150.07 & 2.28 & 3.297 & 1 \\
5101421970 & 150.34 & 2.61 & 2.470 & 3 \\
510838687 & 149.86 & 1.98 & 2.557 & 2 \\
5100556178 & 150.16 & 1.89 & 2.537 & 2 \\
511229433 & 150.11 & 2.41 & 3.256 & 1 \\
511245444 & 150.15 & 2.30 & 3.038 & 1 \\
530048433 & 53.10 & -27.76 & 2.312 & 1 \\
510583858 & 150.05 & 1.86 & 2.417 & 3 \\
511451385 & 150.15 & 2.57 & 2.370 & 2 \\
511025693 & 150.06 & 2.25 & 3.256 & 1 \\
5100997733 & 150.14 & 2.27 & 3.003 & 1 \\
511228062 & 150.07 & 2.41 & 3.354 & 1 \\
5101001604 & 150.18 & 2.24 & 3.157 & 1 \\
510996058 & 150.11 & 2.28 & 2.493 & 1 \\
511001501 & 150.12 & 2.24 & 2.227 & 1 \\
530053714 & 53.04 & -27.73 & 2.436 & 1 \\
\hline\hline
    \end{tabular}
    \tablefoot{\tablefootmark{(a)}{{Right Ascension and Declination.}}
    \tablefootmark{(b)}{{Spectroscopic redshift based on CIII].}}
    \tablefootmark{(c)}{NIR instrument used as: (1)Keck/MOSFIRE, (2)Magellan/FIRE, (3)VLT/X-Shooter}\,\tablefootmark{(d)}{The ID starts with 51 (53) for targets in the COSMOS (ECDFS) field.}}
    \end{scriptsize}
\end{table}

\begin{table}
\caption{Main physical properties of the sample based on SED fitting.}
\label{tab:SED}
    \centering
    \begin{scriptsize}
    \begin{tabular}{|c|c|c|c|c|}\hline\hline
    ID&log(M$_{\star}$)&log(SFR)&E(B-V)$_{\rm{SED}}$&E(B-V)$_{\rm{g}}$\tablefootmark{(a)}\\
    &M$_{\odot}$&M$_{\odot}$ yr$^{-1}$&mag&mag\\
\hline\multicolumn{5}{|l|}{C3-VANDELS sample}\\ \hline
UDS020394 & 8.92$\pm$0.17 & 0.84$\pm$0.10 & 0.08$\pm$0.02 & ... \\
CDFS020954 & 8.82$\pm$0.08 & 1.72$\pm$0.03 & 0.21$\pm$0.01 & ... \\
CDFS023527 & 9.54$\pm$0.11 & 1.53$\pm$0.09 & 0.14$\pm$0.02 & ... \\
UDS021601 & 9.34$\pm$0.10 & 1.01$\pm$0.10 & 0.11$\pm$0.02 & ... \\
CDFS022563 & 8.75$\pm$0.14 & 1.19$\pm$0.07 & 0.14$\pm$0.01 & ... \\
UDS022487 & 9.09$\pm$0.11 & 1.34$\pm$0.15 & 0.14$\pm$0.03 & ... \\
CDFS015347 & 9.11$\pm$0.15 & 1.18$\pm$0.08 & 0.12$\pm$0.01 & ... \\
CDFS019276 & 10.26$\pm$0.07 & 2.27$\pm$0.07 & 0.25$\pm$0.01 & ... \\
UDS020928 & 9.48$\pm$0.09 & 1.78$\pm$0.05 & 0.24$\pm$0.01 & ... \\
CDFS019946 & 8.96$\pm$0.08 & 1.21$\pm$0.06 & 0.14$\pm$0.01 & 0.53$\pm$0.31 \\
UDS020437 & 9.61$\pm$0.09 & 1.42$\pm$0.11 & 0.11$\pm$0.02 & ... \\
CDFS018182 & 9.33$\pm$0.09 & 1.34$\pm$0.07 & 0.08$\pm$0.01 & 0.18$\pm$0.06 \\
CDFS018882 & 9.81$\pm$0.08 & 1.80$\pm$0.08 & 0.24$\pm$0.01 & ... \\
CDFS025828 & 9.57$\pm$0.12 & 2.03$\pm$0.05 & 0.17$\pm$0.01 & ... \\
CDFS022799 & 9.14$\pm$0.05 & 1.86$\pm$0.04 & 0.14$\pm$0.01 & 0.07$\pm$0.05 \\
UDS021398 & 9.25$\pm$0.07 & 1.41$\pm$0.05 & 0.12$\pm$0.01 & 0.49$\pm$0.35 \\
UDS015872 & 9.71$\pm$0.04 & 1.66$\pm$0.04 & 0.23$\pm$0.01 & 0.30$\pm$0.11 \\
\hline\multicolumn{5}{|l|}{C3-VUDS sample}\\ \hline
5100998761 & 7.86$\pm$0.03 & 0.84$\pm$0.04 & 0.04$\pm$0.01 & 0.17$\pm$0.24 \\
5101444192 & 8.76$\pm$0.03 & 1.77$\pm$0.03 & 0.13$\pm$0.01 & ... \\
510994594 & 9.89$\pm$0.08 & 1.31$\pm$0.12 & 0.15$\pm$0.03 & ... \\
5101421970 & 9.78$\pm$0.05 & 1.08$\pm$0.07 & 0.07$\pm$0.02 & 0.21$\pm$0.11 \\
510838687 & 8.87$\pm$0.14 & 1.22$\pm$0.09 & 0.11$\pm$0.01 & ... \\
5100556178 & 8.91$\pm$0.12 & 1.16$\pm$0.08 & 0.11$\pm$0.01 & ... \\
511229433 & 9.43$\pm$0.09 & 1.25$\pm$0.16 & 0.17$\pm$0.03 & ... \\
511245444 & 9.24$\pm$0.12 & 1.31$\pm$0.12 & 0.11$\pm$0.02 & ... \\
530048433 & 9.24$\pm$0.07 & 0.95$\pm$0.06 & 0.06$\pm$0.01 & 0.28$\pm$0.30 \\
510583858 & 9.83$\pm$0.03 & 1.09$\pm$0.06 & 0.09$\pm$0.01 & 0.34$\pm$0.11 \\
511451385 & 8.63$\pm$0.06 & 1.46$\pm$0.05 & 0.20$\pm$0.00 & ... \\
511025693 & 9.18$\pm$0.12 & 1.07$\pm$0.13 & 0.15$\pm$0.02 & ... \\
5100997733 & 10.07$\pm$0.08 & 1.72$\pm$0.14 & 0.23$\pm$0.03 & ... \\
511228062 & 9.56$\pm$0.06 & 1.53$\pm$0.05 & 0.19$\pm$0.01 & ... \\
5101001604 & 9.03$\pm$0.05 & 2.00$\pm$0.03 & 0.22$\pm$0.01 & ... \\
510996058 & 9.11$\pm$0.07 & 1.88$\pm$0.06 & 0.17$\pm$0.01 & 0.31$\pm$0.31 \\
511001501 & 9.21$\pm$0.09 & 1.41$\pm$0.09 & 0.13$\pm$0.01 & 0.34$\pm$0.06 \\
530053714 & 9.72$\pm$0.05 & 0.97$\pm$0.05 & 0.09$\pm$0.01 & 0.12$\pm$0.25 \\
\hline\hline
    \end{tabular}
    \tablefoot{\tablefootmark{(a)}{{Color excess based on Balmer decrement as explained in \ref{sec:fluxes}.}}}
    \end{scriptsize}
\end{table}

\subsection{Rest-frame optical spectroscopy}
We describe the NIR spectroscopy available which was used to select the final sample to be analyzed in this paper. In summary, the rest-optical spectra for VANDELS targets are obtained from the NIRVANDELS survey \citep{Cullen2021} using Keck/MOSFIRE \citep{McLean2012}. While for the VUDS targets, we obtained the rest-optical spectra from different instruments. First, from the public MOSDEF survey \citep{Kriek2015} using Keck/MOSFIRE. We also considered NIR spectroscopy observations with X-Shooter \citep{Vernet2011} for a subsample selected from \cite{Amorin2017}. Finally, we also considered NIR spectroscopy observations with Magellan/FIRE \citep{Simcoe2010}. More details on the selection are presented in the following subsections.  

\subsubsection{MOSFIRE spectroscopy}

From the parent sample of CIII] emitters selected in VANDELS survey with reliable spectroscopic redshift, we cross-match with the catalog of sources in the NIRVANDELS survey \citep{Cullen2021}. This is a Keck/MOSFIRE VANDELS follow-up survey for 35 sources at 2.95 $\leq z \leq$ 3.8 and 10 sources at 2.09 $\leq z\leq$ 2.61. {The details of the observations and the data reduction are described in \cite{Cullen2021}.} The slitwidth was 0.7'', yielding a spectral resolution of $\sim$ 3650 in H and $\sim$3600 in K band, respectively. 19 galaxies in NIRVANDELS {show S/N $> 3$ in CIII]}. Two objects were discarded because they do not show optical emission lines, particularly [OIII]. Our final C3-VANDELS sample contains 17 galaxies in both CDFS and UDS fields.

To build our sample of VUDS galaxies, we use the publicly available MOSDEF survey \citep{Kriek2015}
which is also a Keck/MOSFIRE survey that comprises NIR spectra of $\sim$1500 K-band selected galaxies targeted to lie within three distinct redshift intervals 1.37 $\leq z\leq$1.70, 2.09 $\leq z\leq$2.61 and 2.95 $\leq z\leq$3.80. {The details of the observations and the data reduction are described in \cite{Kriek2015}.}{ The 0.7''-width slit results in a spectral resolution of R= 3000 for J band. For H and K bands, the resolution is the same as in the NIRVANDELS survey}. The MOSDEF survey covers well-studied HST extragalactic legacy fields by the CANDELS and 3D-HST surveys: AEGIS, COSMOS, GOODS-N, GOODS-S, and UDS. We crossmatch the MOSDEF and our VUDS parent sample catalogs to increase the number of galaxies. We prioritized galaxies with S/N CIII]$>3$ and with [OIII] detected. We ended up with a subsample of 11 galaxies in the COSMOS and ECDFS fields. 

\subsubsection{X-shooter Spectroscopy {and data reduction}}
Four additional VUDS galaxies were observed in a follow-up program (Program: 0101.B-0779, PI: Amor\'in R.) with VLT/X-shooter \citep{Vernet2011} {which is a wide-band echelle spectrograph where two dichroic split the light into three arms (UVB, VIS, and NIR), and simultaneously exposures. In this paper, we only used the spectra from the NIR arm}. These galaxies were selected from \cite{Amorin2017} for their intense CIII] emission in deep VUDS spectra. Due to their redshift (at $z\sim 2.4$  and $\sim 3.4$), bright optical emission lines fall within good transmission windows. Observations were done in echelle mode from 2018 May to 2019 March with 900s integrations
using 1.0,0.9,0.9'' (UVB/VIS/NIR) slits for a resolution R=5400,8900, 5600, respectively {in seeing conditions of $\sim$0.6 arcsec}. Each observing block (OB) is $\sim$ 3600s of integration time, and each galaxy has 2-3 OBs. One of the galaxies (VUDS 5101421970) has an additional observing block of 3600s from the program 0103.B-0446(A) (PI: Nakajima, K.) to complete a total of 3 hours of exposure. The NIR region of the X-shooter spans the combined Y, J, H, and K region from 1024–2048 nm. The reduction of each OB is performed using the EsoReflex \citep{Freudling2013} X-shooter pipeline \citep{Modigliani2010},  which provides merged, 2D NIR, visible, and UVB spectra. {With the pipeline, we performed dark subtraction, flat-fielding, flexure correction and 2D mapping, wavelength calibration, and flux calibration with standard stars. } This is the same methodology used in \cite{Matthee2021} where 3 out of 4 of our X-shooter sample are analyzed. To combine the OBs, we use the \texttt{IRAF} \citep{Tody1986} task \texttt{imcombine} with median and $\sigma$-clipping. To extract the 1D spectrum, we use the trace by [OIII]$\lambda\lambda$4959,5007+H$\beta$ that is clearly detected, but the continuum is undetected in any galaxies. 

\subsubsection{FIRE Spectroscopy {and data reduction}}
A follow-up with Magellan/Folded Port Infrared Echellette \citep[FIRE, ][]{Simcoe2010} was carried out for 3 galaxies selected from the VUDS parent sample to have CIII] detection and at $z<3$ so that bright emission lines do not overlap with strong skylines. Observations were conducted on 2022 April/May. FIRE was used in the high-resolution echelle mode. The observations were conducted as follows: the J-band acquisition camera was used to locate a nearby star from which a blind offset was applied to position the science target in the slit. The slits used were either 0.75'', or 1.0'' in width, depending on the seeing {($<$1'')}, {yielding a spectral resolution of R$\sim$5200.} The slits were oriented at the parallactic angle to minimize differential atmospheric refraction. Exposure times of 900s were used for ABBA dither sequences with total integrations ranging from 3 to 4 hr. The readouts were performed with the Sample Up The Ramp mode to minimize overheads. For each science target, one A0V star was observed at a similar airmass for telluric correction. Data were reduced using the publicly available pipeline\footnote{\url{https://github.com/rasimcoe/FIREHOSE}} developed by the instrument team. Unfortunately, the bright emission lines in this subsample were affected by sky emission lines which preclude the kinematic analysis explained in the following sections. {We use this subsample to estimate the flux of emission lines, which are included in the chemical analysis, and to estimate EWs of bright observed lines.} For the VUDS targets, our final C3-VUDS sample is made of 18 galaxies. 

In summary, our final sample combining C3-VANDELS and C3-VUDS samples with different NIR instruments contains 35 CIII] emitters, whose physical properties are described in the following section. {In Table \ref{tab:coordinates}, the coordinates and spectroscopic redshifts based on CIII] of the individual galaxies in our sample are listed with notes on which NIR instrument was used.}

\section{Physical parameters of the final sample}\label{sec:properties}
\subsection{Spectral Energy Distribution (SED) modeling}\label{sec:sed}

\begin{figure}[t!]
    \centering
    \includegraphics[width=\columnwidth]{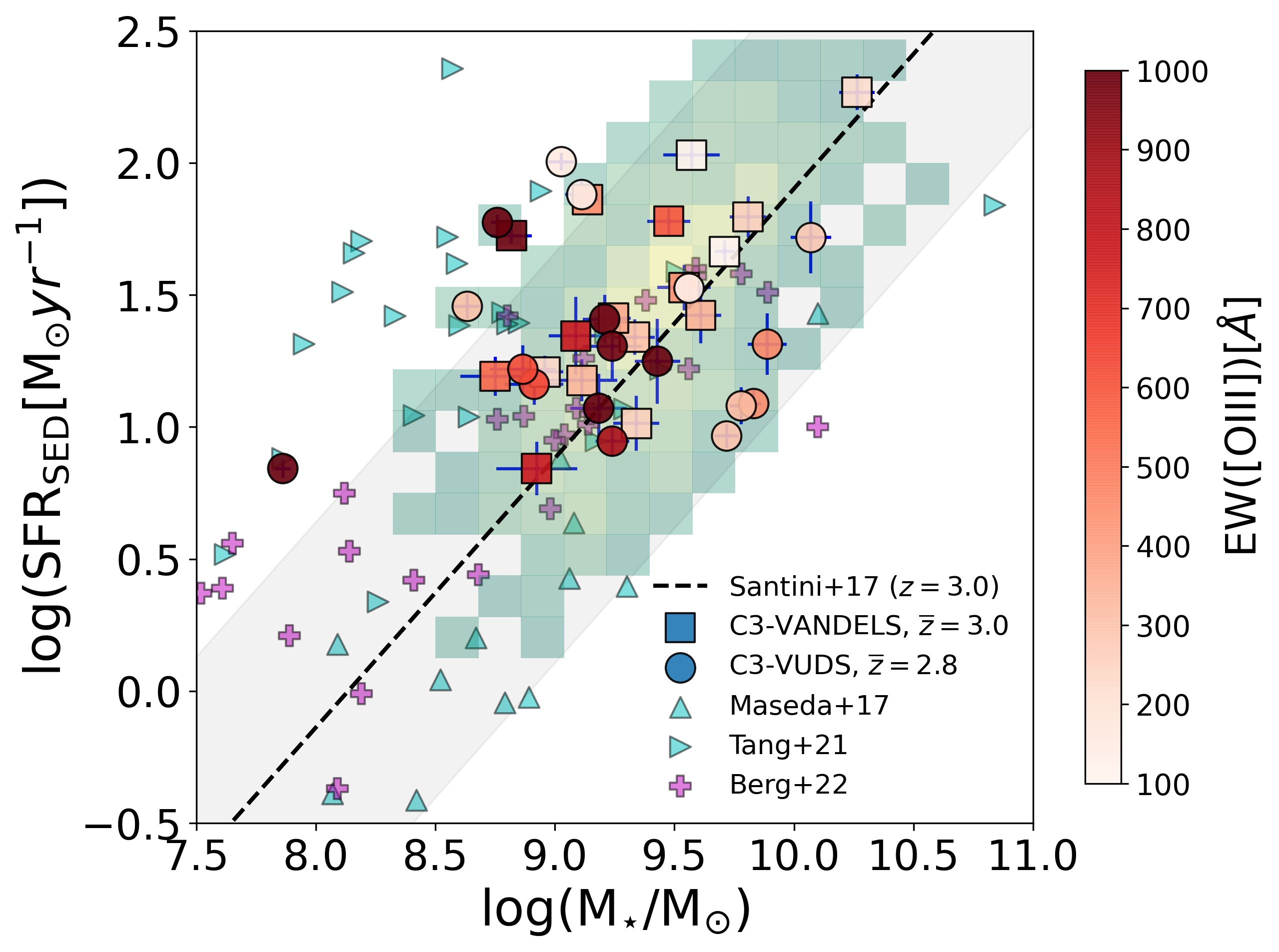}
    \caption{Sample distributed along the star-forming main-sequence at $z\sim3$ color-coded by EW([OIII]). The 2D histogram corresponds to the VANDELS parent sample at the same redshift range, while the black dashed line is the main sequence according to \cite{Santini2017}. The magenta crosses, and cyan triangles are reference samples at low \citep{Berg2022} and intermediate redshifts \citep{Maseda2017,Tang2021}, respectively (see Section \ref{sec:sed}).}
    \label{fig:MS}
\end{figure}

\begin{table*}[t!]
\caption{Observed fluxes and rest-frame EWs of the rest-UV emission lines of our sample}
\label{tab:UVflux}
    \centering
    \begin{scriptsize}
    \begin{tabular}{|c|c|c|c|c|c|c|c|c|c|c}\hline\hline
ID&Ly$\alpha$\tablefootmark{a}&CIV\tablefootmark{a}&HeII$\lambda$1640\tablefootmark{a}&OIII]$\lambda$1666\tablefootmark{a}&CIII]\tablefootmark{a}&EW(Ly$\alpha$)\tablefootmark{b}&EW(CIII])\tablefootmark{b}\\
\hline\multicolumn{8}{|l|}{C3-VANDELS sample}\\ \hline
UDS020394 &12.11$\pm$1.15 &1.41$\pm$0.41 &$<$0.74 &1.25$\pm$0.37 &4.13$\pm$0.64 &23.7$\pm$2.3 &13.6$\pm$2.1 \\
CDFS020954 &4.24$\pm$1.5 &0.61$\pm$0.26 &1.12$\pm$0.5 &1.42$\pm$0.3 &4.25$\pm$0.77 &19.8$\pm$7.0 &9.7$\pm$1.8 \\
CDFS023527 &120.45$\pm$4.62 &$<$0.94 &1.92$\pm$0.29 &2.38$\pm$0.44 &8.83$\pm$1.33 &56.6$\pm$2.2 &6.9$\pm$1.0 \\
UDS021601 &8.7$\pm$1.27 &$<$0.5 &0.67$\pm$0.31 &$<$0.58 &2.73$\pm$0.73 &12.5$\pm$1.8 &6.2$\pm$1.6 \\
CDFS022563 &29.88$\pm$4.73 &$<$0.5 &0.95$\pm$0.38 &0.73$\pm$0.21 &3.13$\pm$0.47 &19.8$\pm$3.1 &6.0$\pm$0.9 \\
UDS022487 &8.61$\pm$2.76 &$<$0.76 &$<$0.75 &2.25$\pm$0.44 &3.77$\pm$0.84 &8.5$\pm$2.7 &4.4$\pm$1.0 \\
CDFS015347 &13.04$\pm$1.47 &$<$0.6 &1.44$\pm$0.38 &1.33$\pm$0.28 &2.38$\pm$0.93 &37.2$\pm$4.2 &4.1$\pm$1.6 \\
CDFS019276 &$<$2.69 &1.66$\pm$0.65 &$<$1.8 &$<$1.11 &7.17$\pm$2.03 &$<$1.1 &3.6$\pm$1.0 \\
UDS020928 &$<$2.24 &1.83$\pm$0.47 &0.78$\pm$0.31 &0.57$\pm$0.28 &3.05$\pm$0.68 &$<$11.4 &3.1$\pm$0.7 \\
CDFS019946 &... &$<$1.49 &$<$1.2 &$<$0.75 &3.02$\pm$0.53 &... &3.0$\pm$0.5 \\
UDS020437 &$<$4.46 &$<$1.84 &$<$0.81 &1.91$\pm$0.58 &3.82$\pm$1.39 &$<$2.3 &3.0$\pm$1.1 \\
CDFS018182 &... &3.94$\pm$1.56 &5.77$\pm$1.07 &1.82$\pm$0.78 &8.41$\pm$1.31 &... &2.9$\pm$0.5 \\
CDFS018882 &-0.88$\pm$0.33 &0.75$\pm$0.21 &$<$0.45 &0.8$\pm$0.23 &2.77$\pm$0.58 &-3.9$\pm$1.5 &2.7$\pm$0.6 \\
CDFS025828 &10.86$\pm$3.08 &$<$1.1 &2.11$\pm$0.72 &$<$1.79 &4.98$\pm$1.15 &4.7$\pm$1.3 &2.5$\pm$0.6 \\
CDFS022799 &... &$<$2.23 &2.74$\pm$1.21 &$<$2.78 &8.95$\pm$1.1 &... &2.5$\pm$0.3 \\
UDS021398 &... &$<$1.39 &1.66$\pm$0.79 &$<$1.14 &2.95$\pm$0.64 &... &1.7$\pm$0.4 \\
UDS015872 &... &$<$1.92 &1.82$\pm$0.69 &$<$0.65 &2.7$\pm$0.66 &... &1.2$\pm$0.3 \\
\hline\multicolumn{8}{|l|}{C3-VUDS sample}\\ \hline
5100998761 &334.39$\pm$8.23 &7.08$\pm$1.33 &5.91$\pm$1.66 &5.02$\pm$0.75 &18.53$\pm$1.56 &151.5$\pm$3.7 &15.3$\pm$1.3 \\
5101444192 &184.78$\pm$2.78 &3.98$\pm$0.44 &2.4$\pm$0.82 &3.02$\pm$0.46 &10.22$\pm$1.05 &327.5$\pm$4.9 &13.1$\pm$1.3 \\
510994594 &103.4$\pm$3.24 &1.92$\pm$0.69 &4.22$\pm$0.76 &$<$0.46 &8.68$\pm$0.92 &118.2$\pm$3.7 &11.5$\pm$1.2 \\
5101421970 &715.43$\pm$21.66 &24.77$\pm$2.91 &11.76$\pm$1.59 &5.39$\pm$1.37 &22.83$\pm$2.11 &243.8$\pm$7.4 &10.6$\pm$1.0 \\
510838687 &19.16$\pm$2.33 &2.71$\pm$1.24 &$<$2.44 &3.69$\pm$0.87 &11.18$\pm$1.52 &23.8$\pm$2.9 &10.2$\pm$1.4 \\
5100556178 &266.96$\pm$5.41 &$<$1.48 &3.36$\pm$0.78 &$<$0.94 &8.88$\pm$0.99 &89.3$\pm$1.8 &8.7$\pm$1.0 \\
511229433 &47.41$\pm$2.03 &0.81$\pm$0.39 &4.61$\pm$0.45 &2.01$\pm$0.39 &8.8$\pm$1.28 &55.9$\pm$2.4 &7.8$\pm$1.1 \\
511245444 &81.83$\pm$2.15 &4.47$\pm$0.88 &1.25$\pm$0.42 &1.61$\pm$0.32 &7.34$\pm$1.96 &67.9$\pm$1.8 &6.2$\pm$1.7 \\
530048433 &165.55$\pm$8.43 &5.82$\pm$2.28 &2.88$\pm$1.23 &2.01$\pm$0.96 &13.3$\pm$2.29 &40.7$\pm$2.1 &5.2$\pm$0.9 \\
510583858 &126.63$\pm$5.55 &3.96$\pm$1.56 &11.56$\pm$3.69 &2.17$\pm$0.58 &7.03$\pm$1.4 &102.6$\pm$4.5 &4.8$\pm$1.0 \\
511451385 &62.34$\pm$3.72 &$<$1.91 &$<$1.93 &1.41$\pm$0.38 &5.28$\pm$1.36 &78.2$\pm$4.7 &4.7$\pm$1.2 \\
511025693 &15.24$\pm$2.22 &$<$1.71 &3.27$\pm$0.86 &2.22$\pm$0.63 &6.92$\pm$1.16 &7.3$\pm$1.1 &4.4$\pm$0.7 \\
5100997733 &7.71$\pm$2.88 &$<$1.68 &$<$2.26 &$<$1.38 &5.35$\pm$1.41 &4.0$\pm$1.5 &4.3$\pm$1.1 \\
511228062 &$<$1.34 &$<$1.23 &$<$0.74 &$<$0.45 &3.97$\pm$1.56 &$<$1.6 &3.9$\pm$1.5 \\
5101001604 &1.93$\pm$0.81 &$<$0.91 &$<$1.37 &$<$1.17 &3.46$\pm$1.56 &9.2$\pm$3.8 &3.4$\pm$1.5 \\
510996058 &18.45$\pm$2.31 &$<$2.19 &$<$1.05 &$<$0.65 &1.28$\pm$0.3 &18.8$\pm$2.4 &2.2$\pm$0.5 \\
511001501 &$<$4.67 &$<$1.53 &$<$1.28 &$<$1.01 &4.59$\pm$0.88 &$<$1.6 &2.1$\pm$0.4 \\
530053714 &$<$2.46 &$<$3.31 &3.17$\pm$1.16 &1.91$\pm$0.73 &3.35$\pm$1.22 &$<$1.2 &1.5$\pm$0.6 \\
\hline\hline
    \end{tabular}
    \tablefoot{
    \tablefoottext{a}{Flux in units $10^{-18}$ erg s$^{-1}$ cm$^{-2}$}\tablefoottext{b}{Rest-frame equivalent width in units \r{A}}}
    \end{scriptsize}
\end{table*}

\begin{table*}[]
    \caption{Observed fluxes and rest-frame EWs of the rest-optical emission lines of our sample}
    \label{tab:opticalflux}
    \centering
    \begin{scriptsize}
    \begin{tabular}{|c|c|c|c|c|c|c|c|}\hline\hline
    ID&[OII]$\lambda$3727\tablefootmark{a}&[OII]$\lambda$3729\tablefootmark{a}&H$\beta$\tablefootmark{a}&[OIII]$\lambda$5007\tablefootmark{a}&H$\alpha$\tablefootmark{a}&EW(H$\beta$)\tablefootmark{b}&EW([OIII]$\lambda$5007)\tablefootmark{b}\\
\hline\multicolumn{8}{|l|}{C3-VANDELS sample}\\ \hline
UDS020394 &$<$3.4 &3.73$\pm$1.7 &9.06$\pm$1.37 &60.77$\pm$3.2 &... &112.8$\pm$17.1 &779.0$\pm$41.1 \\
CDFS020954 &11.22$\pm$1.57 &9.51$\pm$1.5 &18.15$\pm$0.97 &112.83$\pm$12.9 &... &153.5$\pm$8.2 &984.4$\pm$112.6 \\
CDFS023527 &8.94$\pm$1.57 &$<$13.6 &... &122.12$\pm$21.4 &... &... &437.6$\pm$76.7 \\
UDS021601 &... &8.46$\pm$1.87 &6.1$\pm$1.44 &32.2$\pm$1.2 &... &48.5$\pm$11.5 &260.3$\pm$9.7 \\
CDFS022563 &3.78$\pm$1.12 &7.01$\pm$1.12 &5.67$\pm$1.1 &43.17$\pm$1.61 &... &68.9$\pm$13.4 &551.4$\pm$20.6 \\
UDS022487 &17.66$\pm$1.81 &18.21$\pm$1.82 &13.32$\pm$1.4 &122.76$\pm$24.63 &... &82.4$\pm$8.6 &786.9$\pm$157.9 \\
CDFS015347 &3.03$\pm$0.86 &$<$2.44 &3.35$\pm$0.92 &35.4$\pm$1.66 &... &30.9$\pm$8.5 &336.5$\pm$15.8 \\
CDFS019276 &22.7$\pm$3.61 &30.3$\pm$3.6 &37.17$\pm$2.15 &151.27$\pm$24.9 &... &53.1$\pm$3.1 &222.8$\pm$36.7 \\
UDS020928 &13.49$\pm$1.8 &19.33$\pm$2.06 &13.96$\pm$2.42 &136.21$\pm$13.19 &... &59.8$\pm$10.4 &601.1$\pm$58.2 \\
CDFS019946 &... &... &6.77$\pm$1.99 &48.73$\pm$7.83 &31.68$\pm$1.94 &31.5$\pm$9.3 &234.4$\pm$37.7 \\
UDS020437 &18.54$\pm$3.04 &26.44$\pm$3.05 &15.1$\pm$1.0 &105.11$\pm$10.72 &... &48.9$\pm$3.2 &348.3$\pm$35.5 \\
CDFS018182 &... &... &26.53$\pm$1.35 &157.75$\pm$11.43 &88.82$\pm$3.24 &49.8$\pm$2.5 &305.6$\pm$22.1 \\
CDFS018882 &7.67$\pm$1.25 &7.53$\pm$1.24 &7.53$\pm$0.88 &60.3$\pm$3.03 &... &31.4$\pm$3.7 &257.9$\pm$12.9 \\
CDFS025828 &11.67$\pm$2.25 &15.87$\pm$2.42 &4.47$\pm$1.62 &38.66$\pm$1.29 &... &11.9$\pm$4.3 &108.2$\pm$3.6 \\
CDFS022799 &... &... &43.35$\pm$2.07 &235.67$\pm$35.94 &129.48$\pm$2.95 &79.6$\pm$3.8 &451.6$\pm$68.9 \\
UDS021398 &... &... &16.52$\pm$5.61 &125.81$\pm$11.47 &74.42$\pm$3.97 &47.8$\pm$16.2 &377.7$\pm$34.4 \\
UDS015872 &... &... &16.7$\pm$1.64 &66.21$\pm$7.38 &62.61$\pm$2.68 &25.4$\pm$2.5 &102.2$\pm$11.4 \\
\hline\multicolumn{8}{|l|}{C3-VUDS sample}\\ \hline
5100998761 &3.78$\pm$1.89 &$<$5.13 &10.66$\pm$1.64 &103.76$\pm$6.6 &35.09$\pm$6.2 &153.1$\pm$23.6 &1549.8$\pm$98.5 \\
5101444192 &5.66$\pm$2.83 &... &23.19$\pm$4.1 &241.86$\pm$4.59 &... &144.7$\pm$25.5 &1560.8$\pm$29.6 \\
510994594 &$<$9.93 &... &13.04$\pm$1.4 &122.7$\pm$8.85 &... &49.5$\pm$5.3 &463.0$\pm$33.4 \\
5101421970 &8.24$\pm$3.37 &7.91$\pm$3.37 &25.89$\pm$1.57 &168.87$\pm$16.98 &89.04$\pm$7.73 &49.7$\pm$3.0 &322.4$\pm$32.4 \\
510838687 &40.52$\pm$8.34 &$<$16.7 &$<$34.38 &125.5$\pm$3.1 &30.18$\pm$5.72 &$<$6.7 &630.6$\pm$15.6 \\
5100556178 &$<$11.95 &25.47$\pm$5.98 &$<$24.09 &126.42$\pm$4.44 &93.04$\pm$12.51 &$<$4.9 &642.8$\pm$22.6 \\
511229433 &20.92$\pm$6.06 &23.49$\pm$6.06 &39.02$\pm$2.06 &271.06$\pm$14.51 &... &245.1$\pm$13.0 &1715.7$\pm$91.8 \\
511245444 &... &... &... &201.06$\pm$45.9 &... &... &978.9$\pm$223.5 \\
530048433 &20.35$\pm$2.89 &39.34$\pm$6.38 &37.35$\pm$10.95 &243.88$\pm$5.66 &137.42$\pm$3.94 &130.4$\pm$38.2 &870.6$\pm$20.2 \\
510583858 &17.77$\pm$8.05 &$<$16.1 &33.75$\pm$2.33 &233.86$\pm$22.94 &131.38$\pm$11.18 &64.9$\pm$4.5 &445.5$\pm$43.7 \\
511451385 &10.2$\pm$5.01 &$<$10.02 &$<$14.6 &53.19$\pm$1.17 &$<$67.08 &$<$1.9 &304.0$\pm$6.7 \\
511025693 &16.4$\pm$2.38 &16.96$\pm$2.4 &34.89$\pm$1.25 &239.5$\pm$18.03 &... &146.1$\pm$5.2 &1002.9$\pm$75.5 \\
5100997733 &$<$4.71 &14.79$\pm$2.33 &11.58$\pm$2.22 &96.32$\pm$6.97 &... &33.9$\pm$6.5 &279.9$\pm$20.3 \\
511228062 &22.09$\pm$2.15 &17.35$\pm$2.15 &27.23$\pm$2.15 &107.34$\pm$7.94 &... &34.1$\pm$2.7 &134.3$\pm$9.9 \\
5101001604 &$<$3.52 &6.41$\pm$1.76 &8.37$\pm$1.01 &33.14$\pm$1.12 &... &38.3$\pm$4.6 &155.9$\pm$5.3 \\
510996058 &10.7$\pm$1.99 &9.12$\pm$1.99 &10.38$\pm$2.8 &26.61$\pm$2.07 &39.28$\pm$5.86 &50.6$\pm$13.6 &133.4$\pm$10.4 \\
511001501 &42.27$\pm$3.24 &53.89$\pm$3.24 &34.88$\pm$1.89 &185.02$\pm$31.48 &136.42$\pm$2.87 &197.6$\pm$10.7 &1060.6$\pm$180.4 \\
530053714 &10.59$\pm$3.43 &$<$6.86 &27.96$\pm$6.69 &113.72$\pm$2.05 &87.52$\pm$5.02 &74.0$\pm$17.7 &299.5$\pm$5.4 \\
\hline\hline
    \end{tabular}
    \tablefoot{
    \tablefoottext{a}{Flux in units $10^{-18}$ erg s$^{-1}$ cm$^{-2}$}\tablefoottext{b}{Rest-frame equivalent width in units \r{A}}}
    \end{scriptsize}
\end{table*}

We perform a SED fitting to the entire final sample using BAGPIPES \citep{Carnall2018}. The photometric catalog for the C3-VANDELS sample comes from the CANDELS team \citep{Galametz2013,Guo2013} and is the same used for the VANDELS team for targets with HST imaging that are described in \citep{McLure2018,Garilli2021}.{ For the galaxies in the CDFS field, we used the following bands: U-VIMOS, HST: F435W, F606W, F775W, F814W, F850LP, F098M, F105W, F125W, F160W, Ks-ISAAC, Ks-HAWKI, 3.6$\mu$m-IRAC, 4.5$\mu$m-IRAC. For the galaxies in the UDS field, we used the bands: U-CFHT, Subaru:B, V, R, i, z, HST: F606W, F125W, F160W, HAWKI: Y, Ks, WFCAM: J, H, K, 3.6$\mu$m-IRAC, 4.5$\mu$m-IRAC.}

For the C3-VUDS sample, we use the photometric catalog from COSMOS2015 \citep{Laigle2016} for targets in the COSMOS field, while for galaxies in the ECDFS field, we used the photometric catalog used by the VUDS team \citep[e.g.][]{Calabro2017,Ribeiro2017,Lemaux2022} that is obtained from \cite{Cardamone2010,Grogin2011}. {For the galaxies in the COSMOS field, we considered the following bands: u-CFHT, VISTA: Y, J, H, Ks, Subaru: B, V, R, i, z, Y, 3.6$\mu$m-IRAC, 4.5$\mu$m-IRAC. For the galaxies in the ECDFS field, we considered the following bands: MUSYC: U38, U, B, V, R, I, z, J, H, K, Subaru: IA427, IA464, IA484, IA505, IA527, IA574, IA624, IA679, IA709, IA738, IA767, IA827, 3.6$\mu$m-IRAC, 4.5$\mu$m-IRAC.} 

We assume an {exponentially declining} $\tau$-model for the SFH. {We obtain values for the timescale $\tau>5$Gyr, which implies a constant SFH at the redshift range covered by our sample}. For the metallicity, we allow it to vary up to 0.25Z$_{\odot}$, in agreement with the stellar metallicities observed in star-forming galaxies at $z\sim 3$ \citep[e.g.][]{Cullen2019,Calabro2021,Llerena2022}. We include a nebular component {which includes emission lines and nebular continuum emission} in the model given the emission lines that are observed in these galaxies. The nebular component  depends on the ionization parameter ($\log U$) and can vary between $-3$ and $-2$ in our model. We use Calzetti law for the dust reddening determination \citep{Calzetti2000} {with the total extinction $A_V$ free to vary between 0 and 2 mag.} We constrain the ages from 100Myr up to 1Gyr. Our SED fitting leads to differences of $\sim 0.3$ dex towards lower stellar mass and $\sim0.1$ dex towards higher SFR compared with a model with similar SFH but with fixed solar metallicity and without a nebular model. {The larger effect on the SFR offset is the assumption of subsolar metallicity, while the larger effect on the stellar mass offset is due to the inclusion of the nebular model.} The stellar masses and SFRs obtained with the SED fitting are displayed in Fig. \ref{fig:MS} {and reported in Table \ref{tab:SED}}. {We highlight that the SFR is calculated over a timescale of 100Myr}. Our sample ranges $\sim 2.4$ dex in stellar mass from 10$^{7.9}$ to 10$^{10.3}$M$_{\odot}$ and $\sim1.4$ dex in SFR from 6.9 to 185 M$_{\odot}$ yr$^{-1}$. The physical parameters are reported in Table \ref{tab:SED}, including the color excess (E(B-V)$_{\rm SED}$) from the SED model, which ranges from 0.04 to 0.25 mag. {The photometry used, and the resulting SED model are displayed in Fig. \ref{fig:sedC3vandels} and \ref{fig:sedC3vuds} for C3-VANDELS and C3-VUDS samples, respectively.} We perform a SED fitting following the same constraints to our VANDELS parent sample (2D histogram in Fig. \ref{fig:MS}), and we note that they follow the main-sequence (MS) at $z\sim 3$ according to \cite{Santini2017} with stellar mass corrected by a factor of $0.6$ \citep{Madau2014} by adopting the same assumed IMF in this paper.

Regarding our combined samples, we note that most are distributed along the MS, with 7 (5 of which are VUDS galaxies) slightly above 3 times the observed 0.37 dex scatter of the relation. We compare our sample with other works at similar intermediate redshifts \citep{Maseda2017,Tang2021} and with local metal-poor SFGs \citep{Berg2022} that will be used in the following sections. 
\subsection{{Emission-line} fluxes and EWs}\label{sec:fluxes}

\begin{figure*}[t!]
    \centering
    \includegraphics[width=0.95\textwidth]{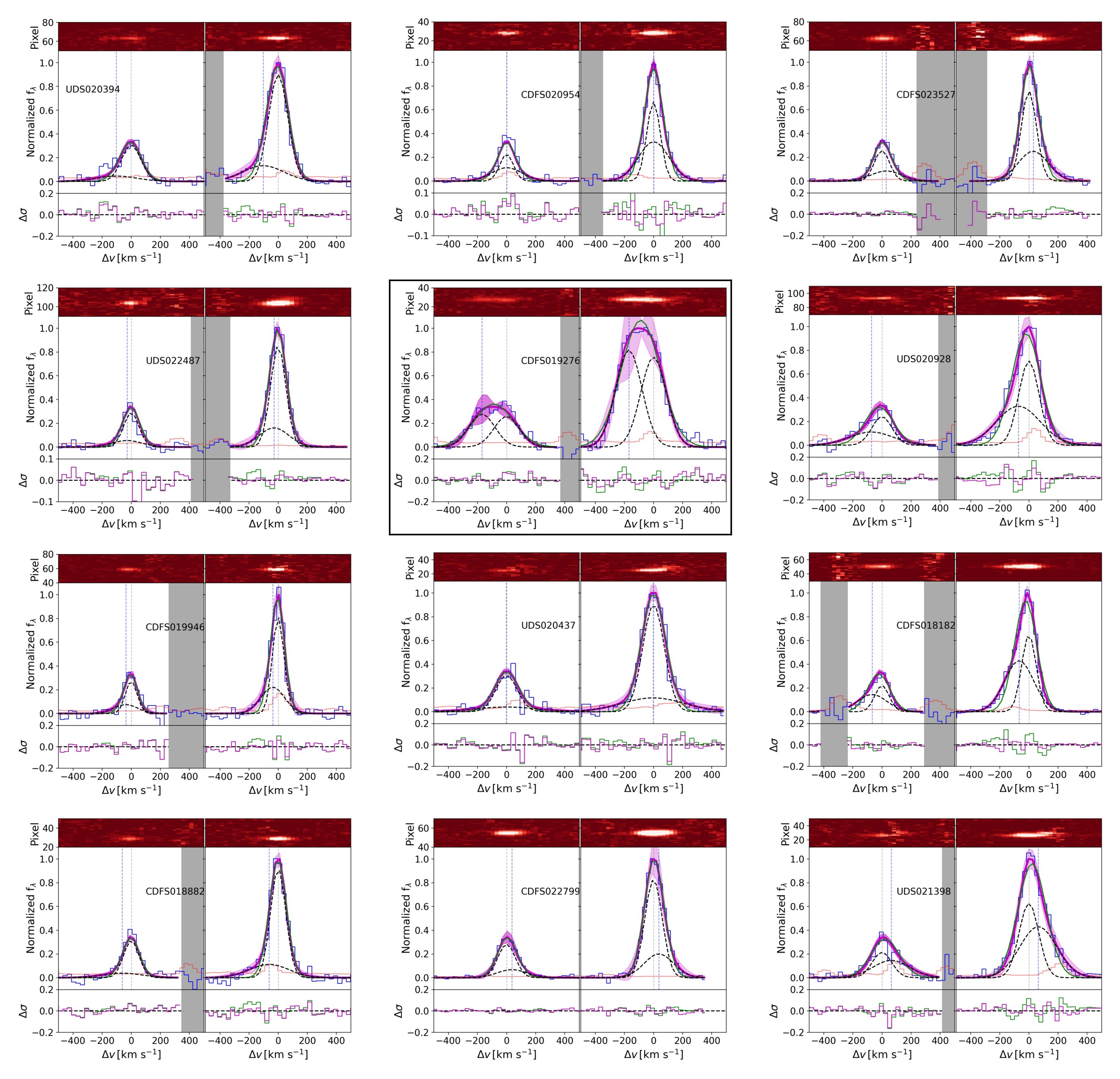}
    \caption{Best-fit of the [OIII] profiles from the C3-VANDELS sample with $\Delta$BIC$>2$. The panels show single galaxies ordered by EW(CIII]) from left-right and top-bottom. \textit{In each panel:} \textit{Top}: The 2D spectrum showing the detected lines. \textit{Middle:} Models for [OIII]$\lambda$4959 (left) [OIII]$\lambda$5007 (right). The blue line is the observed spectrum, while the red line is the error spectrum. The Gaussian lines are normalized to the intensity peak of [OIII]$\lambda$5007. The black dashed lines are the narrow and broad components, and the magenta line is the global fit considering both components. The magenta-shaded region is the 3$\sigma$ uncertainty of the fit. The green line is the single-gaussian fit. The vertical gray line marks the systemic velocity traced by the peak intensity of the narrow component. Meanwhile, the vertical blue line marks the peak intensity of the broad component.  \textit{Bottom:} The residuals ($\Delta\sigma$) for each model are shown with the same colors as described. The gray-shaded regions are masked regions due to sky residuals. The galaxies in the black square show two narrow components.}
    \label{fig:deltabic>2}
\end{figure*}

\begin{figure*}[t!]
    \centering
    \includegraphics[width=0.95\textwidth]{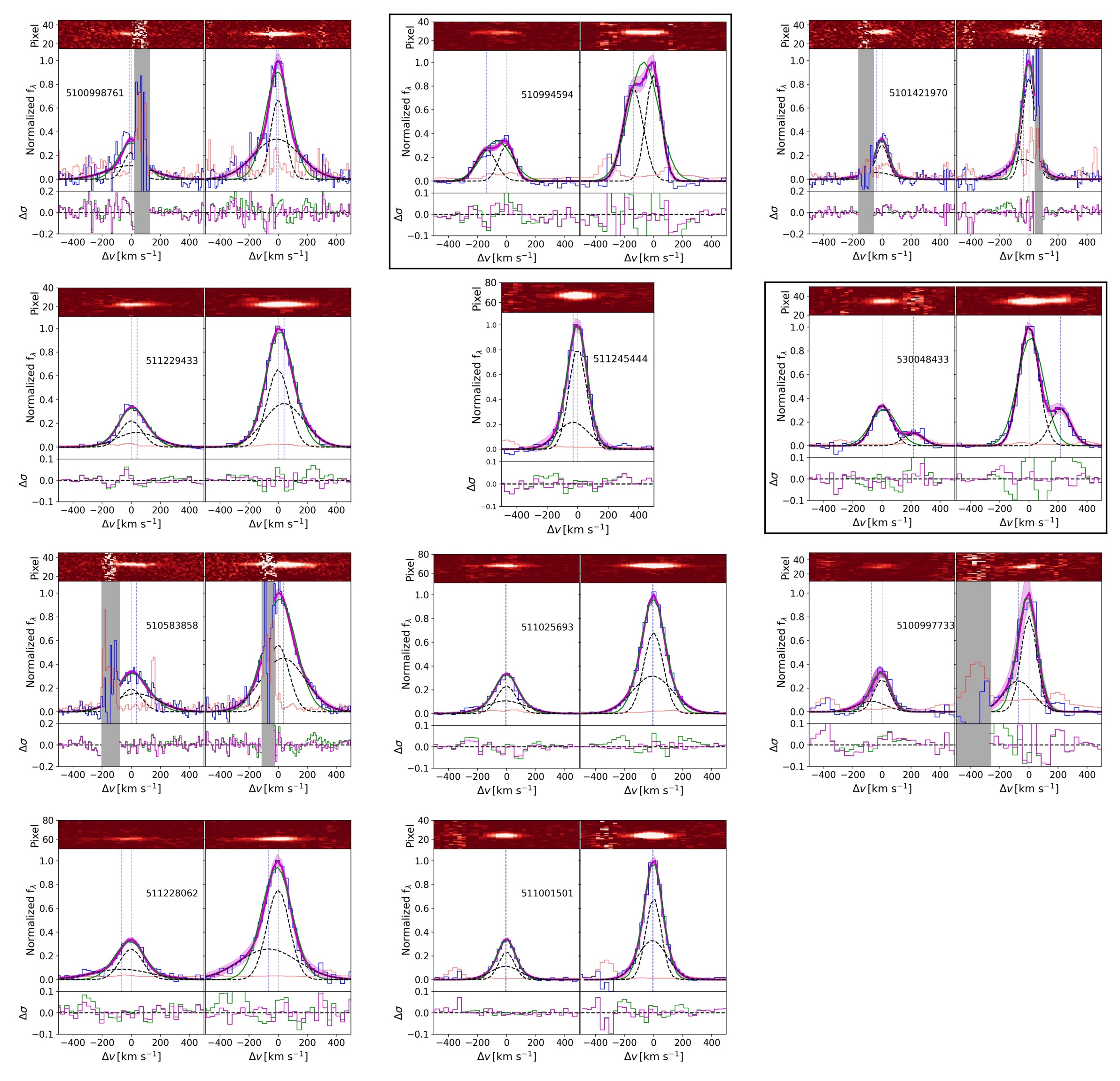}
    \caption{Same as in Fig. \ref{fig:deltabic>2} but for the C3-VUDS sample with $\Delta\rm{BIC}>2$.}
    \label{fig:deltabic>2vuds}
\end{figure*}

\begin{table*}[]
\caption{Kinematic properties of the ionized gas based on [OIII] modeling}
\label{tab:kine}
    \centering
    \begin{scriptsize}
    \begin{tabular}{|c|c|c|c|c|c|c|c|c|c|c|}\hline\hline
    ID&$\sigma_N$\tablefootmark{a}&$\sigma_B$\tablefootmark{a}&$\Delta v_B$&f$_B$\tablefootmark{b}&$\Delta$BIC&r$_H$&$\Sigma_{SFR}$&$\dot{M_{out}}$&$\log\eta$\\
    &km s$^{-1}$&km s$^{-1}$&km s$^{-1}$&&&kpc&M$_\odot$ yr$^{-1}$ kpc$^{-2}$&M$_\odot$ yr$^{-1}$&dex\\
\hline\multicolumn{10}{|l|}{C3-VANDELS sample}\\ \hline
UDS020394 & 55.8$\pm$2.2 & 135.46$\pm$25.61 & -102.27$\pm$2.76 & 0.24$\pm$0.04 & 15.69 & 0.19$\pm$0.05 & 91.17$\pm$64.14 & 20.08$\pm$7.98 & -0.01$\pm$0.25 \\
CDFS020954 & 30.61$\pm$3.24 & 98.41$\pm$9.74 & 1.44$\pm$1.2 & 0.53$\pm$0.1 & 45.47 & 2.94$\pm$0.48 & 0.71$\pm$0.38 & 4.62$\pm$1.27 & -0.92$\pm$0.22 \\
CDFS023527 & 38.27$\pm$3.95 & 96.72$\pm$14.37 & 29.03$\pm$14.29 & 0.39$\pm$0.14 & 11.69 & 3.06$\pm$0.34 & ... & ... & ... \\
UDS021601 & 78.34$\pm$3.71 & ... & ... & 0.0$\pm$0.0 & -3.29 & 1.51$\pm$0.27 & 1.38$\pm$0.77 & ... & ... \\
CDFS022563 & 75.24$\pm$3.61 & ... & ... & 0.0$\pm$0.0 & -2.96 & 0.76$\pm$0.09 & 4.27$\pm$2.06 & ... & ... \\
UDS022487 & 39.02$\pm$3.65 & 95.38$\pm$22.64 & -28.53$\pm$22.38 & 0.27$\pm$0.15 & 4.58 & 1.59$\pm$0.12 & 1.72$\pm$0.79 & 2.21$\pm$1.37 & -1.09$\pm$0.33 \\
CDFS015347 & 72.95$\pm$4.4 & ... & ... & 0.0$\pm$0.0 & -2.52 & 0.37$\pm$0.05 & 14.6$\pm$7.96 & ... & ... \\
CDFS019276 & 80.31$\pm$8.39 & 82.76$\pm$8.76\tablefootmark{x} & -169.59$\pm$26.17 & 0.53$\pm$0.15 & 34.27 & 2.14$\pm$0.06 & 2.79$\pm$1.36 & 22.32$\pm$6.67 & -0.56$\pm$0.25 \\
UDS020928 & 60.09$\pm$4.3 & 145.78$\pm$9.67 & -72.53$\pm$15.9 & 0.5$\pm$0.08 & 46.37 & 1.36$\pm$0.19 & 2.21$\pm$1.15 & 10.94$\pm$3.15 & -0.37$\pm$0.23 \\
CDFS019946 & 18.67$\pm$3.0 & 72.98$\pm$10.86 & -37.17$\pm$17.53 & 0.36$\pm$0.13 & 13.92 & 1.59$\pm$0.15 & 0.84$\pm$0.46 & 1.37$\pm$0.68 & -0.99$\pm$0.31 \\
UDS020437 & 64.94$\pm$3.91 & 210.37$\pm$52.76 & -2.53$\pm$2.02 & 0.27$\pm$0.08 & 9.86 & 3.55$\pm$0.3 & 0.41$\pm$0.17 & 2.31$\pm$0.9 & -1.15$\pm$0.24 \\
CDFS018182 & 34.8$\pm$2.53 & 93.54$\pm$3.4 & -66.83$\pm$8.34 & 0.57$\pm$0.07 & 142.25 & 3.64$\pm$0.19 & 0.14$\pm$0.06 & 1.76$\pm$0.25 & -0.83$\pm$0.19 \\
CDFS018882 & 38.69$\pm$1.53 & 131.58$\pm$19.81 & -61.91$\pm$1.47 & 0.24$\pm$0.04 & 21.75 & 0.55$\pm$0.04 & 13.76$\pm$5.77 & 7.78$\pm$1.94 & -0.52$\pm$0.2 \\
CDFS025828 & 58.65$\pm$2.64 & ... & ... & 0.0$\pm$0.0 & -9.4 & 2.48$\pm$0.23 & 0.43$\pm$0.24 & ... & ... \\
CDFS022799 & 48.45$\pm$2.39 & 93.54$\pm$19.68 & 36.5$\pm$15.84 & 0.29$\pm$0.12 & 15.53 & 2.99$\pm$0.06 & 0.52$\pm$0.21 & 1.41$\pm$0.63 & -1.31$\pm$0.26 \\
UDS021398 & 57.33$\pm$5.16 & 121.47$\pm$11.32 & 64.37$\pm$6.62 & 0.56$\pm$0.08 & 7.57 & 2.21$\pm$0.16 & 0.68$\pm$0.38 & 6.0$\pm$2.31 & -0.54$\pm$0.29 \\
UDS015872 & 67.91$\pm$4.24 & ... & ... & 0.0$\pm$0.0 & 0.63 & 1.59$\pm$0.1 & 1.49$\pm$0.6 & ... & ... \\
\hline\multicolumn{10}{|l|}{C3-VUDS sample}\\ \hline
5100998761 & 45.11$\pm$3.41 & 145.77$\pm$9.99 & -10.85$\pm$6.69 & 0.63$\pm$0.06 & 95.07 & 0.25 & 10.94$\pm$5.2 & 14.56$\pm$3.18 & 0.51$\pm$0.21 \\
5101444192 & 67.05$\pm$2.68 & ... & ... & 0.0$\pm$0.0 & -4.64 & 0.16\tablefootmark{+} & 495.88$\pm$225.56 & ... & ... \\
510994594 & 39.8$\pm$3.08 & 58.13$\pm$5.15\tablefootmark{x} & -139.76$\pm$8.79 & 0.53$\pm$0.06 & 120.65 & 0.41 & 20.67$\pm$9.37 & 23.97$\pm$4.75 & 0.05$\pm$0.2 \\
5101421970 & 36.7$\pm$2.55 & 100.95$\pm$14.48 & -36.99$\pm$15.61 & 0.33$\pm$0.08 & 19.38 & 0.88\tablefootmark{+} & 4.71$\pm$2.07 & 4.84$\pm$1.44 & -0.68$\pm$0.21 \\
510838687 & 68.4$\pm$1.78 & ... & ... & 0.0$\pm$0.0 & -99.9 & 2.06\tablefootmark{+} & 0.52$\pm$0.24 & ... & ... \\
5100556178 & 102.02$\pm$2.32 & ... & ... & 0.0$\pm$0.0 & -99.9 & 0.53\tablefootmark{+} & 23.66$\pm$10.41 & ... & ... \\
511229433 & 66.73$\pm$2.15 & 133.06$\pm$5.44 & 39.32$\pm$2.12 & 0.51$\pm$0.05 & 83.19 & 0.91 & 12.88$\pm$5.57 & 36.84$\pm$5.52 & -0.26$\pm$0.18 \\
511245444 & 45.3$\pm$4.44 & 97.24$\pm$18.26 & -31.17$\pm$20.5 & 0.32$\pm$0.18 & 4.34 & 0.74 & ... & ... & ... \\
530048433 & 57.91$\pm$1.6 & 60.4$\pm$5.27\tablefootmark{x} & 216.22$\pm$5.44 & 0.24$\pm$0.02 & 247.42 & 1.14 & 4.67$\pm$2.39 & 5.81$\pm$1.86 & -0.82$\pm$0.25 \\
510583858 & 70.44$\pm$5.52 & 147.27$\pm$7.68 & 34.21$\pm$7.45 & 0.62$\pm$0.09 & 82.54 & 0.71\tablefootmark{+} & 7.22$\pm$3.53 & 27.85$\pm$5.49 & 0.09$\pm$0.21 \\
511451385 & 93.53$\pm$2.7 & ... & ... & 0.0$\pm$0.0 & -99.9 & 0.24\tablefootmark{+} & ... & ... & ... \\
511025693 & 56.55$\pm$2.48 & 132.1$\pm$7.66 & -6.84$\pm$3.68 & 0.49$\pm$0.06 & 113.15 & 0.84 & 13.71$\pm$6.01 & 29.49$\pm$5.27 & -0.31$\pm$0.19 \\
5100997733 & 42.17$\pm$2.86 & 93.54$\pm$15.86 & -72.62$\pm$-72.62 & 0.37$\pm$0.06 & 16.57 & 2.45 & 0.63$\pm$0.29 & 2.36$\pm$0.96 & -1.0$\pm$0.26 \\
511228062 & 69.14$\pm$3.92 & 196.84$\pm$15.86 & -66.15$\pm$16.49 & 0.47$\pm$0.06 & 68.67 & 0.64 & 22.25$\pm$9.85 & 57.25$\pm$11.52 & 0.01$\pm$0.19 \\
5101001604 & 36.05$\pm$5.07 & ... & ... & 0.0$\pm$0.0 & -11.88 & 0.95 & 5.29$\pm$2.31 & ... & ... \\
510996058 & 52.0$\pm$11.9 & ... & ... & 0.0$\pm$0.0 & -1.1 & 2.71 & 0.4$\pm$0.2 & ... & ... \\
511001501 & 38.92$\pm$3.82 & 92.89$\pm$10.82 & -7.75$\pm$5.2 & 0.48$\pm$0.14 & 27.75 & 0.95 & 4.75$\pm$2.39 & 7.95$\pm$2.72 & -0.53$\pm$0.25 \\
530053714 & 91.19$\pm$2.04 & ... & ... & 0.0$\pm$0.0 & -5.97 & 2.1 & 0.98$\pm$0.47 & ... & ... \\
\hline\hline
    \end{tabular}
    \tablefoot{\tablefootmark{a}{Intrinsic velocity dispersion of the narrow (N) and broad (B) component from the [OIII] line},\tablefootmark{b}{Broad-to-narrow flux ratio},\tablefootmark{+}{For this galaxy, we constrained the size using the i-band HST/F814W image},\tablefootmark{x}{Best-fit with two narrow components}}
    \end{scriptsize}
\end{table*}
\begin{figure}[t!]
    \centering
    \includegraphics[width=\columnwidth]{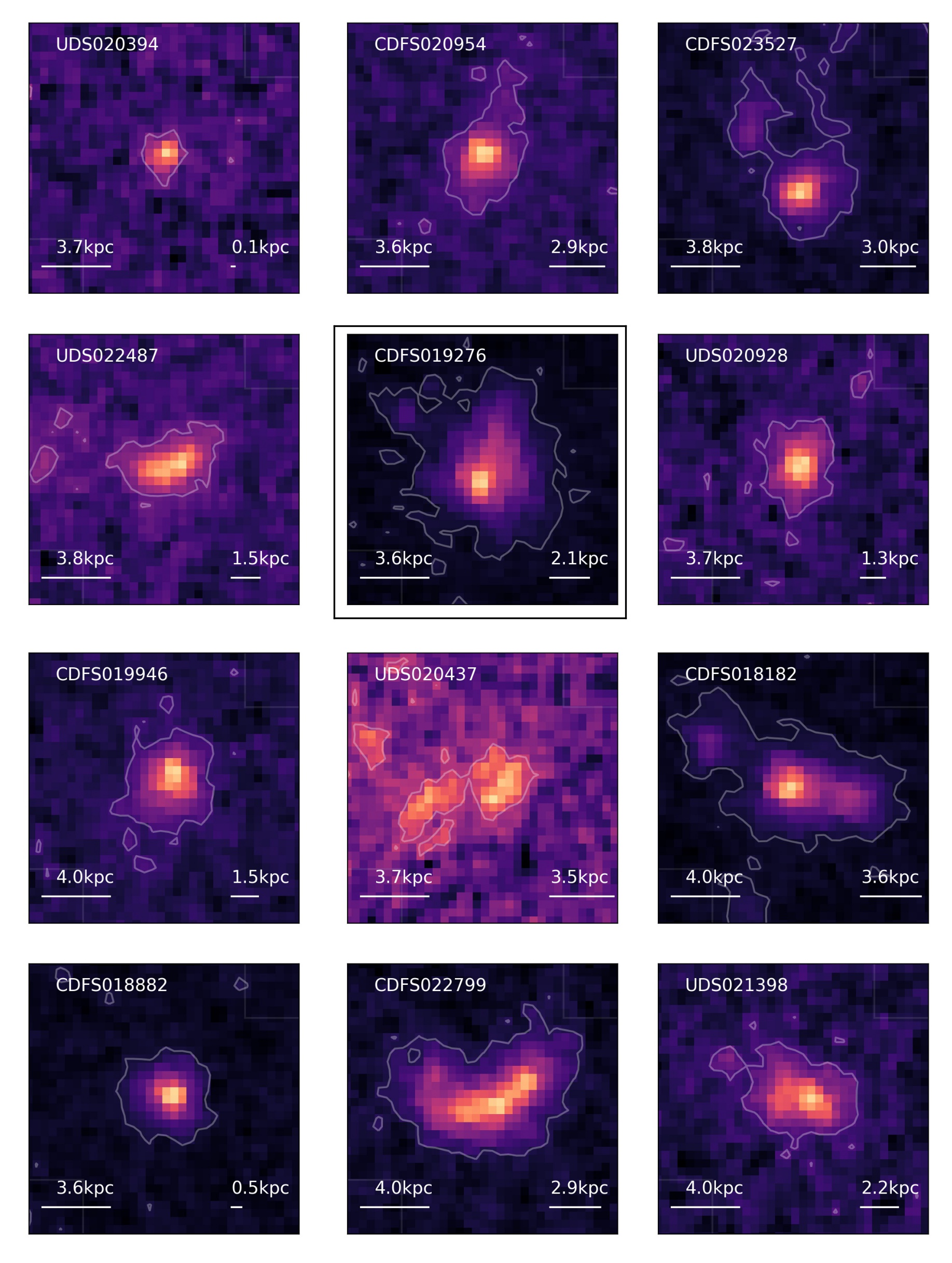}
    \caption{HST/F160W {images} \cite{Koekemoer2011} of the C3-VANDELS sample with $\Delta$BIC$>2$, i.e., the subsample with a broad component in their [OIII] profile. The images are tracing the rest-optical. The white contour is the $3\sigma$ level. The physical scale of 0.5 arcsec at their redshift is shown on the left of each image, while on the right, the effective radius is shown. {The galaxy marked with a black square shows two narrow components in its [OIII] profile.}}
    \label{fig:hstdeltabic>2}
\end{figure}
\begin{figure}[t!]
    \centering
    \includegraphics[width=\columnwidth]{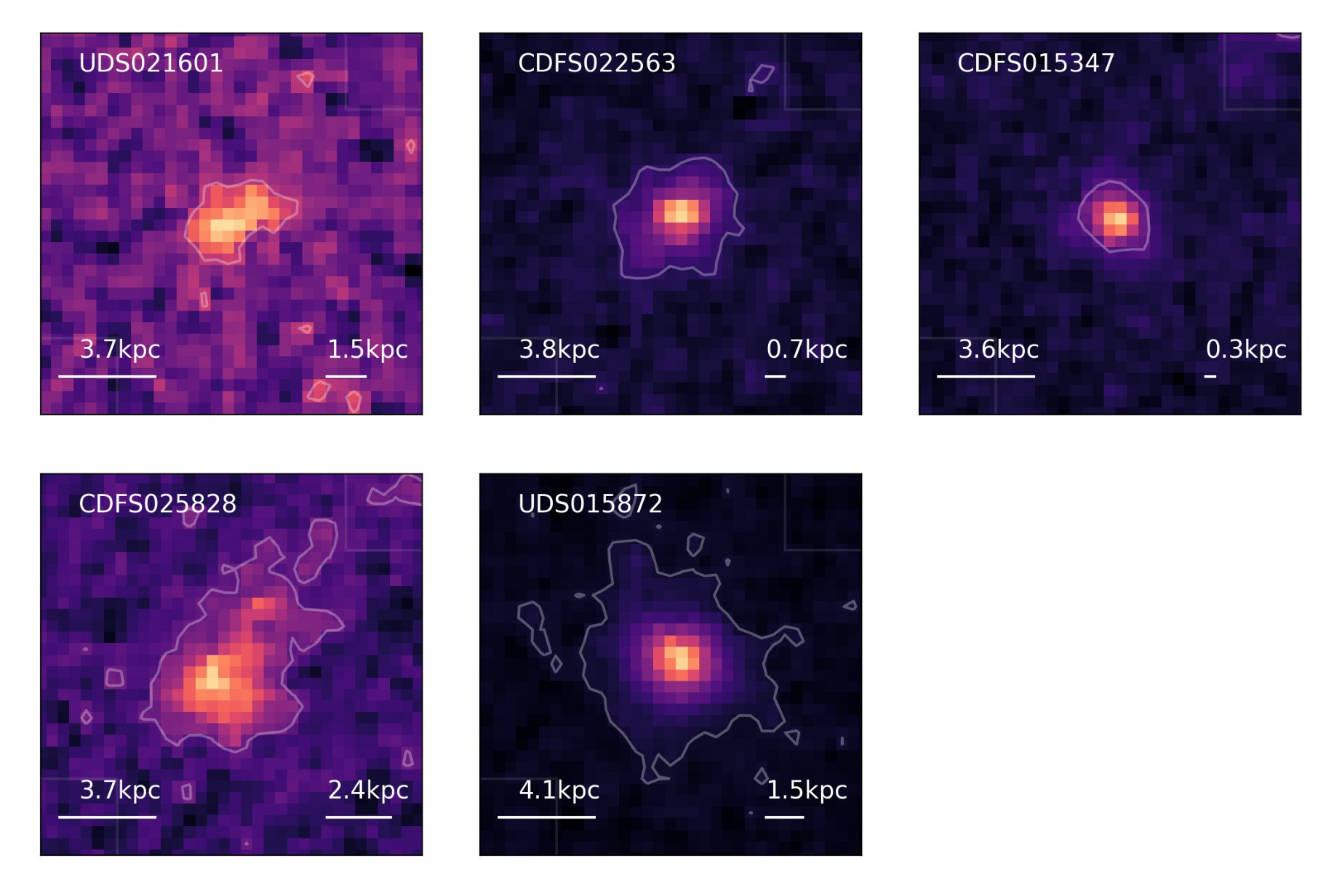}
    \caption{The same as in Fig. \ref{fig:hstdeltabic>2} but for the C3-VANDELS sample without features of a broad component in their [OIII] profile.}
    \label{fig:hstdeltabic<2}
\end{figure}

\begin{figure*}[t!]
    \centering
    \includegraphics[width=0.9\columnwidth]{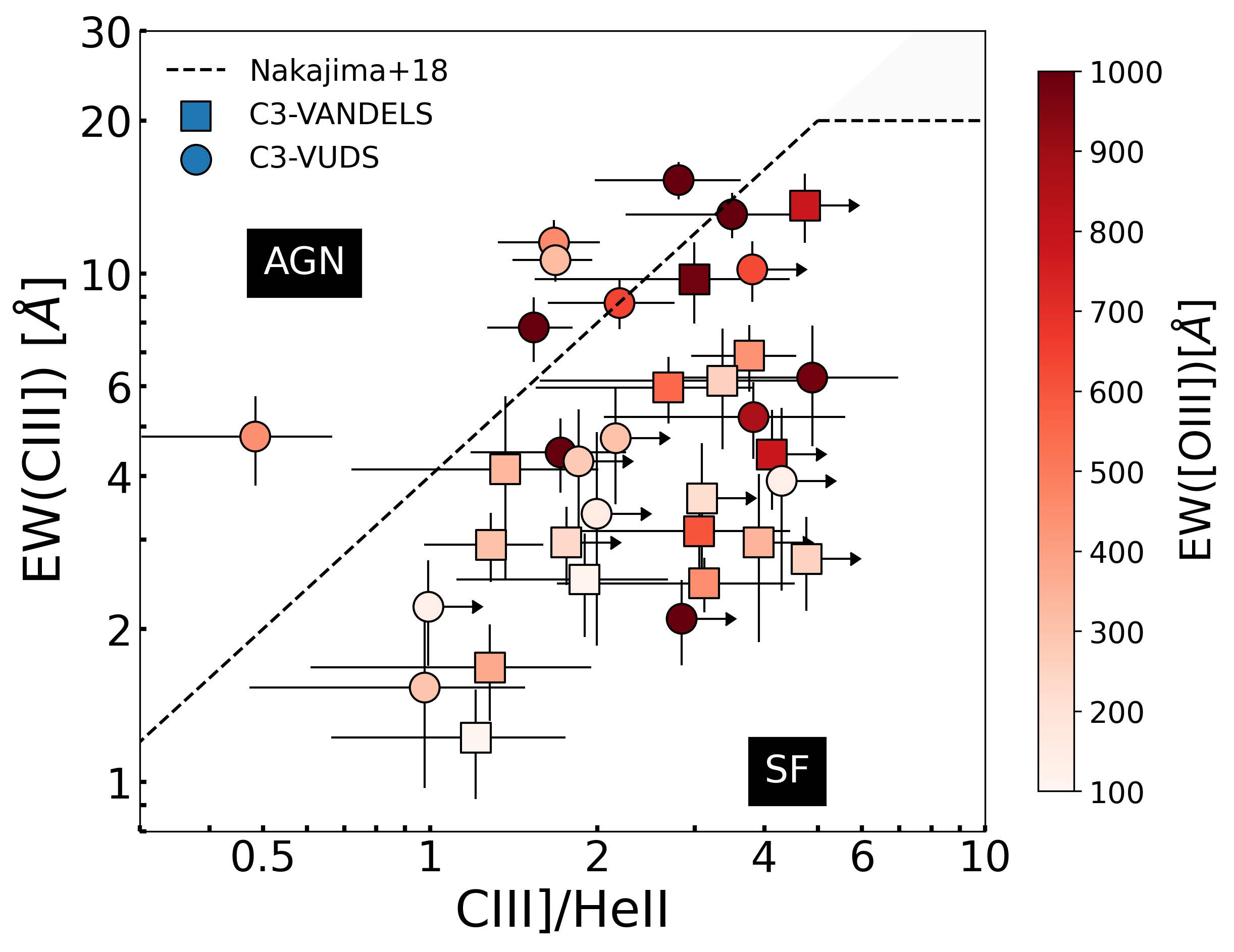}\,
    \includegraphics[width=0.9\columnwidth]{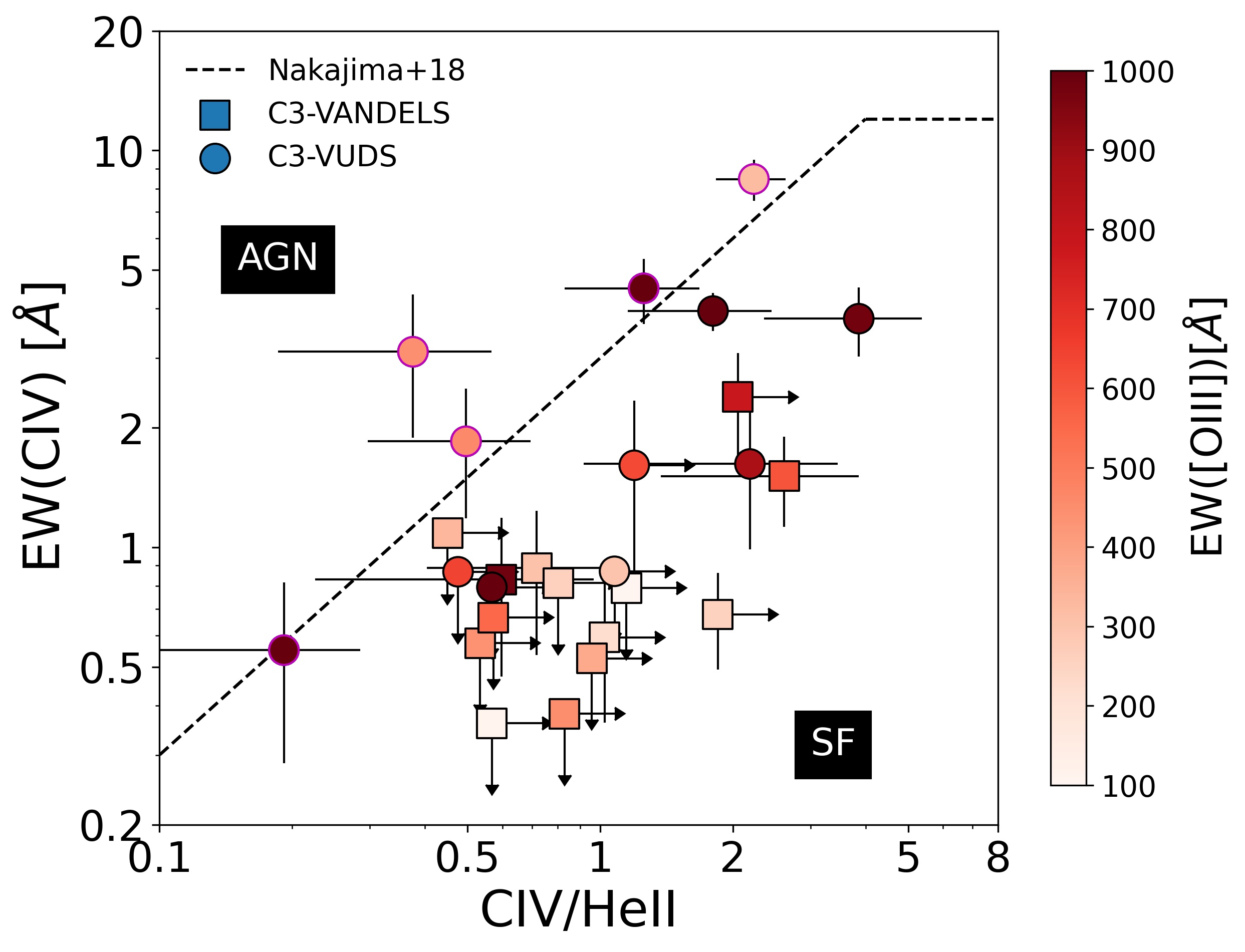}
    \caption{UV diagnostic diagrams for our sample based on the EWs of CIII] (\textit{left}) and CIV (\textit{right}). In both panels, our sample is color-coded by EW([OIII]), and the black dashed lines are the demarcation between AGN (on the left) and SF (on the right) according to \cite{Nakajima2018}. {On the right panel, the symbols with magenta edges are galaxies classified as AGN according to the EW(CIII])-CIII]/HeII diagram.}}
    \label{fig:UVBPT}
\end{figure*}

\begin{figure*}[t!]
    \centering
    \includegraphics[width=0.9\columnwidth]{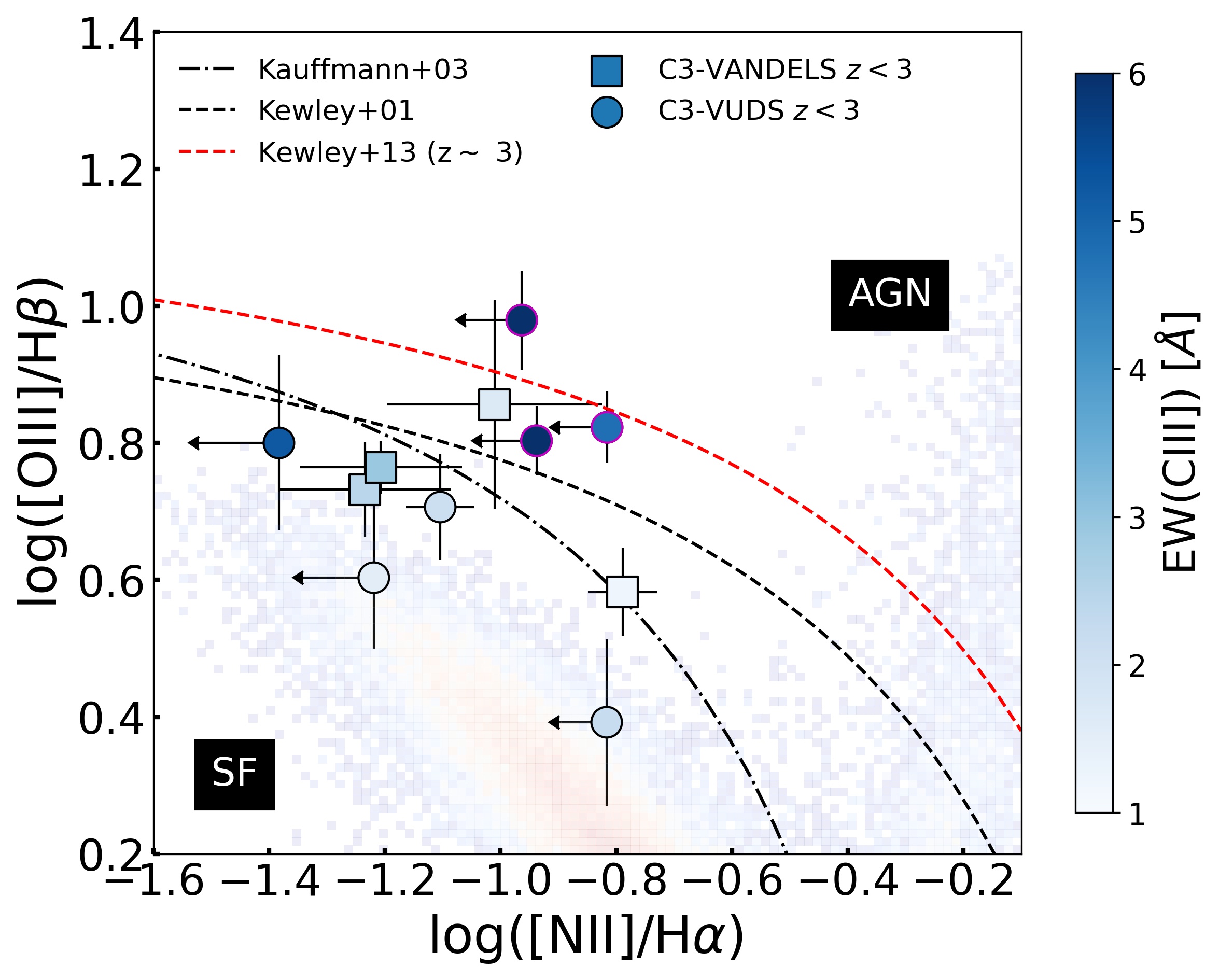}\,\includegraphics[width=0.9\columnwidth]{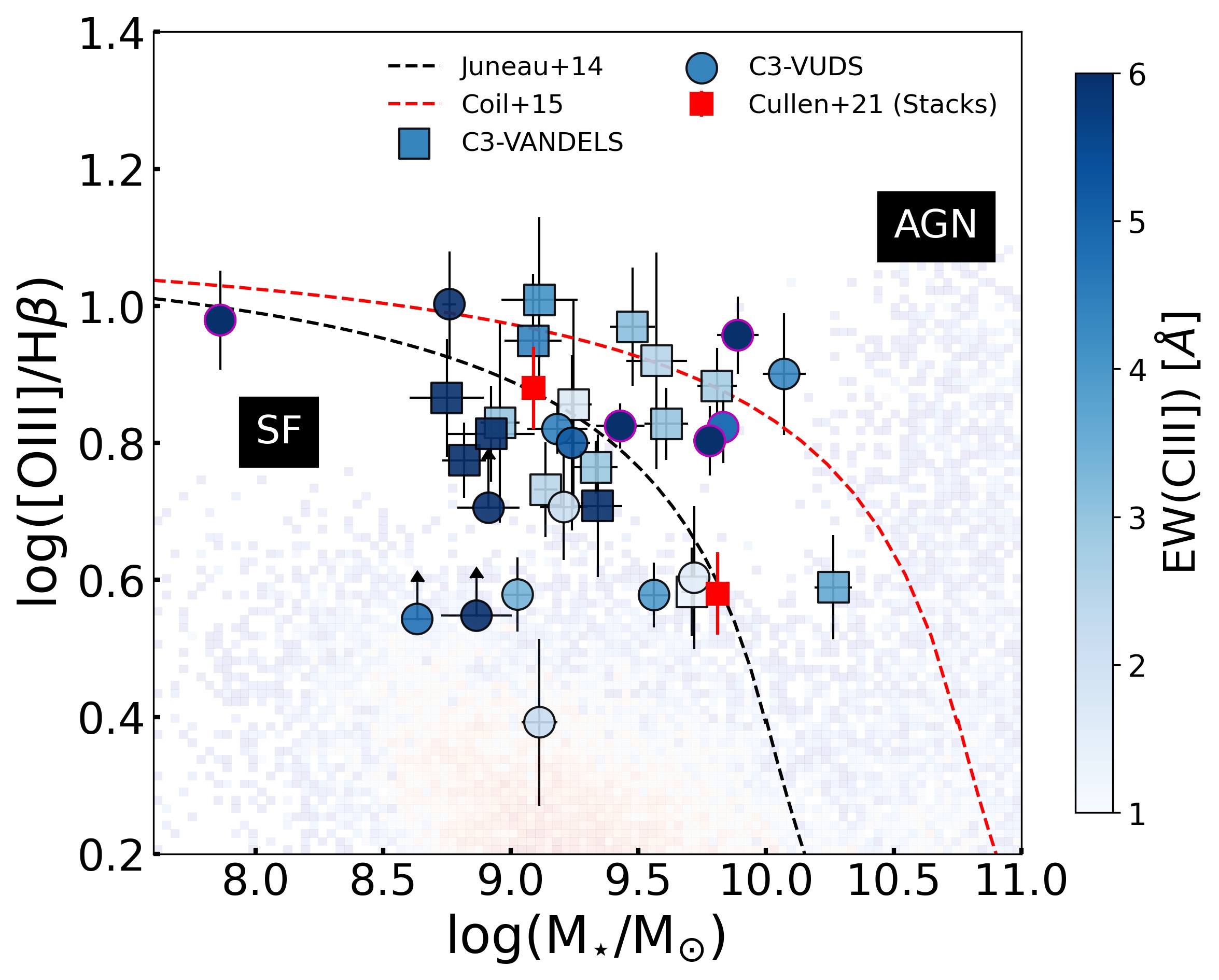}
    \caption{\textit{Left:} Classical BPT diagram \citep{BPT1981} for our subsample at $z<3$, color-coded by EW(CIII]). The black dashed lines are the typical local AGN/SF demarcation lines from \cite{Kewley2001,Kauffmann2003}. The red dashed line is the demarcation at $z\sim3$ according to \cite{Kewley2013}. \textit{Right:} Mass-Excitation diagram for our entire sample color-coded by EW(CIII]). The black and red dashed lines are the AGN/SF demarcation at low-$z$ \citep{Juneau2014} and $z\sim2.3$ \citep{Coil2015}, respectively. As a reference, we include the stacks from \cite{Cullen2021}, from which our VANDELS subsample is a subset. {On both panels, the symbols with magenta edges are galaxies classified as AGN according to UV diagnostic diagrams in Fig. \ref{fig:UVBPT}.}}
    \label{fig:optBPT}
\end{figure*}

Line flux and EW were measured individually in each galaxy. We measure the fluxes of UV lines using single Gaussians and {a linear component to include the} local continuum. We consider Ly$\alpha$, CIV, HeII$\lambda$1640, OIII]$\lambda$1666, and CIII] in the set of lines to be measured. We considered the CIII] width as the maximum width of the lines, which is in the range of $\sim 250-300$ km s$^{-1}$ (FWHM$\sim 15-18$\r{A}) and depends on each galaxy. The systemic redshift is based on the CIII] peak. {Given that the CIII] doublet is not resolved in our observations, we assume an air wavelength of 1907.05\r{A} for the average peak.}  The observed fluxes and rest-frame EW of the observed lines are reported in Table \ref{tab:UVflux}. {The uncertainties are estimated directly from the Gaussian fitting based on the covariance matrix using non-linear least squares.} For galaxies in the C3-VANDELS sample at $z<3$, Ly$\alpha$ is not measured since it is not included in the VANDELS spectral range.

On the other hand, the observed fluxes of rest-optical lines are reported in Table  \ref{tab:opticalflux}. The fluxes are computed from a single Gaussian component (except for [OIII]$\lambda\lambda$4959,5007, which is explained later). {A local continuum is set to zero because it is not detected above 1 $\sigma$}. We consider [OII]$\lambda$3727, [OII]$\lambda$3729, H$\beta$, [OIII]$\lambda\lambda$4959,5007, H$\alpha$, and [NII]$\lambda$6583 to be measured. {The spectra of the galaxies in the C3-VANDELS and C3-VUDS samples are displayed in Fig.  \ref{fig:spectraC3vandels} and \ref{fig:spectraC3vuds}, respectively.}

For [OIII]$\lambda\lambda$4959,5007 we consider a more detailed modeling. We fit the [OIII]$\lambda\lambda$4959,5007 profiles with two Gaussian components. In the other rest-optical lines, unfortunately, the S/N is lower, and a similar analysis can not be performed. We fit the doublet [OIII]$\lambda\lambda4959, 5007$ simultaneously using LMFIT \citep{NewvilleLMFIT}, {considering a wavelength range from 4939 to 5027\r{A} to include only the doublet.} We fix the ratio between both components to 1:2.98 \citep{Storey2000}. We mask the region in the spectra between both lines and the regions with strong sky residuals. As an initial guess, we include one narrow (100km s$^{-1}$) and one broad component (150km s$^{-1}$), {and both widths are free to vary}. We constrain the kinematics, assuming that the width of the components in each line is the same, as well as their peak velocities and ratios. We also perform a single Gaussian model to compare with and to test the improvement of the model with two Gaussians based on the Bayesian Information Criterion (BIC), {which is a statistical measure used to compare models based on their fit to the data and complexity. According to this criterion, the model having the lowest BIC is the preferred model. When the difference between the BIC values of the two models is greater than 2, it is rated as positive evidence against the significance of the model with the higher BIC \citep{Fabozzi2014}. We choose the BIC criterion as it prevents the selection of an over-fit model because it penalizes models with more free parameters.} To compare, we estimate the $\Delta \rm{BIC}=BIC_{\rm single-gaussian}-BIC_{\rm double-gaussian}$ and consider that the second component is statistically needed if $\Delta \rm{BIC}>2$ \citep{Fabozzi2014}. A similar criterion is adopted for the complex line profiles of strong Ly$\alpha$ emitters \citep[LAEs, ][]{Matthee2021} and Green Pea (GP) galaxies \citep{Bosch2019,Hogarth2020} {based on $\chi^2$ minimization}. {We find that 23 (65\%) out of the 35 galaxies in the sample show two kinematic components in their [OIII]$\lambda\lambda$4959,5007 profiles. In the remaining 12 galaxies, only one single component is statistically justified.} We only consider that the second component is broad if there is a difference of one spectral resolution element between the widths of the two Gaussian components. Otherwise, we consider that there are two narrow components in the double Gaussian model. The results from the fitting are shown in Fig. \ref{fig:deltabic>2} and \ref{fig:deltabic>2vuds} for the subsample with two components in the C3-VANDELS and C3-VUDS samples, respectively. For [OIII], we report in Table \ref{tab:opticalflux} the global flux (double Gaussian model) or the flux of the single-gaussian model if the second component is not justified. The kinematic information of the fits is reported in Table \ref{tab:kine}. We consider the intensity peak of the narrower component as the systemic velocity.

Regarding the rest-EW, the SED model obtained from photometry is used to determine the continuum of the rest-optical emission lines of our sample. {Since we are interested in qualitative trends with the EWs, we do not perform an additional aperture correction for the continuum.} For a particular line with intensity peak at $\lambda_{\rm peak}$, the continuum is obtained as the average between the continuum at $\lambda_{\rm peak}-20$ [\r{A}] and $\lambda_{\rm peak}+20$[\r{A}]. The measured EWs for $H\beta$ and [OIII] are reported in Table \ref{tab:opticalflux}. For the rest-UV lines, we use the local continuum measured directly from the spectra, which is detected in all galaxies in the final sample. We check that the UV continuum from the SED fitting model is consistent with the continuum measured in the spectra. For example, for the C3-VANDELS sample, the difference for the continuum for CIII] between the SED model and the local continuum measured directly from the spectra is, on average, $\sim$14\%. This difference implies a difference in the EW of $\sim$0.6\r{A}, on average, which is smaller than the typical error of 0.93\r{A} in the EWs.   

Most of the targets (20 out of 35) in our sample are at $z>3$, which means that H$\alpha$ is not included in the observed NIR spectral range. For this reason, we can not determine the nebular extinction (E(B-V)$_{\rm g}$) based on Balmer decrement for the entire sample. For the galaxies (12 out of 35) with detected H$\alpha$ and H$\beta$ (S/N$>2$), we estimated the nebular attenuation assuming H$\alpha$/H$\beta=2.79$ under Case B Approximation for T$_{\rm e}=15000$K and $n_{\rm e}=100$ cm$^{-3}$ \citep{EPM2017} and considering the Cardelli law \citep{Cardelli1989}. The obtained E(B-V)$_{\rm g}$ values are reported in Table \ref{tab:SED}. On the other hand, we have an estimation of the stellar extinction, which is obtained from the SED fitting. Most of our galaxies have low dust content with E(B-V)$_{\rm SED}<0.25$ mag (see Table \ref{tab:SED}). From this subsample, we extrapolate the nebular extinction for the entire sample based on their SED extinction. The best linear fit leads to ${\rm E(B-V)}_{\rm g}=0.75\times {\rm E(B-V)}_{\rm SED}+0.19$. We use this extrapolation only where E(B-V)$_{\rm g}$ is not determined directly by the Balmer decrement.  We correct the observed fluxes (reported in Tables \ref{tab:UVflux} and \ref{tab:opticalflux}) assuming the \cite{Reddy2015} law with $R_V=2.505$, based on other works with UV emission lines \citep[e.g.][]{Mingozzi2022}.

\subsection{SFR surface density}\label{sec:Sigma}
We also estimate the instantaneous SFR(H$\alpha$) using the relation $\log\rm{SFR}(H\alpha)=\log ( L(H\alpha)[\rm{erg s}^{-1}])-41.27$ which assumes the \cite{Chabrier2003} IMF \citep{Kennicutt2012} and the luminosities are corrected by dust reddening as explained in Sec. \ref{sec:fluxes}. For the cases where H$\alpha$ is not available, we use H$\beta$ assuming the same theoretical ratio H$\alpha$/H$\beta=2.79$. Since H$\alpha$ is only available for a small subsample, and H$\beta$ is fainter and has lower S/N compared to [OIII], a multiple Gaussian components analysis is not possible for the entire sample. For this reason, we assume the same broad-to-narrow flux (f$_B$) ratio from [OIII] to take into account only the flux from the narrow H$\alpha$ (or H$\beta$) to estimate the SFR, excluding the contribution from the broad component. 

We use the above values to derive the instantaneous SFR surface density ($\Sigma_{\rm SFR}$) defined as $\Sigma_{\rm SFR}=\dfrac{\rm{SFR}(H\alpha)}{2\pi r_H^2}$, where $r_H$ is the effective radius of the galaxy measured in the H-band. We obtained the effective radius directly from the literature based on HST imaging. For the C3-VANDELS sample, they are obtained directly from the CANDELS catalog \citep{vanderWel2012}, while for the VUDS sample they are obtained from \cite{Ribeiro2016}. In both cases, they used GALFIT \citep{Peng2002,Peng2010} and HST/F160W images to fit Sérsic profiles to obtain the effective radius. For the C3-VUDS subsample, there are 6 out of 18 galaxies for which the H-band HST image is not available. They are identified in Table \ref{tab:kine}. In those cases, we consider the effective radius using HST/F814W images from the same \cite{Ribeiro2016} catalog following consistent methodology. 

We note that at the redshift range covered for our sample, the F814W images trace the rest-frame UV-continuum by mostly massive stars, while the F160W images trace the rest-frame optical, which includes older stars and extended gas. From the C3-VUDS galaxies with both images available (12 galaxies), we find that the effective radius from the H-band image is typically a factor 1.4 greater than in the i-band. Still, in a few cases (3 out of 12), the differences can be up to a factor of $\sim3$. When only HST/F814W is available, we correct the effective radius by a factor 1.4.  

In Fig. \ref{fig:hstdeltabic>2}, we display the HST/F160W images of the C3-VANDELS sample with two components in their [OIII] profile. While in Fig. \ref{fig:hstdeltabic<2}, we show the same but for the C3-VANDELS sample with a single Gaussian model. The images for the C3-VUDS sample are shown in Fig.  \ref{fig:hstdeltabic>2vuds} and \ref{fig:hstdeltabic<2vuds} for the galaxies with and without broad component, respectively. {We note that in both cases, most of the galaxies (25 out of 35) tend to be compact, i.e., they show smaller sizes compared with the Size-Mass Relation at $z\sim2.75$ of late-type galaxies \citep[][]{vanderWel2014}. They do not show clear evidence of mergers. }

\section{Results}\label{sec:results}
\subsection{Ionizing sources: Diagnostic diagrams}\label{sec:BPT}
In this section, we analyze the ionization source of the galaxies in our sample. First, we use the UV diagnostic diagram based on the EW(CIII]) vs CIII]/HeII$\lambda$1640 flux ratio and EW(CIV]) vs CIV/HeII$\lambda$1640 flux ratio. The results are shown in Fig. \ref{fig:UVBPT}. It is shown that most of the galaxies in the sample are consistent with being ionized by massive stars {according to the demarcation lines in \cite{Nakajima2018}}. For some galaxies, these ratios are lower limits due to the fact that HeII$\lambda$1640 (or CIV) is not detected in their spectra, and then 2$\sigma$ upper limits are determined. It can be seen that the more extreme [OIII] emitters tend to fall in the upper right of the diagrams, suggesting high EW(CIII]) and EW(CIV) but not very strong HeII$\lambda$1640 that could be an indication of AGN contribution.

We find that all the galaxies in the C3-VANDELS sample are consistent with pure stellar photoionization, but there are few (5 galaxies in each diagram) C3-VUDS galaxies that are in the AGN region or near the boundary between both regions {(symbols with magenta edges on the right panel in Fig. \ref{fig:UVBPT})}. In particular, two of these galaxies (5100998761 and 5101421970) show also unusually high $T_{\rm e} >20000$K {(see Section \ref{sec:temperature})}, and some non-thermal contribution {cannot be ruled out} according to these diagnostics. Such high $T_{\rm e}$ are now observed in young SFGs in the EoR ($z=5.33-6.93$) with intense EW([OIII]$\lambda\lambda$4959,5007+H$\beta$)$\sim 1000$\r{A}\,\citep{Matthee2022} {and have been previously reported in extremely metal-poor starbursts in the local Universe \citep{Kehrig2016}}.

Recently, the C IV/C III] ratio has been proposed as a potential indirect indicator to constrain the LyC escape fraction \citep{Schaerer2022,Saxena2022}. We also find that most of the galaxies are consistent within the errors with CIV/CIII]$<0.7$, which indicates they are not strong LyC leakers and are likely to show f$_{\rm esc}<0.1$. Only one VUDS galaxy (5101421970) shows CIV/CIII]$>0.7$, which in the UV diagnostic diagrams is classified as AGN. 

We also explore Baldwin, Phillips \& Terlevich \citep[BPT,][]{BPT1981} optical diagnostics to verify the SF nature of these galaxies. In Fig. \ref{fig:optBPT} {(left panel)} we show the classical BPT diagram ([NII]$\lambda$6583/H$\alpha$ vs [OIII]/H$\beta$) for the subsample of galaxies at $z<3$ for whom H$\alpha$ is included in K band. It can be seen that this subsample falls within demarcation lines for SF regions and is broadly offset from the typical excitation of local SFGs. We notice that the subsample at $z<3$ are among the fainter CIII] emitters in the entire sample with EW(CIII])$<3$\r{A}, in particular the ones with [NII]$\lambda$6584 detected. The more extreme CIII] emitters (with EWs$>4$\r{A}) tend to show upper limits in [NII] and high [OIII]/H$\beta$ ratios. 

Finally, {in Fig. \ref{fig:optBPT} (right panel)}, we explore the Mass-Excitation \citep[MEx, ][]{Juneau2014} diagram for the entire sample (except for the galaxies that do not have $H\beta$ in the spectral range). Similarly to the BPT diagram, we find that they are also consistent with SF. The more extreme CIII] emitters tend to fall in the low mass region (roughly $<10^{9.5}$M$_{\odot}$) and high [OIII]/H$\beta>4$, which indicates a highly excited ISM. The two particular extreme galaxies with possible AGN contribution are in the SF region in this diagram (one of them shows the smaller stellar mass in the sample). The above analysis suggests that combining UV and optical lines may give us more clues on their nature, particularly for extremely metal-poor SF galaxies, which may display very high electron temperatures. {Hereafter, we consider that the galaxies in our sample are dominated by SF with a highly excited ISM and do not show clear evidence of AGN contribution based on their emission lines.} 

\subsection{Ionization parameter}\label{sec:U}
{We also study the ionization properties of our sample of galaxies by means of a tracer of the ionization parameter $\log {\rm U}$. Given the limited wavelength coverage of our spectra, we use the 
$\dfrac{[OIII]\lambda\lambda4959,5007}{[OII]\lambda\lambda3727,3729}$ ratio  
as a proxy for $\log {\rm U}$ \citep{Kewley2019}. {We find [OIII]/[OII] values within 1 and 10, which imply a high ionization parameter between roughly $\log {\rm U}=$-3 and  $\log {\rm U}=$-2 \citep{Reddy2023mosdef}.}} 

\begin{figure}[t!]
    \centering
    \includegraphics[width=0.9\columnwidth]{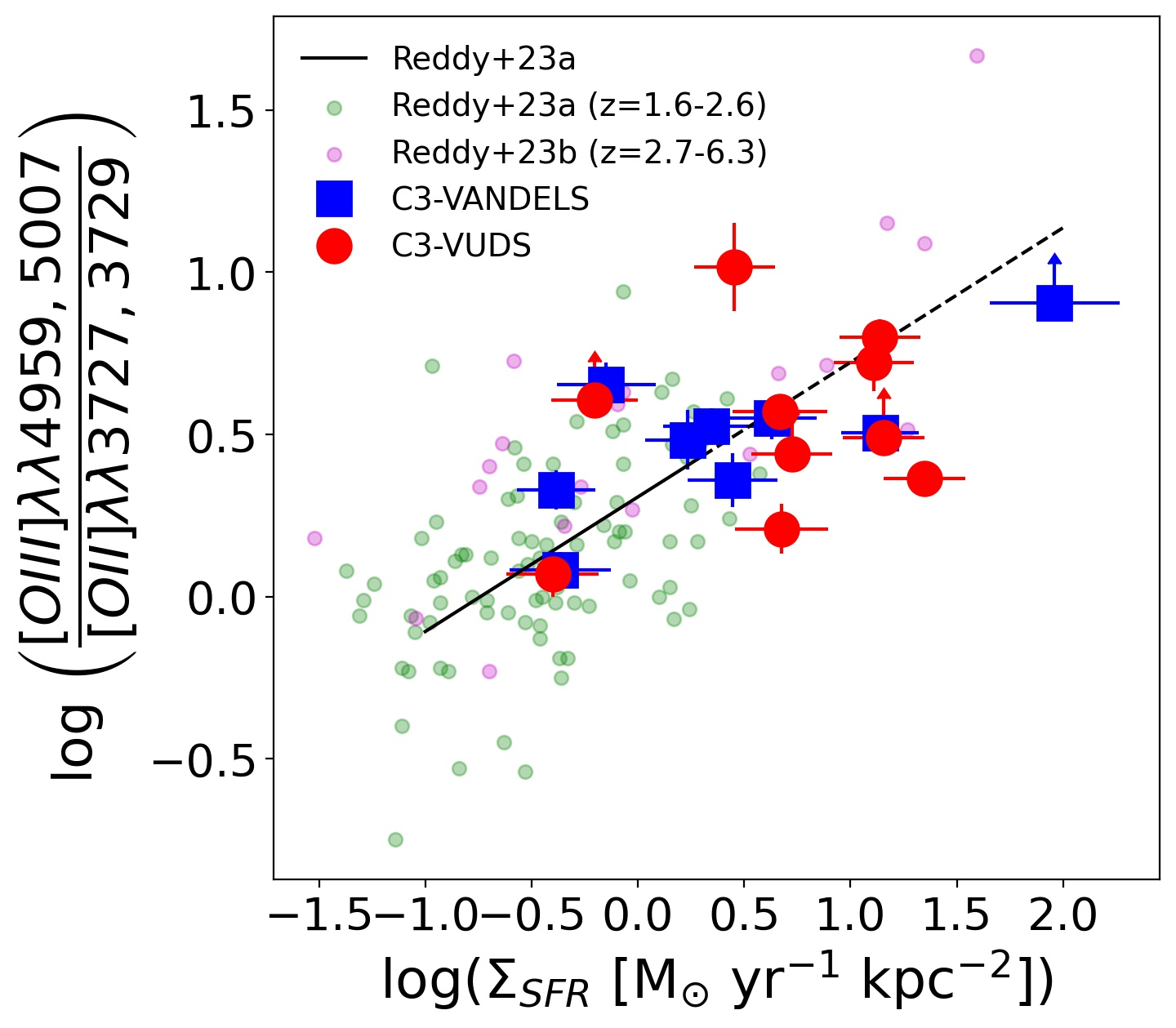}
    \caption{{Variation of [OIII]/[OII] with $\Sigma_{\rm SFR}$. Individual galaxies with detected [OII] are shown with red and blue symbols for the C3-VUDS and C3-VANDELS samples, respectively. We also include lower limits based on upper limits on [OII]. The green and magenta small circles are galaxies from \cite{Reddy2023mosdef} at $z=1.6-2.6$ and \cite{Reddy2023jwst} at $z=2.7-6.3$, respectively. The black solid line is the relation presented in \cite{Reddy2023mosdef} and its extrapolation up to $\Sigma_{\rm SFR}=100$ M$_{\odot}$ yr$^{-1}$ kpc$^{-2}$ is in dashed black line.}}
    \label{fig:sigma_o32}
\end{figure}

{In Fig. \ref{fig:sigma_o32}, we present the relation for [OIII]/[OII] and $\Sigma_{\rm SFR}$ found by \citet[][]{Reddy2023mosdef,Reddy2023jwst} for SFGs in the MOSDEF survey and others at higher redshift, suggesting that galaxies with more compact and violent star formation show more extreme ionization conditions. We find that our VANDELS and VUDS galaxies follow the same trend and lie closer to the $z>3$ galaxies than the ones at $z<2.6$ in the above works. Despite our sample galaxies lying mostly in the SF main sequence at $z\sim$3 (Fig. \ref{fig:MS}), their ionization properties and emission line EWs appear more extreme than their counterparts at lower redshift. }

\subsection{EW relations}\label{sec:EW}

{For UV lines}, we find that our sample covers a range of EW(CIII]) from $\sim1$\r{A} to $\sim15$\r{A} with a mean value of 5.6\r{A} ($\sigma=3.7$\r{A}). Six galaxies 
(17\%) show EW(CIII])$>10$\r{A},  {similar to the EW values} observed in galaxies at $z>6$ \citep[e.g.][]{Stark2017,Hutchison2019}. 
{Our sample includes 
14 galaxies at $z\gtrsim$3 with Ly$\alpha$ included in our spectral range and with EW(Ly$\alpha$)$>20$\r{A}, five of them reaching EW(Ly$\alpha$)$>100$\r{A}. }

{For optical lines, we find strong [OIII]$\lambda\lambda$4959,5007 emission, with} 
EW([OIII]) spanning from 102\r{A} to 1715\r{A} with a mean value of 563\r{A} ($\sigma=420$\r{A}), well within the typical EWs defined for Extreme Emission Line Galaxies (EELG) at lower redshifts \citep[e.g.][]{Amorin2014,Amorin2015,Calabro2017,PM2021}. 
{Only 5 galaxies (14\%) of the sample show  EW([OIII])$>1000$\r{A}, which are typical values found for $z>6$ EoR galaxies from photometric data \citep[e.g.][]{Endsley2021} and more recently with JWST spectroscopy \citep{Matthee2022}. }

\begin{figure}[t!]
    \centering
    \includegraphics[width=\columnwidth]{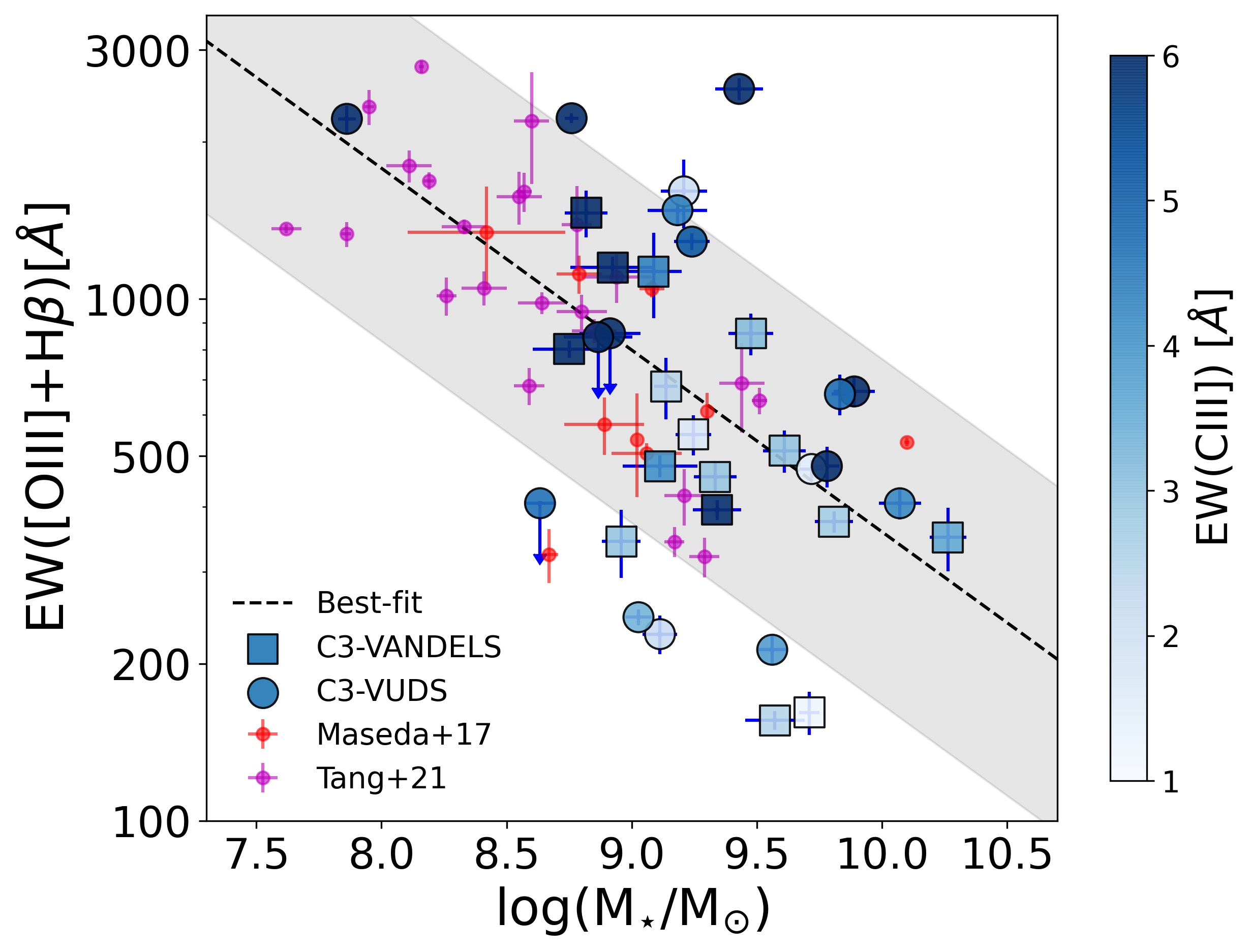}
    \caption{Relation between EW([OIII]$\lambda\lambda$4959,5007+H$\beta$) and stellar mass. Our sample is color-coded by EW(CIII]). The small red and magenta circles are literature samples at intermediate redshifts from \cite{Maseda2017} and \cite{Tang2021}, respectively. The black dashed line is the best fit (slope -0.34) with our data and the gray shaded region is the 1$\sigma$ observed scatter of 0.33 dex.}
    \label{fig:EW-mass}
\end{figure}

\begin{figure}[t!]
    \centering
    \includegraphics[width=\columnwidth]{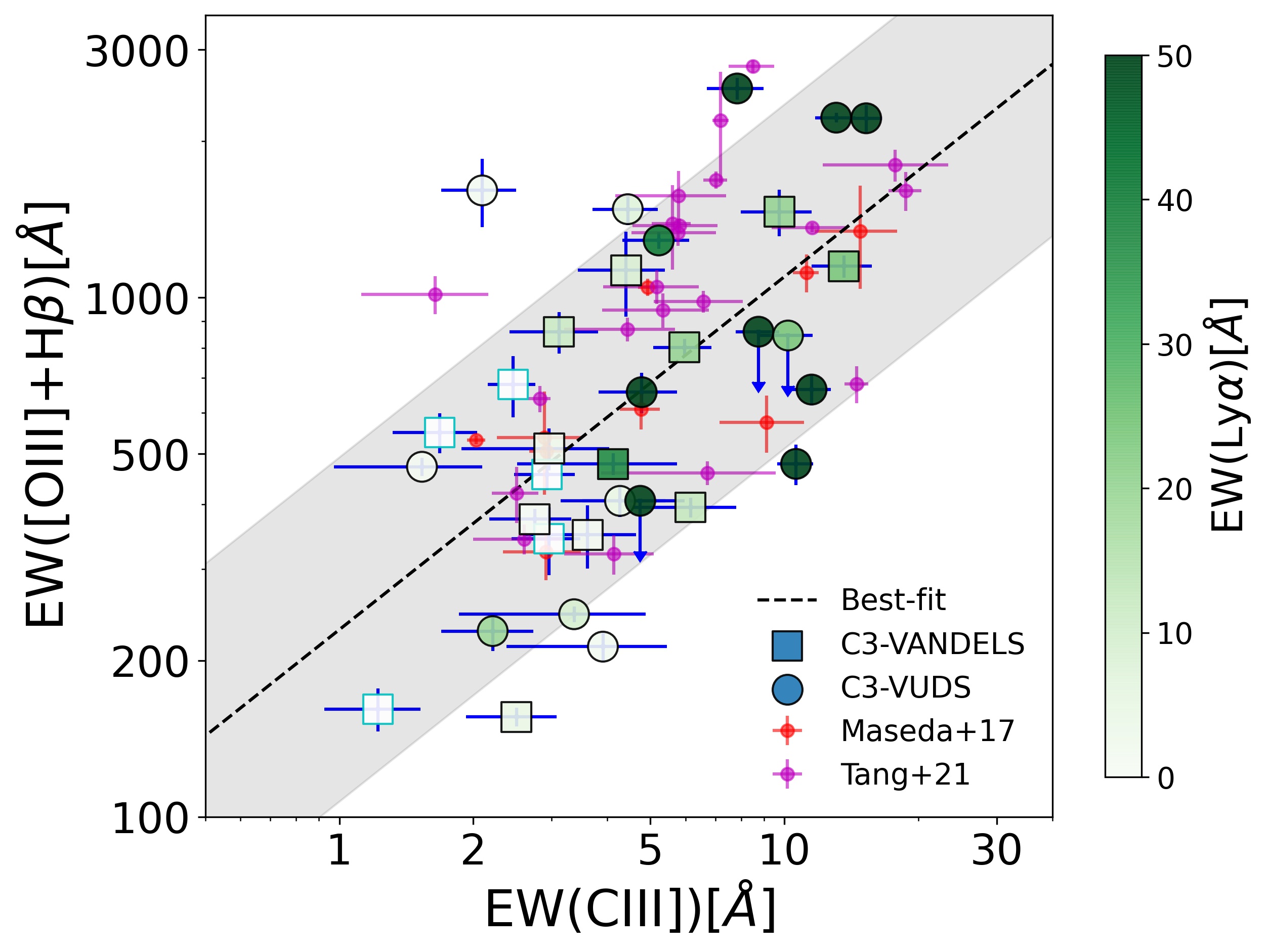}
    \caption{Relation between EW([OIII]$\lambda\lambda$4959,5007+H$\beta$) and EW(CIII]). Our sample is color-coded by EW(Ly$\alpha$). {The square symbols with cyan edges are galaxies in the C3-VANDELS at $z<3$, for which Ly$\alpha$ is not in the spectral range.} The small red and magenta circles are literature samples at intermediate redshifts from \cite{Maseda2017} and \cite{Tang2021}, respectively. The black dashed line is the best fit (slope 0.68) with our data and the gray shaded region is the 1$\sigma$ observed scatter of 0.33 dex.}
    \label{EW-EW}
\end{figure}

We explore the correlation between stellar mass and [OIII]+H$\beta$ EWs in Fig. \ref{fig:EW-mass}. 
{We find that low-mass galaxies tend to show higher EWs, following a similar trend found in literature, both at $z\sim$2-3 \citep[e.g.][]{Maseda2017,Tang2021} and in reionization galaxies} \citep{Endsley2021}. 
{The most extreme EWs in our sample correspond to galaxies with EW(CIII]) $>$5 \r{A} and EW([OIII]$\lambda\lambda$4959,5007+H$\beta$)$>$ 500 \r{A}, as shown in Fig. \ref{EW-EW}, which are still rare at $z\sim3$ but becomes} the norm towards 
reionization \citep[e.g.][]{Smit2014,DeBarros2019,Endsley2021,Sun2022a,sun2022,Matthee2022}.
{The  trend in Fig.~\ref{EW-EW} also shows that the strong [OIII] and CIII] emitters at $z>3$ are typically those with larger EW in Ly$\alpha$ ($\gtrsim$20\AA). However, a few galaxies with weak Ly$\alpha$ emission are found among the strong [OIII] emitters, in agreement with previous works \citep{LeFevre2019,Du2020}.}

{The above results can be interpreted in the context of the galaxies ionizing photon production efficiency, which correlates with H$\alpha$ and [OIII] EWs \citep{Tang2019}. Even for extreme [OIII]+H$\beta$ emitters that  efficiently produce ionizing photons, the emerging Ly$\alpha$ line will not be necessarily intense due to its resonant nature, which makes it sensitive to dust content, neutral hydrogen column density and their spatial distribution \citep[e.g.][]{Hayes2015}. Indeed, our subsample of strong [OIII] emitters with lower EW(Ly$\alpha$) are found to show higher dust extinction. 
These objects are likely found in a very early phase after the onset of star formation where the young ($<$2-3Myr) massive stars are still embedded in their dense and dusty birth clouds and did not have enough time to clear channels through the ISM via feedback (e.g., winds, supernovae), and then Ly$\alpha$ photons are trapped while LyC photons are absorbed \citep{Naidu2022}. }

\subsection{Electron densities and temperatures }\label{sec:temperature}
The two components of the doublet [OII]$\lambda\lambda3727,3729$ are detected and unaffected by strong sky residuals in 15 galaxies of our entire sample. For this subsample, we estimate the electron density using {the {\verb|getTemDen|} task in \pyneb\, assuming T$_e=15000$K {and performing 100 Monte Carlo simulations in order to include the uncertainties of the observed fluxes. We find a wide range of electron densities from $\sim 47$ to $\sim 1261$ cm$^{-3}$, with a mean value of $560$ cm$^{-3}$.} {These values are within the range of values observed at similar redshifts \citep[e.g.][]{Sanders2016density,Reddy2023mosdef}. Given the uncertainties (we obtained a mean value of $\sim 270$ cm$^{-3}$) in the estimated n$_e$ values  due to the low S/N of [OII]$\lambda\lambda3727,3729$, we report densities in Table \ref{tab:abundance} but in the following sections used only the mean value of the entire sample.}}

In order to estimate the electron temperature we used the OIII]$\lambda$1666/[OIII]$\lambda$5007 ratio where available. OIII]$\lambda$1666 is detected with S/N$>2$ in 21 galaxies in our sample. We use the {\verb|getTemDen|} task in \pyneb\, {assuming $n_e=560$cm$^{-3}$}. Given that an offset between the $T_{\rm e}$([OIII]) found using this ratio and the commonly used ratio 
[OIII]$\lambda$4363/[OIII]$\lambda$5007 has been found in local galaxies, we correct downwards the obtained temperature by the typical difference of $-0.025$ dex obtained for CLASSY SFGs in \citet{Mingozzi2022}, which leads to differences of $\sim-1000$K. We report these temperatures in Table \ref{tab:abundance}. As discussed in \cite{Mingozzi2022}, one of the intrinsic reasons that may explain the offset between both methods is that the ISM is patchy. The UV light is visible only through the less dense (and/or less reddened) regions along the line of sight, while the optical may be arising also from denser (and/or more reddened regions). We find T$_{\rm e}$([OIII])$>1.3\times10^4$K with some objects having very high temperatures but not exceeding  $2.5\times10^4$K, as reported in Table \ref{tab:abundance}. We find a mean value {of $1.77\times10^4$K} for the entire sample with typical errors of $0.2\times10^4$K, which were calculated with 100 MonteCarlo simulations taking into account a normal distribution of the fluxes of the used lines to estimate T$_{\rm e}$([OIII]). {Our results are consistent with the early results from the JWST/NIRSpec, based on the first direct detection of the faint auroral line [OIII]$\lambda$4363 at $z\sim 8$, where T$_{\rm e}$ from $1.2\times 10^4$ and up to $2.8 \times 10^4$K are estimated, with the T$_{\rm e}$-based metallicities ranging from extremely metal-poor (12+log(O/H)$<$7) to about one-third solar \citep{Schaerer2022b,Trump2022,Curti2022,Nakajima2022}.}

We note that the OIII]$\lambda$1666/[OIII]$\lambda$5007 ratio is obtained from two different instruments, which may result in possible systematic errors due to flux calibration and dust attenuation corrections, given the large wavelength separation of the lines. First, we checked that the spectra are consistent with the photometry based on the SED model to avoid flux-matching issues. 
For example, we find that the difference between the mean continuum between 1450 and 1500\r{A} in the SED model and the observed spectrum is on average $\sim$10\% of the observed flux which is lower than the scatter in this spectral region, which is on average 25\% of the observed flux. Additionally, we check that changing the dust attenuation law does not affect our results. For instance, if we consider the \cite{Calzetti2000} law, the mean $T_{\rm e}$ changes by only 76 K, which is negligible compared with the typical uncertainties. 

\subsection{Oxygen and Carbon abundance}\label{sec:abundance}
\begin{figure}[t!]
    \centering
    \includegraphics[width=\columnwidth]{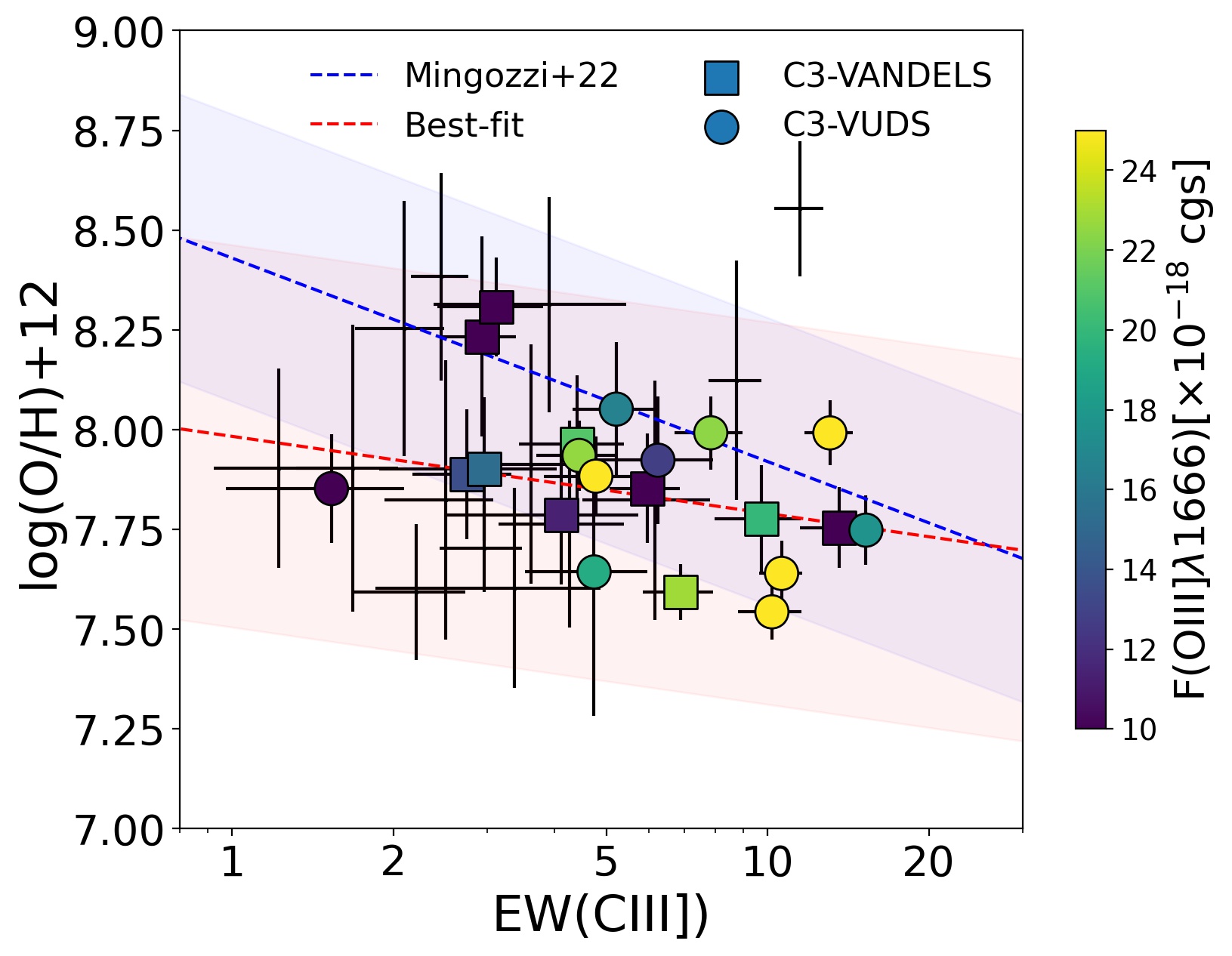}
    \caption{Relation between gas-phase metallicity and EW(CIII]) for our sample. The blue dashed line and {blue shaded region} corresponds to the relation found in \cite{Mingozzi2022} at low $z$ {and their observed 2$\sigma$ scatter.} The red dashed line is our best fit and {the observed 2$\sigma$ scatter is the red shaded region.} The sample es color-coded by OIII] flux in the cases where this line is detected at S/N$>2$. In the other cases, only errorbars are shown.}
    \label{fig:OH EW}
\end{figure}

\begin{table*}[]
    \caption{Electron density, temperature and chemical abundances estimated for our sample}
    \label{tab:abundance}
    \centering
    \begin{scriptsize}
    \begin{tabular}{|c|c|c|c|c|c|c|c|}\hline\hline
    ID&n$_e$[cm$^{-3}$]&t$_e$[$10^4$K]&$\log$(O/H)$_\pyneb$+12\tablefoottext{a}&$\log$(O/H)$_{\rm{SLC}}$+12\tablefoottext{b}&$\log$(O/H)$_{\rm{SLC-UV}}$+12\tablefoottext{c}&$\log$(O/H)$_\hcmuv$+12\tablefoottext{d}&$\log$(C/O)\\
    &&&(1)&(2)&(3)&(4)&\\\hline
\hline\multicolumn{8}{|l|}{C3-VANDELS sample}\\ \hline
UDS020394 & ... &1.85$\pm$0.2 &7.74$\pm$0.1 &... &... &7.74$\pm$0.1 &-0.49$\pm$0.19 \\
CDFS020954 & 1048$\pm$324 &1.83$\pm$0.15 &7.78$\pm$0.13 &8.08$\pm$0.12 &7.61$\pm$0.06 &7.78$\pm$0.13 &-0.69$\pm$0.17 \\
CDFS023527 & ... &1.93$\pm$0.16 &... &... &7.65$\pm$0.04 &7.39$\pm$0.09 &-0.66$\pm$0.12 \\
UDS021601 & ... &<1.81 &... &... &... &7.85$\pm$0.32 &-0.44$\pm$0.35 \\
CDFS022563 & 272$\pm$38 &1.82$\pm$0.18 &7.85$\pm$0.14 &8.11$\pm$0.12 &7.64$\pm$0.05 &7.85$\pm$0.14 &-0.51$\pm$0.2 \\
UDS022487 & 570$\pm$155 &1.89$\pm$0.2 &7.96$\pm$0.17 &8.14$\pm$0.12 &... &7.96$\pm$0.17 &-0.9$\pm$0.15 \\
CDFS015347 & ... &2.47$\pm$0.27 &7.78$\pm$0.18 &... &7.46$\pm$0.11 &7.78$\pm$0.18 &-0.92$\pm$0.14 \\
CDFS019276 & 259$\pm$107 &<1.59 &... &8.18$\pm$0.12 &... &7.84$\pm$0.29 &-0.31$\pm$0.2 \\
UDS020928 & 179$\pm$73 &1.29$\pm$0.2 &8.31$\pm$0.12 &8.12$\pm$0.12 &7.67$\pm$0.07 &8.31$\pm$0.12 &-0.34$\pm$0.15 \\
CDFS019946 & ... &<2.81 &... &... &... &7.76$\pm$0.08 &-0.48$\pm$0.23 \\
UDS020437 & 239$\pm$83 &1.79$\pm$0.2 &7.9$\pm$0.18 &8.19$\pm$0.12 &... &7.9$\pm$0.18 &-0.73$\pm$0.23 \\
CDFS018182 & ... &1.38$\pm$0.15 &... &... &7.59$\pm$0.05 &8.23$\pm$0.23 &-0.4$\pm$0.22 \\
CDFS018882 & 830$\pm$294 &1.88$\pm$0.23 &7.89$\pm$0.16 &8.13$\pm$0.12 &... &7.89$\pm$0.16 &-0.64$\pm$0.18 \\
CDFS025828 & 328$\pm$123 &<3.02 &... &8.28$\pm$0.12 &... &7.94$\pm$0.3 &-0.73$\pm$0.21 \\
CDFS022799 & ... &<1.22 &... &... &... &8.4$\pm$0.18 &-0.67$\pm$0.14 \\
UDS021398 & ... &<2.03 &... &... &... &7.78$\pm$0.2 &-0.57$\pm$0.29 \\
UDS015872 & ... &<1.54 &... &... &... &7.92$\pm$0.32 &-0.33$\pm$0.22 \\
\hline\multicolumn{8}{|l|}{C3-VUDS sample}\\ \hline
5100998761 & ... &2.15$\pm$0.12 &7.75$\pm$0.09 &... &7.64$\pm$0.03 &7.75$\pm$0.09 &-0.49$\pm$0.09 \\
5101444192 & ... &1.64$\pm$0.06 &7.99$\pm$0.08 &... &7.64$\pm$0.04 &7.99$\pm$0.08 &-0.56$\pm$0.11 \\
510994594 & ... &<1.22 &... &... &... &8.36$\pm$0.16 &0.02$\pm$0.18 \\
5101421970 & 1201$\pm$1339 &1.98$\pm$0.21 &7.64$\pm$0.08 &7.95$\pm$0.13 &7.6$\pm$0.03 &7.64$\pm$0.08 &-0.27$\pm$0.08 \\
510838687 & ... &2.14$\pm$0.24 &... &... &... &7.59$\pm$0.14 &-0.69$\pm$0.16 \\
5100556178 & ... &<1.38 &... &... &... &8.11$\pm$0.27 &-0.24$\pm$0.14 \\
511229433 & 468$\pm$352 &1.46$\pm$0.08 &7.99$\pm$0.09 &8.05$\pm$0.12 &7.59$\pm$0.04 &7.99$\pm$0.09 &-0.58$\pm$0.09 \\
511245444 & ... &1.42$\pm$0.12 &... &... &7.7$\pm$0.06 &8.07$\pm$0.21 &-0.35$\pm$0.1 \\
530048433 & 47$\pm$25 &1.41$\pm$0.2 &8.05$\pm$0.17 &8.11$\pm$0.12 &7.71$\pm$0.06 &8.05$\pm$0.17 &-0.21$\pm$0.22 \\
510583858 & ... &1.61$\pm$0.14 &7.88$\pm$0.1 &... &7.45$\pm$0.06 &7.88$\pm$0.1 &-0.53$\pm$0.16 \\
511451385 & ... &2.44$\pm$0.32 &... &... &... &7.73$\pm$0.62 &-0.6$\pm$0.17 \\
511025693 & 510$\pm$217 &1.52$\pm$0.13 &7.94$\pm$0.09 &8.03$\pm$0.12 &7.56$\pm$0.06 &7.94$\pm$0.09 &-0.65$\pm$0.17 \\
5100997733 & ... &<1.94 &... &... &... &7.83$\pm$0.19 &-0.52$\pm$0.32 \\
511228062 & 1261$\pm$281 &<1.29 &... &8.18$\pm$0.12 &... &8.24$\pm$0.16 &-0.29$\pm$0.17 \\
5101001604 & ... &<2.83 &... &... &... &7.7$\pm$0.14 &-0.62$\pm$0.28 \\
510996058 & 1061$\pm$398 &<2.14 &... &8.28$\pm$0.12 &... &7.51$\pm$0.21 &-0.41$\pm$0.29 \\
511001501 & 195$\pm$62 &<1.39 &... &8.23$\pm$0.12 &... &8.1$\pm$0.39 &-0.43$\pm$0.35 \\
530053714 & ... &1.39$\pm$0.15 &7.85$\pm$0.14 &... &7.45$\pm$0.12 &7.85$\pm$0.14 &-0.68$\pm$0.2 \\
\hline\hline
    \end{tabular}
    \tablefoot{\tablefoottext{a}{Metallicity from \pyneb\, method}\tablefoottext{b}{Metallicity from Strong-Line Calibration method with O32 parameter \citep{EPM2021} }\tablefoottext{c}{Metallicity from corrected UV Strong-Line Calibration method with He2-O3C3 parameter   \citep{Mingozzi2022}}\tablefoottext{d}{Metallicity from \hcmuv\, corrected according by a factor 0.3 dex when $\log$(O/H)$_\pyneb$ is not available.}}
    \end{scriptsize}
\end{table*}

The study of the abundances of heavy elements in the ISM of galaxies provides precious insights into the physical processes responsible for their formation and how the relative importance of such processes has changed across cosmic time \citep[see the review by][]{Maiolino2019}. Due to the line production mechanisms, nebular UV emission can be used to also directly calculate the physical and chemical conditions under which {emission lines} are produced. For instance, while the oxygen abundance (O/H) is the standard measure of a gas-phase metallicity in galaxies, CIII] provides a path to estimate the C abundance, which is a non-$\alpha$ element. Because C is primarily produced in lower mass stars than O, the injection of C and O to the ISM occurs on different time scales, providing a probe of the duration, history, and burstiness of the star formation \citep[e.g.][]{Berg_2019}. {Also, the relative abundance of C and O can be relatively unaffected by hydrodynamical processes, like outflows of enriched gas \citep[e.g.][]{Edmunds1990}.} In particular, the line intensity ratio C III]/O III]1666 has been used to estimate the relative C/O abundances at different redshifts \citep[e.g.][]{Garnett1995,Shapley2003,Erb2010,Steidel2016,Berg2016,PM2017,Berg_2019,Amorin2017,Llerena2022}. Moreover, the combination of UV+optical emission lines has been explored to constrain T$_{\rm e}$ using the OIII]$\lambda$1666/[OIII]$\lambda$5007 ratio \citep{PM2017}, and then to estimate gas-phase metallicity, in particular for those cases where the optical auroral line {[OIII]$\lambda$4363} is not available due to weakness or spectral coverage.

We derive the C/O abundance based on photoionization models using {the public} version 4.23\footnote{\url{http://home.iaa.csic.es/~epm/HII-CHI-mistry-UV.html}} of \hcmuv\, \citep{PM2017} {using as an ionizing source for the models the POPSTAR synthetic SEDs \citep{Molla2009}
and assuming the relations between metallicity and excitation for EELGs assumed by the code when no auroral emission line is provided.} {The code \hcmuv\, performs a Bayesian-like calculation that compares extinction-corrected UV emission line fluxes and their uncertainties with the prediction of a grid of models covering a wide range of values in O/H, C/O, and logU. The code calculates C/O as the average of the $\chi^2$-weighted distribution of the C/O values in the models. Then, C/O is fixed in the grid of models, and both O/H and log U are calculated as the mean of the model input values in the $\chi^2$-weighted distribution. The $\chi^2$ values for each model are derived from the quadratic relative differences between the observed and predicted emission-line ratios. The uncertainties of the derived parameters are calculated as the quadratic addition of the weighted standard deviation and the dispersion of the results after a Monte Carlo simulation.}

As input, we use CIII], OIII]$\lambda\lambda$1661,1666, H$\beta$, and [OIII]. {We do not include CIV in the input since CIV is only detected in 42\% of our sample. }{We consider the ratio OIII]$\lambda$1661/OIII]$\lambda$1666=0.44, based on photoionization models \citep{Gutkin2016}. Due to the spectral resolution of $\sim 7$\r{A} at 1666\r{A} for the C3-VUDS sample, the OIII]$\lambda\lambda$1661,1666 is blended, and then the measured flux corresponds to the doublet. On the other hand, for the C3-VANDELS sample, the spectral resolution is $\sim 3$\r{A} at 1666\r{A}, we multiplied by a factor 1.44 the measured OIII]$\lambda$1666 flux to account for the doublet.} 
We include the flux of the doublet OIII]$\lambda\lambda$1661,1666 in the input in the cases where the line is detected with S/N$>$2. For the other lines, in the case of an upper limit, the 2$\sigma$ limit is considered as input in the code with an error of 1$\sigma$, {following the methodology described in \cite{PM2023}}. The obtained C/O values are, on average, $\sim 0.11$ dex higher than that obtained if using the C3O3 calibration proposed by \cite{PM2017}. 

We find a mean value {of $\log$(C/O)$=-0.52$ ($\sim54\%$ solar)} with a typical error of 0.15 dex which is consistent with the typical value of SFGs at $z\sim 3$ based on stacking \citep{Shapley2003,Llerena2022}. Our results of $\log$(C/O) range from {$-0.90$ (23\% solar) to $-0.15$ (128\% solar)} and are consistent with the wide range of C/O values observed in local Blue Compact Dwarf (BCD) galaxies \citep{Garnett1995,Garnett1997,Kobulnicky1997,Kobulnicky1998,Izotov1999,Thuan1999,Berg2016,Senchyna2021} and Giant HII regions \citep{Garnett1995,Kurt1995,Garnett1999,Mattsson2010,Senchyna2021}. The lowest values in our sample are also within the range of values reported of log(C/O)=-0.83$\pm$0.38 at metallicity $<2$\% solar for EoR galaxies at $z\sim 8$ from JWST spectra and is consistent with the large dispersion in log(C/O) observed in $z\sim0-2$ low-metallicity dwarf galaxies without evidence of an evolution in the C/O versus O/H relationship \citep{Arellano2022}. Even a lower value log(C/O)=-1.01$\pm$0.22 has been reported at $z=6.23$ which is consistent with the expected yield from core-collapse supernovae, indicating negligible carbon enrichment from intermediate-mass stars \citep{Jones2023}.

For the gas-phase metallicity derivation, we first use the \hcmuv\, code with the same input as described before. We then compare these results with the results obtained from an alternative methodology using \pyneb\,\citep{Luridiana2012,Luridiana2015}.
To derive the total oxygen abundance we use the approximation ${\rm O/H}=\dfrac{O^+}{H^+}+\dfrac{O^{2+}}{H^+}$. To estimate $O^{2+}/H^+$, we use the \verb|getIonAbundance| task in \pyneb\, assuming the corrected electron temperature, the global [OIII] flux {and n$_{\rm e}=560$ cm$^{-3}$.} To estimate $O^+/H^+$ we use the same task with [OII]$\lambda$3727 where available (otherwise, we used [OII]$\lambda$3729 if available). To estimate T$_{\rm e}$[OII], we assume the relation T$_{\rm e}$([OII])$=$ $\dfrac{2}{{\rm T}_{\rm e}([{\rm OIII}])^{-1}+0.8}$ based on photoionization models to infer T$_{\rm e}$([OII]) from T$_{\rm e}$([OIII]) \citep{EPM2017}. The total oxygen abundance log(O/H)$_{\pyneb}$ with this method is reported in Table \ref{tab:abundance}.  

A comparison of the above two methods finds an offset {of $\sim 0.4$ dex (a factor of 2.5)} towards lower metallicity for the results based on HCm-UV. For the galaxies where the lack of detected lines the \pyneb\, method is not possible, we correct upwards the gas-phase metallicity from \hcmuv\, by a factor 0.4 dex. These values are reported in column 4 in Table \ref{tab:abundance}. 
 We find a mean value {of $\log(O/H)_{\hcmuv}+12=7.91$ (17\% solar)} with values ranging from 7 to 73\% solar. 
Finally, we compare our metallicity derivation with the expected values using the EW(CIII])-metallicity calibration  proposed by \citet{Mingozzi2022} for local analogs, which is displayed in Fig. \ref{fig:OH EW}. For most galaxies in our sample, our results are consistent within $2\sigma$ with the  local relation (blue dashed line), which has an observed scatter of $0.18$ dex. Our best fit (red dashed line) shows an offset towards lower values compared to the local relation and the observed scatter is {0.24 dex}.

We compare our estimations with the results using Strong-Line Calibration (SLC) from pure optical and UV lines. For example, if we use the O32=log$\left(\dfrac{{\rm [OIII]}\lambda\lambda4959,5007}{{\rm [OII]\lambda3727}}\right)$ calibration from \cite{EPM2021}, we find that the obtained metallicities tend to {be $\sim$0.19 dex} offset towards higher metallicities. The values found using this calibrator are reported in column 2 in Table \ref{tab:abundance}. A slightly higher mean offset of {$\sim 0.30$ dex} is found using the \cite{Bian2018} O32 calibration. Moreover, if we use calibrations based only on UV lines, for instance, He2-O3C3 \citep{Byler2020} based on the ratios $\log\left(\dfrac{{\rm OIII]}\lambda1666}{{\rm CIII]}}\right)$ and 
$\log\left(\dfrac{{\rm HeII]}\lambda1640}{{\rm CIII]}}\right)$, we find metallicities that tend to {be $\sim 0.42$ dex} lower than our estimated metallicites. This mean offset towards lower metallicities is reduced {to $\sim 0.30$ dex} if we used the proposed corrected He2-O3C3 calibration in \cite{Mingozzi2022} where offsets between UV and optical methods have been observed in local galaxies \citep{Mingozzi2022}. The values found using this calibrator are reported in column 3 in Table \ref{tab:abundance}.

Hereafter, {we consider for our discussions the metallicities derived combining UV and optical emission lines in Table \ref{tab:abundance} column (4)  since this method allows us to estimate gas-phase metallicities for the entire sample of galaxies}.

\subsection{Ionized gas kinematics}\label{sec:kinematics}

\begin{figure}[t!]
    \centering
    \includegraphics[width=0.85\columnwidth]{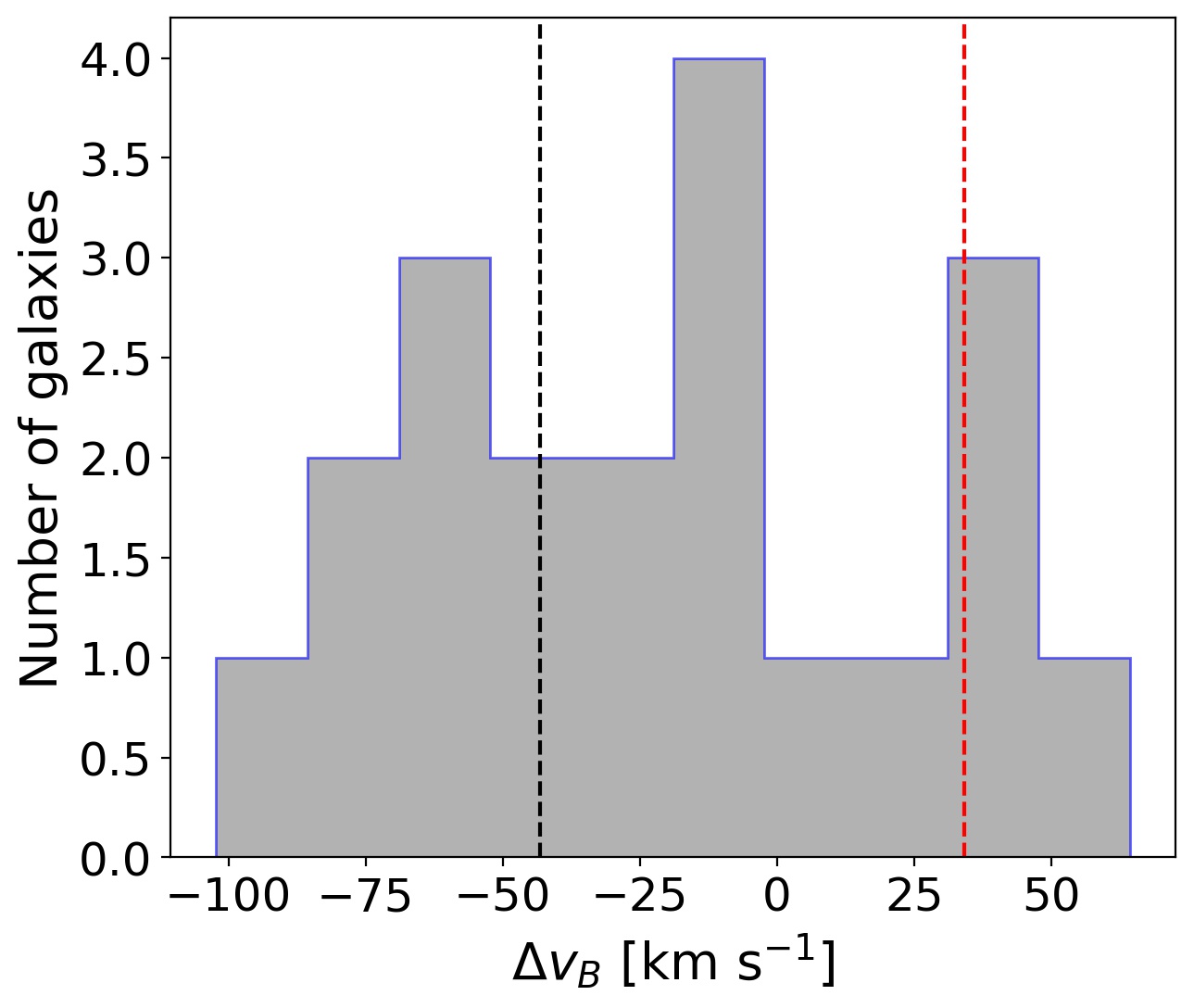}
    \caption{{Distribution of the velocity shift $\Delta v_{\rm B}$ of the broad component in our sample. The black and red dashed lines mark the mean values when the broad component is blue- and red-shifted, respectively.} }
    \label{voff}
\end{figure}

In this paper, we interpret the narrow component in the [OIII] profile as the {warm gas tracing} virial motions within SF regions while the broad component is interpreted as the turbulent outflowing gas with a velocity in reference to the systemic velocity 
\citep[cf, e.g.][]{Amorin2012,Arribas2014,Freeman2019,Hogarth2020}. The outflowing gas can be SF- or AGN-driven depending on the source of energy. From observations, SF-driven winds are closely coupled to SF properties such as $\Sigma_{\rm SFR}$ whereas AGN-driven winds are strongly correlated with stellar mass and are rare in low-mass galaxies \citep{Forster2019}. In the literature, broad components have been observed in AGN with large velocity dispersion $>200$km s$^{-1}$ \citep{RodriguezdelPino2019,Forster2019}.

In order to obtain the intrinsic velocity dispersion of the [OIII] lines we subtracted the instrumental ($\sigma_{ins}$) and thermal ($\sigma_{ther}$) widths in quadrature. For the former, we considered the resolution in the H-band of $R=3600$ (for MOSFIRE) and $R=5600$ (for X-shooter) which corresponds to $\sigma_{ins}\sim35$km s$^{-1}$ and $\sim23$km s$^{-1}$, respectively. For the latter, {we assume the same formula as in \cite{Hogarth2020} and} assume the mean T$_{\rm e}$ value obtained for the sample of 1.77$\times$10$^4$K, which leads to $\sigma_{ther}\sim0.8$km s$^{-1}$. As described in Sec. \ref{sec:fluxes}, we assume that the second component is broad if there is a difference greater than 35km s$^{-1}$ with respect to the narrower component which coincides with the global peak of the profile. 
From the 23 galaxies with two kinematic components, we find that 3 of them do not show a broad component but they show two narrow instead. These galaxies are marked in Table \ref{tab:kine}. In these cases, the complex profile is more likely to be tracing a merger instead of a turbulent outflowing gas but it is not totally clear from HST images where obvious companions are not observed (see {images} marked by the black square in Fig.   \ref{fig:hstdeltabic>2} and \ref{fig:hstdeltabic>2vuds}). We exclude these 3 galaxies from the interpretation of outflowing gas. 

Regarding the intrinsic velocity dispersion ($\sigma_{\rm vel}$) of the components, we find that the narrow component shows values ranging from $\sim$ 19 to $\sim 102$ km s$^{-1}$, with a mean value of 57 km s$^{-1}$ (typical error of 3.7 km s$^{-1}$) for the entire sample. If we consider only the galaxies that show broad components, the $\sigma_{\rm vel}$ of the narrow component are up to 70 km s$^{-1}$ with a mean value of 48 km s$^{-1}$. Meanwhile, the broad components range from 73 to 210 km s$^{-1}$ with a mean value of 121 km s$^{-1}$ (typical error of 34 km s$^{-1}$). The uncertainties are estimated directly from the Gaussian fitting {based on the covariance matrix using non-linear least squares}. {We note that the continuum non-detection implies that the true intensity and width of the broader components fit in [OIII] should be considered as possible lower limits.}

In 14 of these galaxies, the broad component is blue-shifted from the systemic velocity while in 6 of them, this component is redshifted, with a mean velocity shift ($\Delta v_{\rm B}$) of $-43$ and $34$ km s$^{-1}$, respectively {as shown in Fig. \ref{voff}. The velocity shift shows typical errors of 7 km s$^{-1}$.} In 5 of these galaxies, the velocity shift is $<10$km s$^{-1}$ that is lower than the NIR resolution, then they may show outflows very close to the systemic velocity. We find that in both cases, the broad component shows a mean velocity dispersion $\sigma_{\rm B}= 118-124$ km s$^{-1}$ which is in the range of typical widths for ionized outflows from SFGs at similar redshifts with comparable EWs \citep[e.g.][]{Matthee2021} and in local analogs \citep[e.g.][]{Amorin2012,Bosch2019,Hogarth2020}. {We highlight that none of the galaxies in our sample show the extremely broad wings ($\sigma\sim255$ km s$^{-1}$) observed in local giant HII regions dominated by supernovae feedback \citep{Castaneda1990}.} {Regarding the broad-to-narrow flux ratio (f$_{\rm B}$), our sample covers values from 24\% to 63\% with a mean value of 43\% ($\sigma=$0.13).}

For the three galaxies of the C3-VUDS sample observed with X-shooter, we find consistent results with the kinematic analysis presented in \cite{Matthee2021}, in which the authors considered only our 3 galaxies with $z<3$. 
Additionally, the resolved Ly$\alpha$ profiles of these galaxies were analyzed in \cite{Naidu2022}. For example, in the Ly$\alpha$ profile of 510583858, an intense blue peak is reported, which suggests the presence of an inflow of gas \citep{Yang2014}. This interpretation is in line with that from the kinematics of the ionized gas, where we find a redshifted broad component for the optical lines. For the other two galaxies included in \citet{Naidu2022}, they show Ly$\alpha$ profiles with weaker blue peaks compared with the dominant red peak, which is indicative of outflowing material. 

{\subsubsection{Outflow interpretation for the broad emission}}
Here we note that the blueshifted broad component can be interpreted as an outflow on the near side of the galaxy or as an inflow on the far side, and the opposite for the redshifted components. Moreover, the dust distribution may also play a role in the interpretation given that dust may be blocking the flowing \citep{Arribas2014}. Given the low dust content in our sample galaxies, we assume that in both cases, blue or redshifted, the broad component is primarily tracing outflowing material, and the shift depends on the geometry and the particular line of sight.

In order to derive the maximum velocity ($v_{\rm max}$) of the unresolved outflow, we follow the same approach as in \citet[][]{Hogarth2020,Avery2021,Concas2022} and references therein. We consider $v_{\rm max}=|\Delta v| +2\times\sigma_B$, in order to include both blue or redshifted cases {as outflowing gas where the maximum velocity will be in the line profile wings dominated by the outflow \citep[see for instance][]{RodriguezdelPino2019,Lutz2020}}. We find a mean maximum velocity of $283$km s$^{-1}$, ranging from 183 to 460 km s$^{-1}$. 

For the C3-VANDELS sample, we also perform a qualitative comparison with the kinematics analysis based on rest-UV absorption lines described in \cite{Calabro2022}. They found that the bulk ISM velocity along the line of sight is 60$\pm$10 km s$^{-1}$ for low ionization gas, consistent with the mean value found in our sample. However, for the maximum velocity, they find a mean value of 500km s$^{-1}$. The higher value compared with our results may be explained by the low resolution of the VANDELS spectra, which leads to higher unresolved line widths. Another possible explanation {is that nebular emission lines may trace denser gas than ISM absorption lines \citep[e.g.][]{Marasco2022}}. 
{In this case, the emission lines would trace smaller-scale outflows near the star cluster and with lower velocities, while the UV absorption lines would trace large-scale outflows with a galactic extent and with larger velocities} \citep[][and references therein]{Chisholm2016}.

In a galaxy-by-galaxy comparison using the bulk velocity traced by the combined fit of low-ionization lines (SiII$\lambda$1260, SiII$\lambda$1526, CII$\lambda$1334, and AlII$\lambda$1670), we find that in 8 (out of the 11 C3-VANDELS galaxies with detected broad component) the bulk velocity and the offset of the broad component agrees in the direction of the flow (i.e. if they are blue or redshifted). In 2 cases, we find a broad component that is blueshifted while the bulk velocity is redshifted but they are consistent within the uncertainties. Only in one case, did we find a redshifted broad component and a negative bulk velocity. Our results suggest that the broad component of [OIII] is indeed tracing flowing material that roughly agrees to be blue or redshifted in both absorption and emission lines. 

{\subsubsection{Outflow velocities and star formation properties}}
Now, we relate the maximum velocity of the outflow with the global properties of our sample. We find a weak positive correlation between the maximum velocity and the {galaxy} stellar mass ({Pearson correlation coefficients $\rho$:0.22}, 0.91$\sigma$ significance). This trend suggests that more massive galaxies can power faster outflows compared with less massive galaxies. {Regarding the relation  with SED-derived SFR, we find no correlation ($\rho\sim 0$,  0.36$\sigma$ significance). }
The correlation is stronger {($\rho$=0.16, 0.66$\sigma$ significance)} when we consider the instantaneous SFR traced by H$\beta$ (or H$\alpha$). The slope of the best fit is 0.06, which is in agreement with the trend found in \cite{Arribas2014} for a sample of local luminous and ultra-luminous infrared galaxies (U/LIRGs) at galactic and sub-galactic (i.e., star-forming clumps) scales. This suggests that longer timescale SFR is not tracing the actual outflow, while the instantaneous SFR is a more real tracer of the outflow and the maximum velocity that is reachable. {This might be the reason why only a marginal (at 2$\sigma$ level) correlation was found between maximum velocity and SFR in \cite{Calabro2022}.}

\begin{figure}[t!]
    \centering
    \includegraphics[width=\columnwidth]{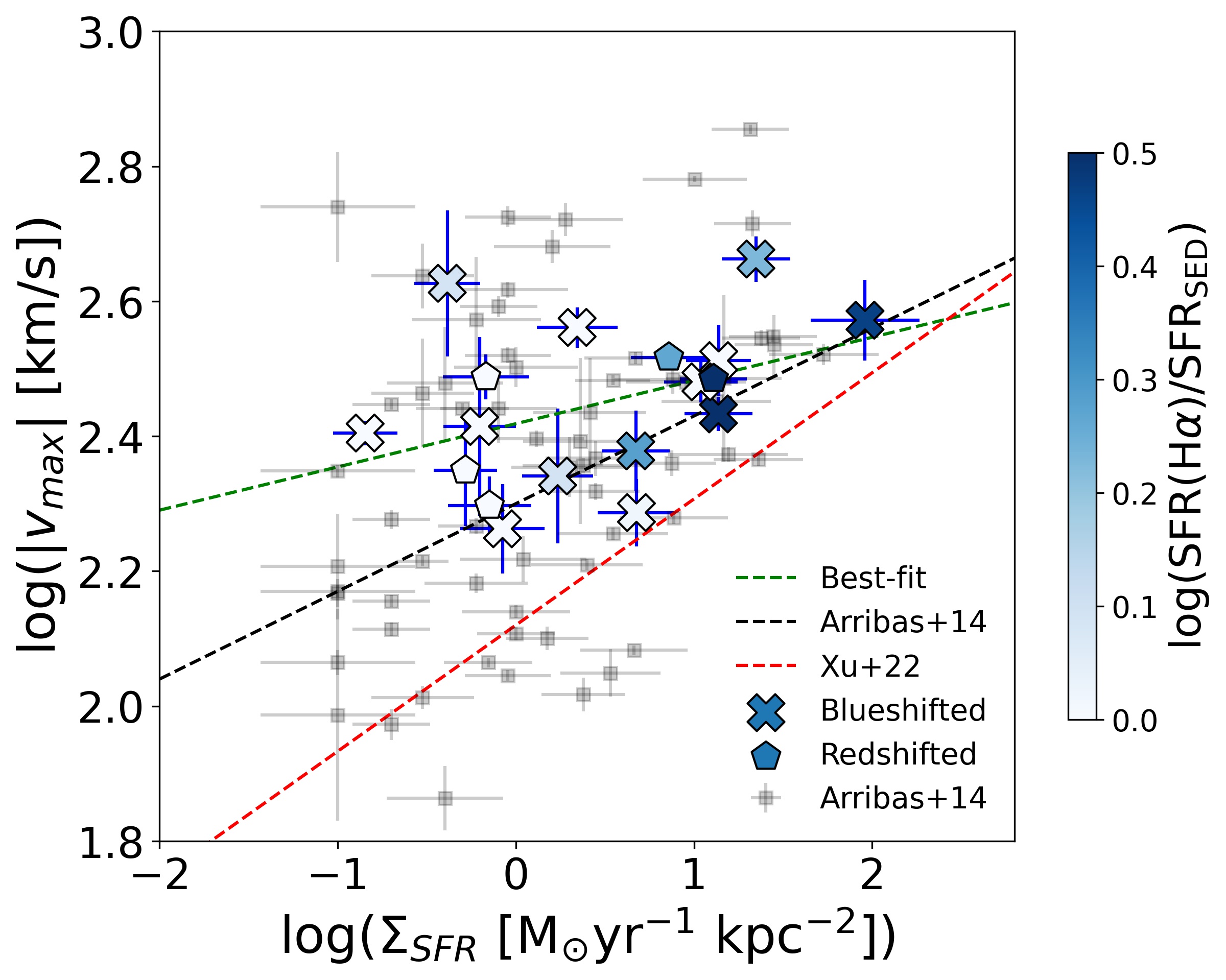}
    \caption{Relation between outflow velocity and $\Sigma_{SFR}$. Our sample is color-coded by the SFR(H$\beta$)/SFR$_{\rm SED}$ ratio. Our sample is divided by galaxies with blue (cross symbol) and redshifted (pentagon symbol) broad components. The black dashed line corresponds to the best slope from \cite{Arribas2014}, and the black squares are their observed data. The green dashed line is our best fit. The red dashed line is the best fit from \cite{Xu2022classy}. {Two galaxies with broad components (CDFS023527 and 511245444) are excluded from this plot since H$\alpha$ and H$\beta$ are not included in the observed spectral range.}}
    \label{vout}
\end{figure}

In Fig. \ref{vout}, we explore the relationship between the maximum velocity of the outflow and the $\Sigma_{\rm{SFR}}$. A relation between both properties is expected as explained by simple models \citep[e.g.][]{Heckman2015,Xu2022classy}. The warm ionized outflow is described by a collection of clouds or filaments driven outward by the momentum transferred by the very hot gas of the stellar ejecta from the starburst, which creates a fast-moving wind that accelerates the ambient gas. We find a weak positive correlation ($\rho$=0.40,  1.69$\sigma$ significance) {which is consistent with what was also found from absorption lines \citep[e.g.][]{Calabro2022}. This suggests that even though they might trace different regions, they might be triggered by the exact mechanisms and with a similar origin.} Interestingly, we find that our best fit shows a shallower slope of 0.06$\pm$0.03 (green dashed line) compared to the slope of 0.13 (scaled black dashed line) found in \cite{Arribas2014}. Here, we use the instantaneous SFR traced by the narrow component of H$\beta$ (or H$\alpha$), assuming the same ratio observed in [OIII]. Based on UV absorption lines, a similar trend is observed with a {steeper} slope of 0.18 (red dashed line) found in the CLASSY survey with local metal-poor galaxies \citep{Xu2022classy}.  

Additionally, we use the SFR(H$\beta$)/SFR$_{\rm SED}$ ratio as a proxy of the burstiness of the galaxy since both SFR estimators trace different timescales, with SFR(H$\beta$) more sensitive to younger ages. {We find a mean SFR(H$\beta$)/SFR$_{\rm SED}$ ratio of 1.4 with values ranging from 0.4 to 5.1. Half of the sample shows SFR(H$\beta$) higher than SFR$_{\rm SED}$.} We find that the galaxies with the larger {burstiness} tend to show also high $\Sigma_{\rm SFR}$, which is displayed in the color-codes in Fig. \ref{vout}. 

\section{Discussions}\label{sec:discussion}

\subsection{Mass-Metallicity {and Fundamental Metallicity relations}}\label{sec:MZR}

\begin{figure}[t!]
    \centering
    \includegraphics[width=\columnwidth]{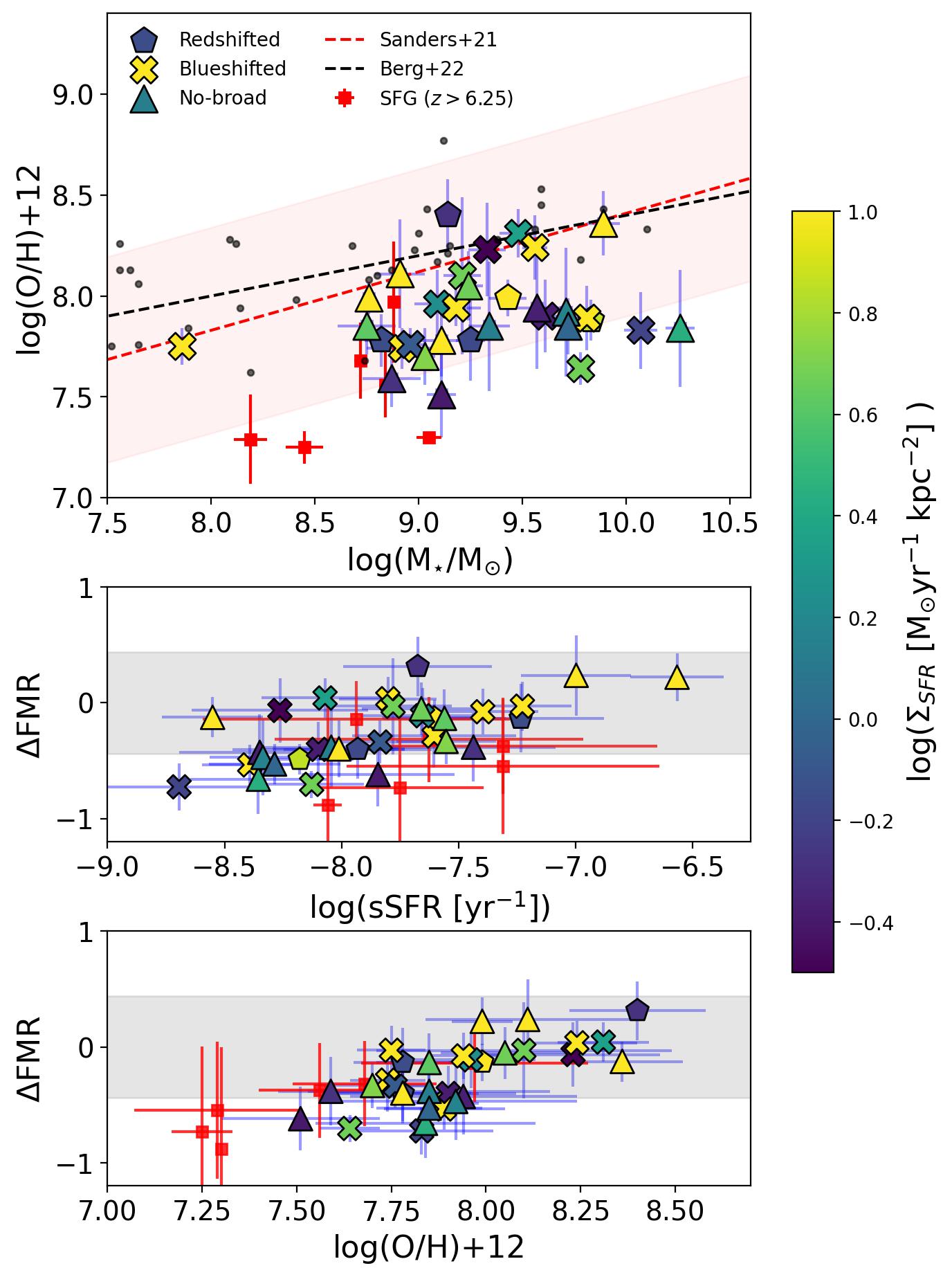}
    \caption{\textit{Top:} Mass-Metallicity relation: The red dashed line is the MZR of SFGs at $z\sim 3.3$ from \cite{Sanders2021} and the red shaded region is the observed 3$\sigma$ scatter. For comparison, metal-poor local galaxies from \cite{Berg2022} are included as black circles and the corresponding MZR as the black dashed line. The red squares are SFGs at $z>6.25$ from \cite{Matthee2022} (stack) and \cite{Heintz2022} (single {lensed} galaxies). \textit{Middle (Bottom):} Differences between the metallicity from FMR and our metallicity estimations as a function of sSFR (metallicity). The gray shaded region is the {2$\sigma$} observed scatter of 0.22 dex for the FMR at $z\sim 3.3$ \citep{Sanders2021}. {In all panels, our sample is color-coded by $\Sigma_{\rm SFR}$, and the triangle symbols are the galaxies that do not show broad components. At the same time, pentagons and crosses are galaxies in our sample that show red- and blue-shifted broad components, respectively.}}
    \label{mzr}
\end{figure}

The gas-phase metallicity of galaxies encodes information on how physical processes drive the evolution of galaxies. The emergence of scaling relations, such as MZR, plays a fundamental role in understanding galaxy formation processes and constraining physical models of galaxy evolution. 
{As galaxies grow, feedback processes are responsible for shaping the MZR, as they promote the dispersion and removal of a significant fraction of metals from the star-forming regions into the CGM. The selective loss of newly synthesized heavy elements could be particularly important in low-mass galaxies due to their shallow potential wells \citep[e.g.][]{Tremonti2004,Andrews2013}. Besides, the slope and normalization of the MZR may evolve with redshift, with high-$z$ galaxies showing lower metallicity for a given stellar mass \citep[e.g.][]{Sanders2021}. Simulations  explain this evolution invoking the higher gas fractions of galaxies at higher redshifts} \citep{Torrey2019}. 

{In Figure \ref{mzr}, we show that our subset of galaxies with reliable H$\beta$ luminosities follow the MZR built using MOSDEF main-sequence galaxies at similar redshift by \citet{Sanders2021} and the $z\sim$0 relation for local analogs of similar stellar mass in the CLASSY survey \citep{Berg2022}. Our sample is mostly within 3$\sigma$, the intrinsic scatter of these relations, with a few galaxies showing lower metallicity. We interpret this apparent offset as a likely selection effect, i.e., our sample includes strong emission-line galaxies which tend to have higher SFR, more extreme ionization conditions, and lower metallicity than the sample used to establish the MZR we are using for reference. Indeed, the high log([OIII]/H$\beta$)$>0.6$ of our sample is roughly the higher excitation in the lower-mass bin used for the MZR in \cite{Sanders2021}. This is also evident from Fig.~\ref{fig:sigma_o32}, showing that our sample has larger [OIII]/[OII] ratios than MOSDEF galaxies used for the determination of the MZR at $z\sim$3.} {Compared to strong [OIII] emitters at higher redshifts, we find our galaxies following  a similar trend to that found for a more extreme subset of EoR galaxies \citep{Matthee2022,Heintz2022}. }

The scatter of galaxies around the median MZR correlates {with SFR, i.e the FMR \citep[][]{Mannucci2010,Curti2020}, which implies that 
at fixed stellar mass, galaxies with higher SFR show lower metallicities. Observationally, the FMR appears invariant with redshift up to $z \sim$ 3.5 \citep[e.g.][]{Curti2020,Sanders2021}, although low-mass starbursting galaxies may show clear deviations \citep{Amorin2014,Calabro2017}. The shape of the FMR is found moduled by the age of the stellar populations, with younger  ($<150$ Myr) and more bursty SF showing lower metallicity \citep{Duarte2022}. Models and simulations find that the FMR results from the smooth evolution of galaxies in a quasi-equilibrium state, which is regulated by inflows and outflows over time \citep{Lilly2013,Nelson2019TNG}. The strength of the implied SFR-metallicity relation is found dependent on feedback via the shape of the star formation history, particularly for low-mass starbursting galaxies \citep{Torrey2018}.}

{In order to explore the position of our sample in the FMR, we compare gas-phase metallicities obtained with the expected FMR values. We define the parameter}  $\Delta\rm{FMR}=\log({\rm O/H})-\log({\rm O/H})_{\rm{FMR}}$, where $\log({\rm O/H})_{\rm{FMR}}$ is the relation found in \cite{Sanders2021} that depends on the stellar mass and SFR.  {While most of the galaxies in our sample are consistent (within the uncertainties in the gas-phase metallicity) with the observed {$2\sigma$} scatter of the relation} ({middle and} bottom panel in Fig. \ref{mzr}), {we find a weak negative trend between $\Delta$FMR and specific SFR (sSFR=SFR/M$_{\star}$) similar to that shown in galaxies at $z>6.25$ with {comparable} sSFR \citep[e.g.][]{Matthee2022,Heintz2022}. However, no dependence is found with $\Sigma_{\rm SFR}$.} {Although we find the above trend appears independent of the method used to estimate the metallicity, we should take this result with caution as the uncertainties in metallicity are still  large.} 

{Indeed, it is worth noticing that some caveats may affect the above comparisons}. First, stellar masses in our work are obtained assuming sub-solar stellar metallicities, which could lead to differences of up to $\sim$0.3 dex compared to assuming solar values as in \cite{Sanders2021}, as mentioned in Sec. \ref{sec:sed}. {Second, the different methods used to derive metallicities may lead to possible systematic differences. We use a Te-consistent method based on the comparison of UV and optical emission line ratios with predictions from photoionization models. As described in Section \ref{sec:abundance}, metallicity differences of up to $\sim$0.3 dex can be found when a sole strong-line calibration is used instead. Acknowledging these caveats, rather than absolute values we are interested in the emerging trends of such comparisons.}

{In conclusion}, we find that our sample follows the MZR at $z\sim$3 with {a subset of them showing} slightly lower metallicities. {These galaxies also show lower metallicities than expected} from the FMR, which {could be due to the relatively extreme ionization properties of their young and intense SF regions}. Recent gas accretion fueling a compact starburst and the intense stellar feedback it produces, i.e. inflows and outflows, respectively, are physical mechanisms that could naturally drive the observed offset in the MZR and FMR towards lower metallicities. {Probing relative abundances (such as C/O or N/O) which depend on the SF history of galaxies \citep[e.g.][]{Vincenzo2018,Berg_2019} can be a useful tool to explore this hypothesis.}  

\subsection{The C/O-O/H relation}\label{sec:co-oh}
\begin{figure*}[t!]
    \centering
    \includegraphics[width=\columnwidth]{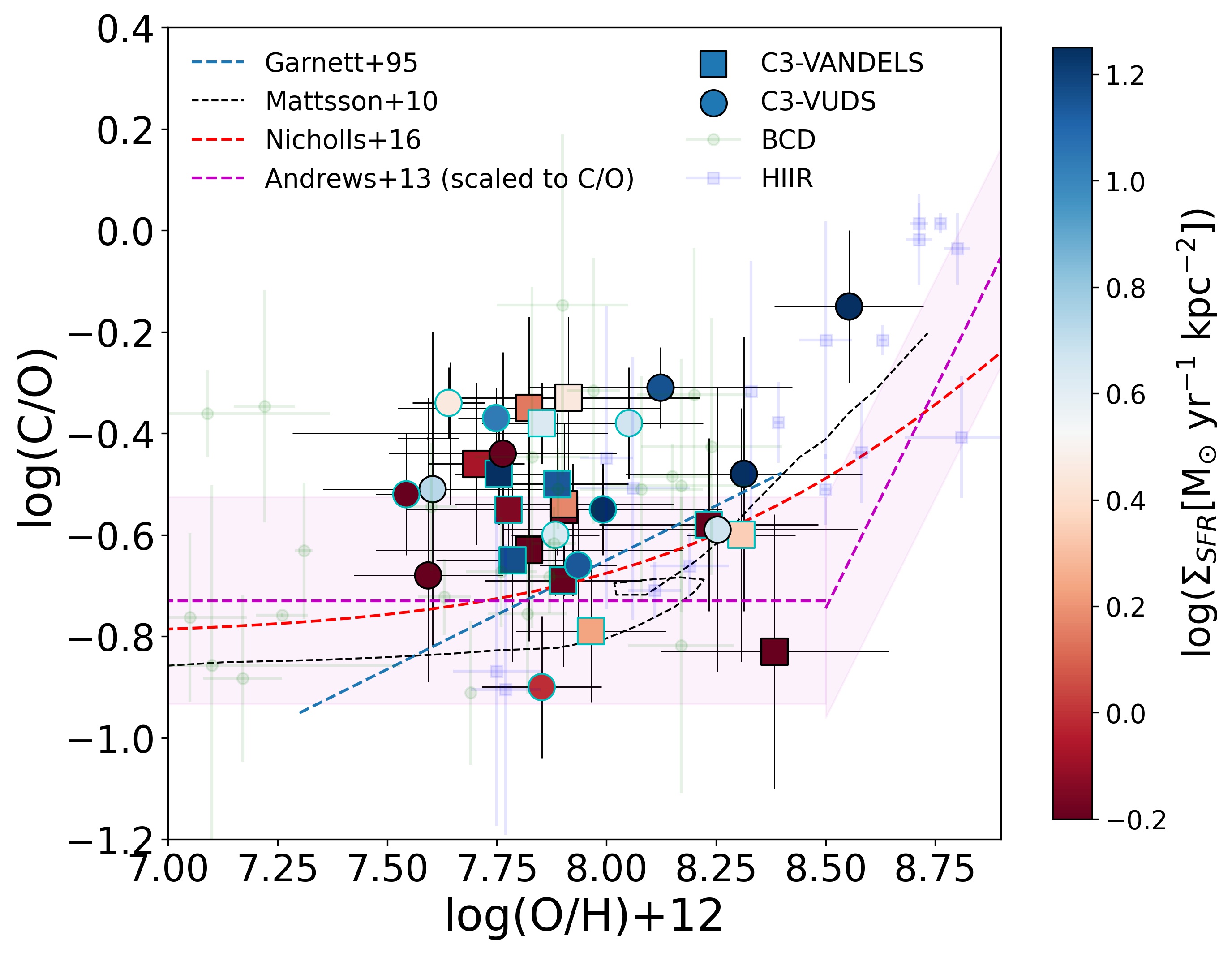}\,\includegraphics[width=\columnwidth]{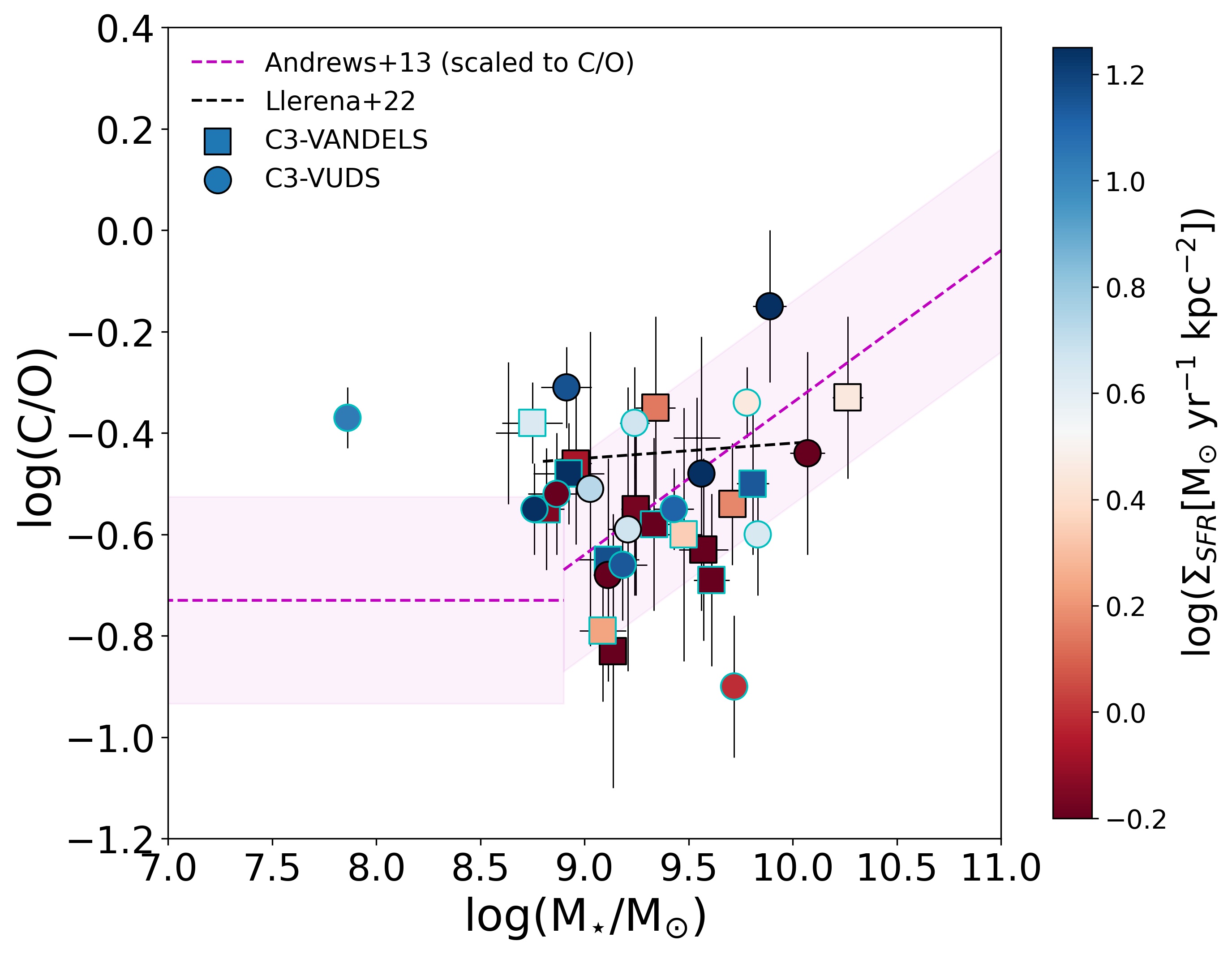}
    \caption{{Left panel:} C/O-O/H relation. Our sample is color-coded by $\Sigma_{\rm SFR}$. The symbols with cyan edges are galaxies with limits in OIII]$\lambda$1666. We compare our results with local BCD galaxies \citep[small green circles,][]{Garnett1995,Garnett1997,Kobulnicky1997,Kobulnicky1998,Izotov1999,Thuan1999,Berg2016,Senchyna2021} and HII regions \citep[small blue squares][]{Garnett1995,Kurt1995,Garnett1999,Mattsson2010,Senchyna2021}. We also show chemical evolution models from the literature as dashed lines with colors described in the legend \citep{Garnett1995,Mattsson2010,Nicholls2017}. {The magenta dashed line is the scaled N/O-mass  relation in \cite{Andrews2013} assuming a constant C/N factor based on \cite{Berg_2019}. The magenta-shaded region is the 1$\sigma$ uncertainty considering the observed scatter in the relation and the conversion factor. \textit{Right panel}: Relation between C/O and stellar mass. Symbols are the same as in the left panel. In this panel, the black dashed line is the relation presented in \cite{Llerena2022} at $z\sim$3 based on stacking.}}
    \label{co-oh}
\end{figure*}

The C/O {abundance} may provide us with general trends in the evolutionary state of a galaxy and its ISM. 
Models \citep[e.g.][]{Henry2000,Molla2015,Mattsson2010} and observations of local galaxies \citep{Garnett1995,Berg2016,Berg_2019} show that C/O increases with metallicity for galaxies with $Z\gtrsim$20\% solar.  
This trend can be explained because C is primarily produced by the triple-$\alpha$ process in both massive and low- to intermediate-mass stars but, in massive stars, carbon arises almost exclusively from 
metallicity-dependent stellar winds, mass loss, and ISM enrichment, which are larger at higher metallicities \citep{Henry2000}. {Instead, in younger metal-poor systems the delayed release of C (mostly produced by low- and intermediate-mass) relative to O (produced almost exclusively by massive stars) appears as the driver of the observed trend }\citep{Garnett1995}. However, {the large dispersion of C/O values over a large range in metallicity found for galaxies at $z\sim$0-2 ($\sim0.2$ dex) suggests that the C/O abundance is largely sensitive to other factors, such as the detailed SFH, with longer burst durations and lower star formation efficiencies corresponding to low C/O ratios} \citep{Berg_2019}.

{In Figure~\ref{co-oh} (left panel), we show our galaxy sample in the C/O-O/H plane. 
We find no apparent increase of C/O with metallicity, as models predict \citep[e.g.][]{Mattsson2010,Nicholls2017}, but a large scatter of C/O values around metallicity $\sim10-20$\% solar. Despite the uncertainties in both C/O and O/H, part of this scatter could be physical, as discussed in previous works showing similar findings  at low and high redshifts \citep[e.g.][]{Amorin2017,Berg_2019,Llerena2022}. 
In Fig.~\ref{co-oh}, we do not find a clear correlation of the C/O scatter with $\Sigma_{\rm SFR}$ or sSFR, which may suggest a more local effect affecting the C/O-O/H relation at low metallicity, a topic that will be addressed in a future, more specific study. Finally, these results also suggest some caution in using C/O as an indicator of metallicity, as it may be subject to large uncertainty and possible selection effects.} 

{On the other hand, the right panel in Fig. \ref{co-oh} shows that our galaxies appear consistent with an increase of C/O with stellar mass. In order to compare with local galaxies, we re-scaled the N/O-stellar mass relation reported in \cite{Andrews2013} and  assumed a constant C/N conversion \citep{Berg_2019} to obtain the relation shown by the magenta dashed line. 
Our sample shows a large scatter in C/O for a given stellar mass, roughly consistent with \citet{Llerena2022} for CIII] emitters at $z\sim$3 using stacking.   
However, they appear to follow the trend expected for their stellar mass, suggesting that a fraction of their C/O may have a secondary origin. This is not seen in the O/H-C/O plane. While we do not find a correlation with outflow velocities and  $\Sigma_{\rm SFR}$, one possible interpretation for these trends is that a recent metal-poor inflow may produce a dilution of O/H while keeping the C/O as large as expected for their stellar mass. To explore further this and other possible interpretations larger representative samples over a wide range of   mass and metallicity are needed. }
JWST spectroscopy will certainly help to reduce uncertainties in the chemical abundance of high-z galaxies, thus providing a more robust interpretation of the C/O-O/H relation for samples at $z\gtrsim$\,4 \citep[e.g.][]{Arellano2022}. 
{In conclusion, the low metallicity and relatively large C/O abundance of our sample suggest that these galaxies are in a relatively active and early phase of chemical enrichment. The complex interplay between metal content and stellar feedback will be discussed in the following sections.}

\begin{figure}[t!]
    \centering
    \includegraphics[width=1\columnwidth]{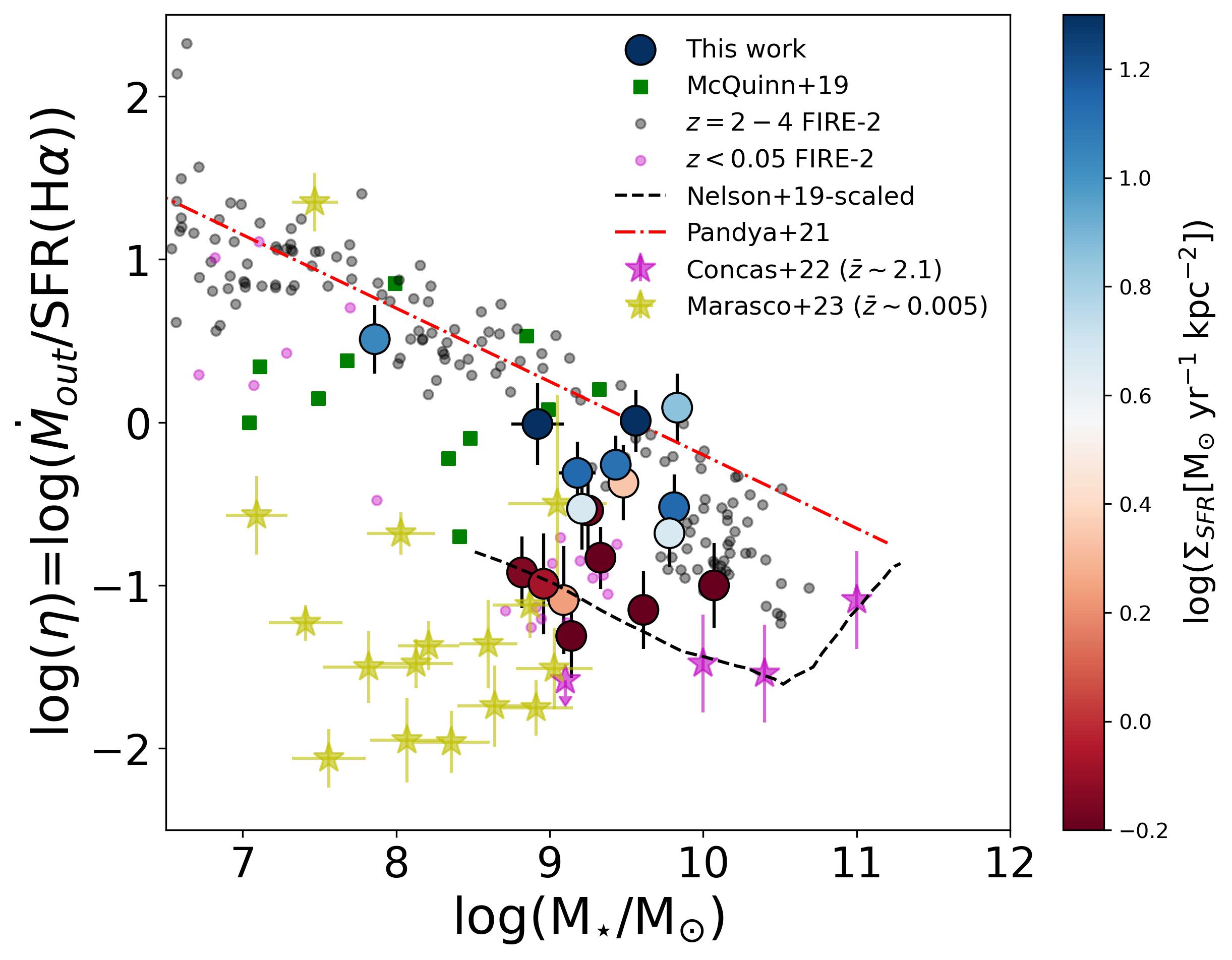}\\
    \includegraphics[width=1\columnwidth]{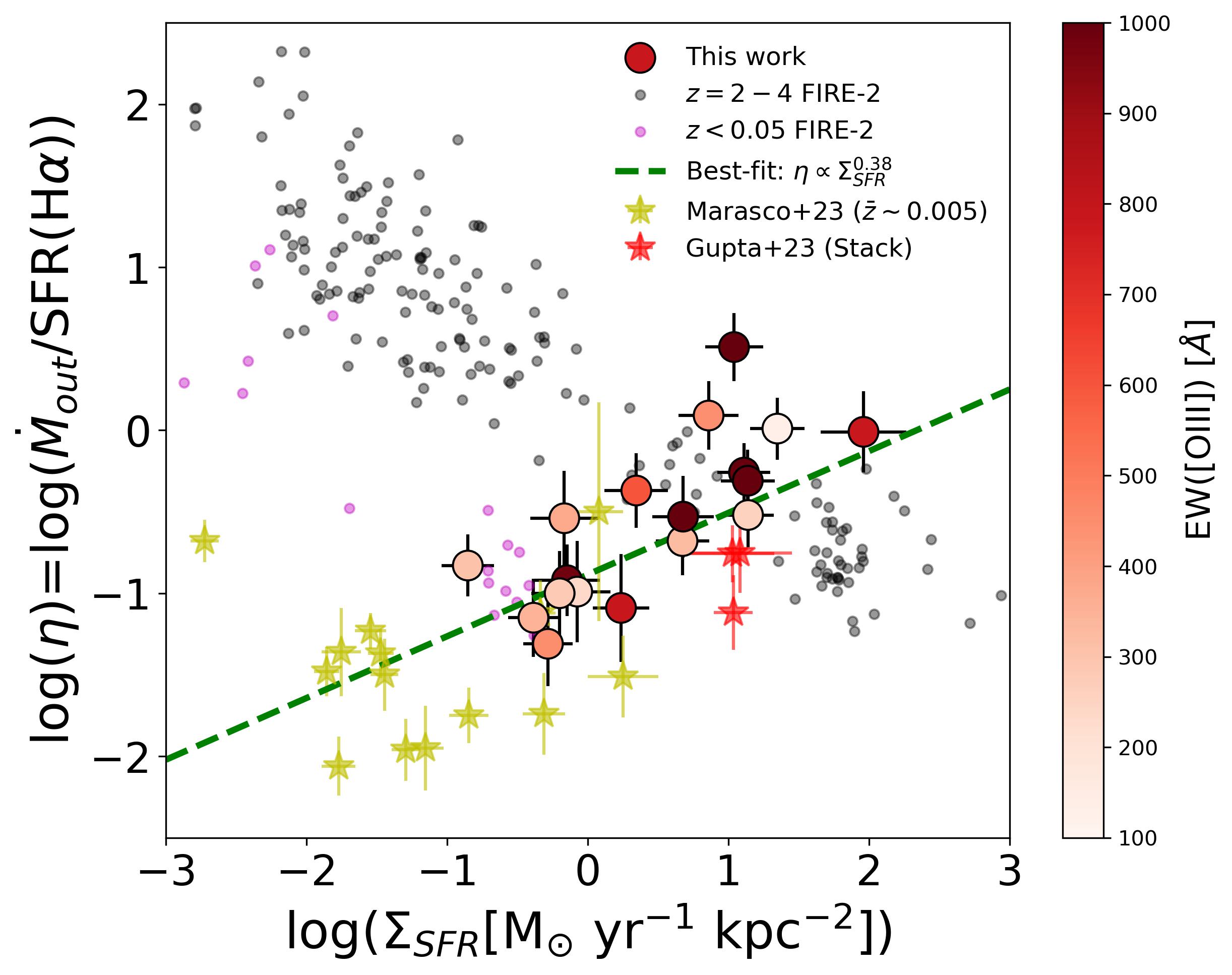}
    \caption{\textit{Top panel}: Relation between mass loading factor and stellar mass. Our sample is color-coded by $\Sigma_{\rm SFR}$. The black dashed line is the re-scaled relation found in simulations according to \cite{Nelson2019}. The dotted-dashed line is the relation found in \cite{Pandya2021}. {The small black (magenta) circles are single simulated haloes from \cite{Pandya2021} at redshift range 2-4 ($<0.05$)}. 
    We also include observational results from \cite{McQuinn2019,Concas2022,Marasco2022}. \textit{Bottom panel:} Relation between mass loading factor and $\Sigma_{\rm SFR}$. Symbols are the same as in the top panel, but our sample is color-coded by EW([OIII]). The red stars are stacks at $z=3-4$ from \cite{Gupta2022}. {The green dashed line is the best fit for $\log(\Sigma_{\rm SFR}[M_{\odot}$yr$^{-1}$kpc$^{-2}])>-3$ including our sample and the observed galaxies and stacks from literature.}}
    \label{fig:loading}
\end{figure}

\subsection{{Outflow properties and their relation with star formation rate density}}\label{sec:outflows}

One important parameter in understanding the effects of outflows in the properties of galaxies is the mass-loading factor ($\eta=\dot{M}_{\rm out}/$SFR) which measures how efficiently outflows remove gas from the galaxy relative to the formation of stars. To be consistent with previous works, we follow the same models in \cite{Concas2022} to calculate the mass loading factor $\eta$. In the case of a multiconical or spherical outflow and a constant outflow velocity, the mass outflow rate $\dot{M}_{\rm out}$ from the H$\alpha$ line is given by
\begin{equation}
    \dot{M}_{\rm out}=1.02\times 10^{-9}\left(\dfrac{v_{\rm max}}{\rm{km s}^{-1}}\right)\left(\dfrac{M^{\rm{H\alpha}}_{\rm out}}{\rm{M}_{\odot}}\right)\left(\dfrac{\rm{kpc}}{R_{\rm out}}\right)C\,[\rm{M}_{\odot}\, \rm{yr}^{-1}],\label{eq:mass-out-rate}
\end{equation}
where the factor $C$ depends on the assumed outflow
history, $R_{\rm out}$ is the radius of the outflow, {$v_{\rm max}$ is the maximum outflow velocity defined in Sec. \ref{sec:kinematics}} and $M^{\rm{H\alpha}}_{out}$ is the mass outflow which is given by
\begin{equation}
    M^{\rm{H\alpha}}_{\rm out}=3.2\times 10^5\left(\dfrac{L^{\rm{H\alpha}}_B}{10^{40}\rm{erg s}^{-1}}\right)\left(\dfrac{100 \rm{cm}^{-3}}{n_{\rm e}}\right)\rm{M}_{\odot},\label{eq:mass-out}
\end{equation}
where $L^{\rm{H\alpha}}_B$ is the dust-corrected luminosity of the outflow (broad) component of H$\alpha$. To estimate $L^{\rm{H\alpha}}_B$ we use the dust-corrected H$\beta$ luminosity and assumed the broad-to-narrow flux ratio from [OIII]. We do not {use} the model-based [OIII] since it depends on the metallicity  \citep[see][]{Concas2022} and {we prefer to keep the mass outflow rate independent of the method used to derive metallicity. 
However, we checked that the values obtained by using the two lines differ by $0.28$ dex, with higher $\dot{M}_{\rm out}$ values when using the [OIII] line.} This difference is {also} discussed in \cite{Concas2022} {and} they find agreement only when a higher metallicity of 50\% solar is assumed.

From Eq. \ref{eq:mass-out}, we find { the outflow mass ranging from 10$^{7.0}$ to 10$^{8.03}\rm{M}_{\odot}$, with a mean value} of 10$^{7.58}\rm{M}_{\odot}$. Since our data do not allow a reliable estimate of n$_e$ for the outflow component (i.e. using only the broad emission of  [OII]$\lambda\lambda$3727,3729), we assumed an outflow mean electron density of { 380 cm$^{-3}$, 
which is based on  the stacking of 33 galaxies with SF-driven outflows \citep{Forster2019}. This assumption makes our estimate consistent with other works using the same value  \citep[e.g.][]{Davies2019, Concas2022,Gupta2022}}. {If we use the mean (global) density of 560 cm$^{-3}$ obtained for our sample, the resulting $\eta$ values are 0.16 dex lower, which is smaller than the mean uncertainties (0.23 dex), it does not affect our results and conclusions.}

For the mass outflow rate, we assume a constant outflow rate that starts at $-t=-{\rm R}_{\rm out}/v_{\rm max}$ that leads to $C=1$, as in other works \citep{Concas2022}. We assume that the outflow radius ${\rm R}_{\rm out}$  is the effective radius measured in the F160W band in WFC3/HST, as described in Section \ref{sec:Sigma}, {which has values ranging from 0.2 to 3.6 kpc, with a mean ${\rm R}_{\rm out}=$1.6 kpc}. We assume that the outflow velocity is the maximum velocity (see Table \ref{tab:kine}, Section \ref{sec:kinematics}). We find mass outflow rates ranging from 1.4 to 57 $\rm{M}_{\odot}$ yr$^{-1}$, with a mean $\dot{M}_{\rm out}=$13.3 $\rm{M}_{\odot}$ yr$^{-1}$,  {nearly half the mean SFR derived for the sample}. 
As a comparison, we find that our sample is in agreement with the relation M$_{\rm out}$-SFR reported in \cite{Avery2021} at the more extreme SFR values, which indicates that SF is the driver of the outflow. They found a linear slope of $\dot{M}_{\rm{out}}\propto \rm{SFR}^{0.97}$ for integrated outflows on local MaNGA galaxies using H$\alpha$ to measure outflows. This relation is also consistent with the values reported in \cite{Marasco2022} for a sample of 19 nearby systems above the local MS and that shows lower values of SFRs and $\dot{M}_{\rm out}$ than our sample. We also note that some of our galaxies show higher values than those found in \cite{Xu2022classy} for a sample of local dwarf galaxies using UV absorption lines with similar high SFRs. 

In order to estimate $\eta$, we considered the instantaneous SFR using Balmer lines, as explained in Sec. \ref{sec:Sigma}. We report the obtained values in Table \ref{tab:kine}. We find a mean value of {$\eta=0.54$} 
with values ranging from 0.05 to 3.26. Our mean value is higher than the estimation {of $\eta\sim0.2$} using stacking from a sample of SFGs at $z=3-4$ with EW(H$\beta$+[OIII]$\lambda$5007)$>$600\r{A} \citep{Gupta2022} and for a local GP galaxy \citep{Hogarth2020}. {On the other hand, our mean value is lower than the typical $\eta$ of 1.5 for neutral and low-ionization gas of SFGs at $z\sim3$ derived from absorption lines \citep[e.g.][]{Calabro2022}.} 

In Figure~\ref{fig:loading} (top panel), we display the relation between the mass loading factor and stellar mass for our sample galaxies and compare it with similar {theoretical and observational} results from the literature. {Our galaxies show a very large range of $\eta$ values in a limited range of $M_{\star}$. While this precludes a rigorous study of possible trends in $\eta$ with $M_{\star}$ at $z\sim3$, we find that, for a given stellar mass, galaxies show more than one order of magnitude scatter in $\eta$ values according to their star formation rate surface density.  }
{We find a clear trend in the scatter with galaxies with more compact star formation, i.e., $\Sigma_{\rm SFR} \gtrsim$10$M_{\odot}$yr$^{-1}$kpc$^{-2}$, showing larger $\eta$ for a given $M_{\star}$. Instead, galaxies with $\Sigma_{\rm SFR}\lessapprox10$M$_{\odot}$yr$^{-1}$kpc$^{-2}$ show on average $\sim$1 dex lower $\eta$ values. While the former appears roughly consistent with the trend predicted by FIRE-2 simulations \citep[dotted-dashed lines,][]{Pandya2021}, the latter appears more consistent with the scaled (by a factor $\sim 1/141$ lower) trend predicted by the Illustris-TNG simulations \citep{Nelson2019}. }
 
\begin{figure*}[t!]
    \centering
    \includegraphics[width=0.95\columnwidth]{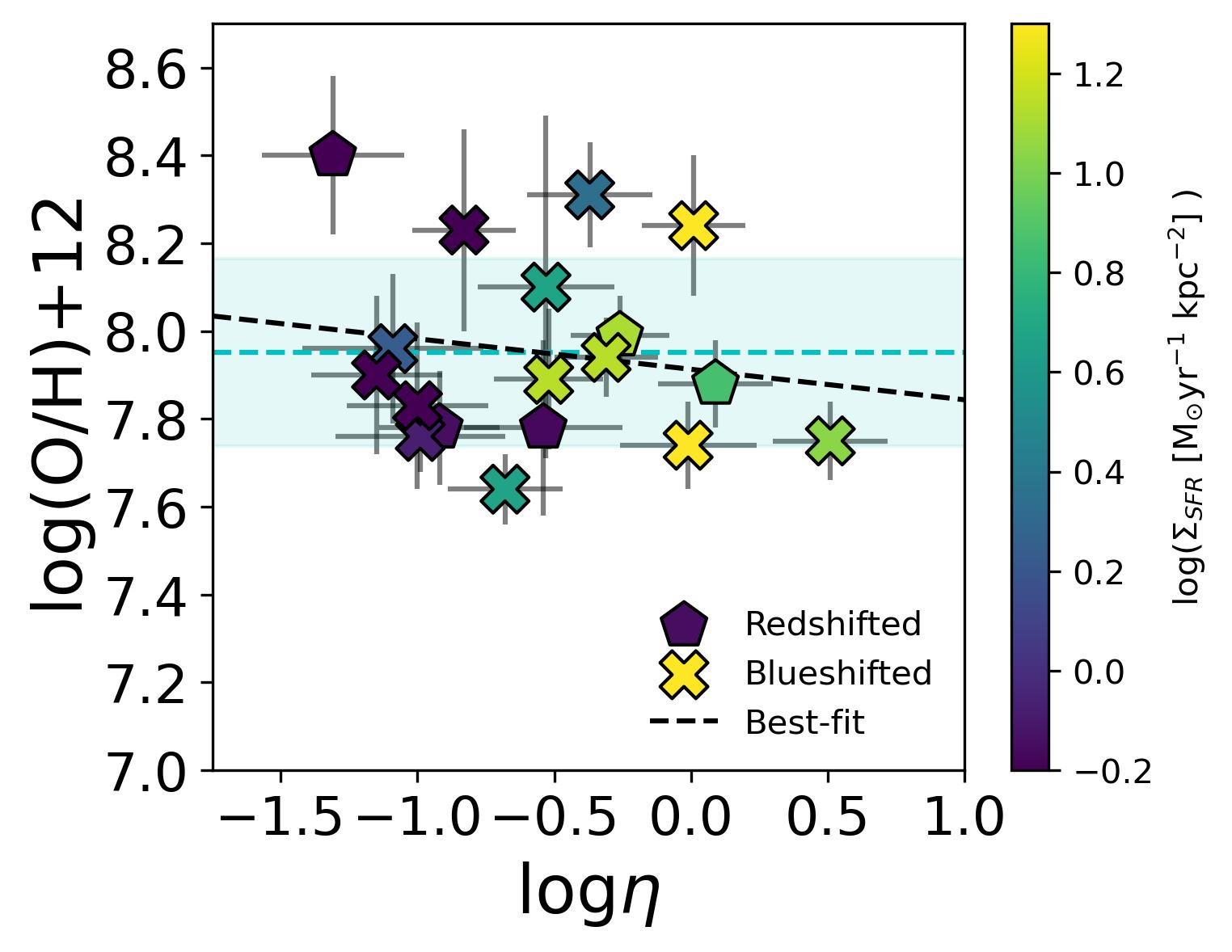}\,\includegraphics[width=0.95\columnwidth]{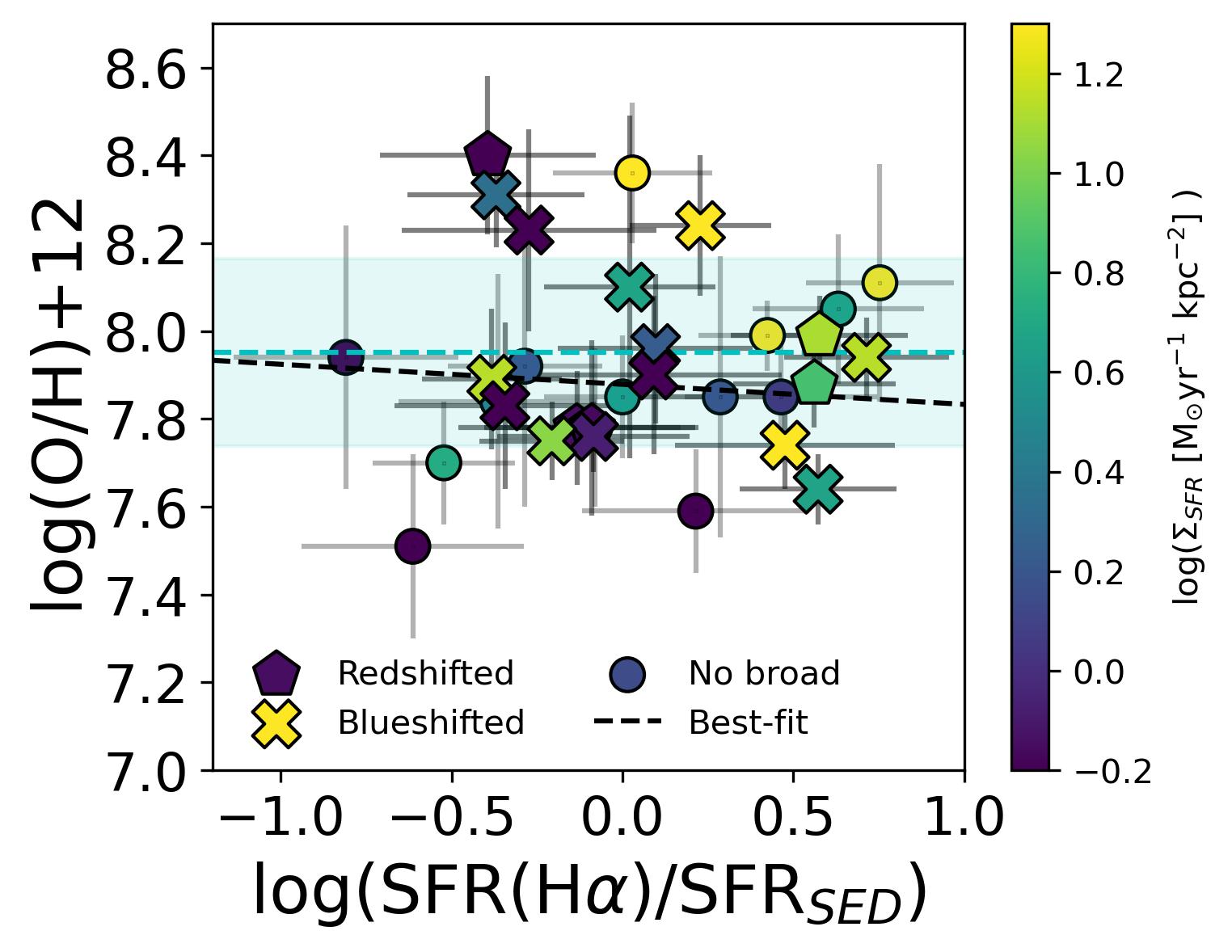}\\
    \includegraphics[width=0.95\columnwidth]{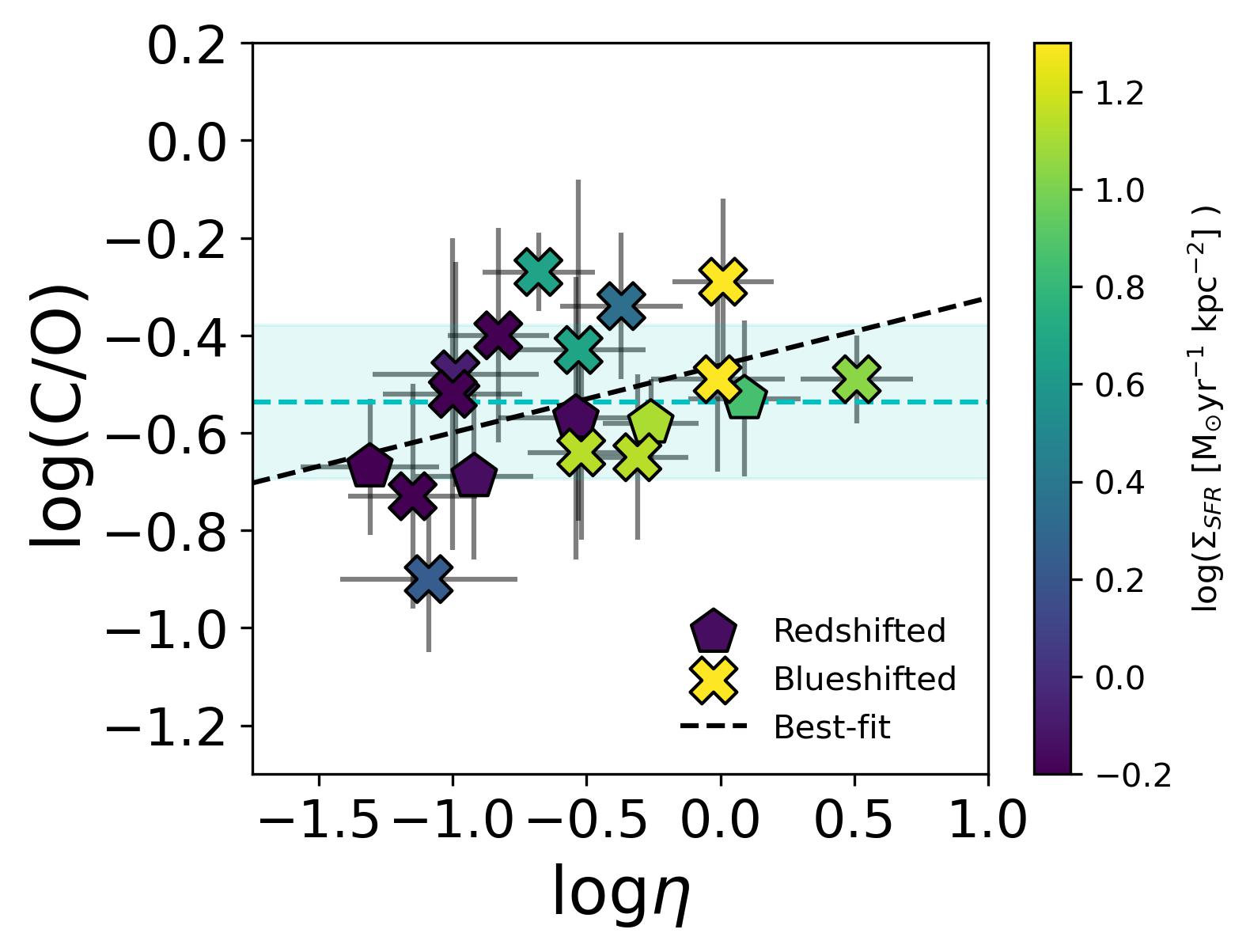}\,\includegraphics[width=0.95\columnwidth]{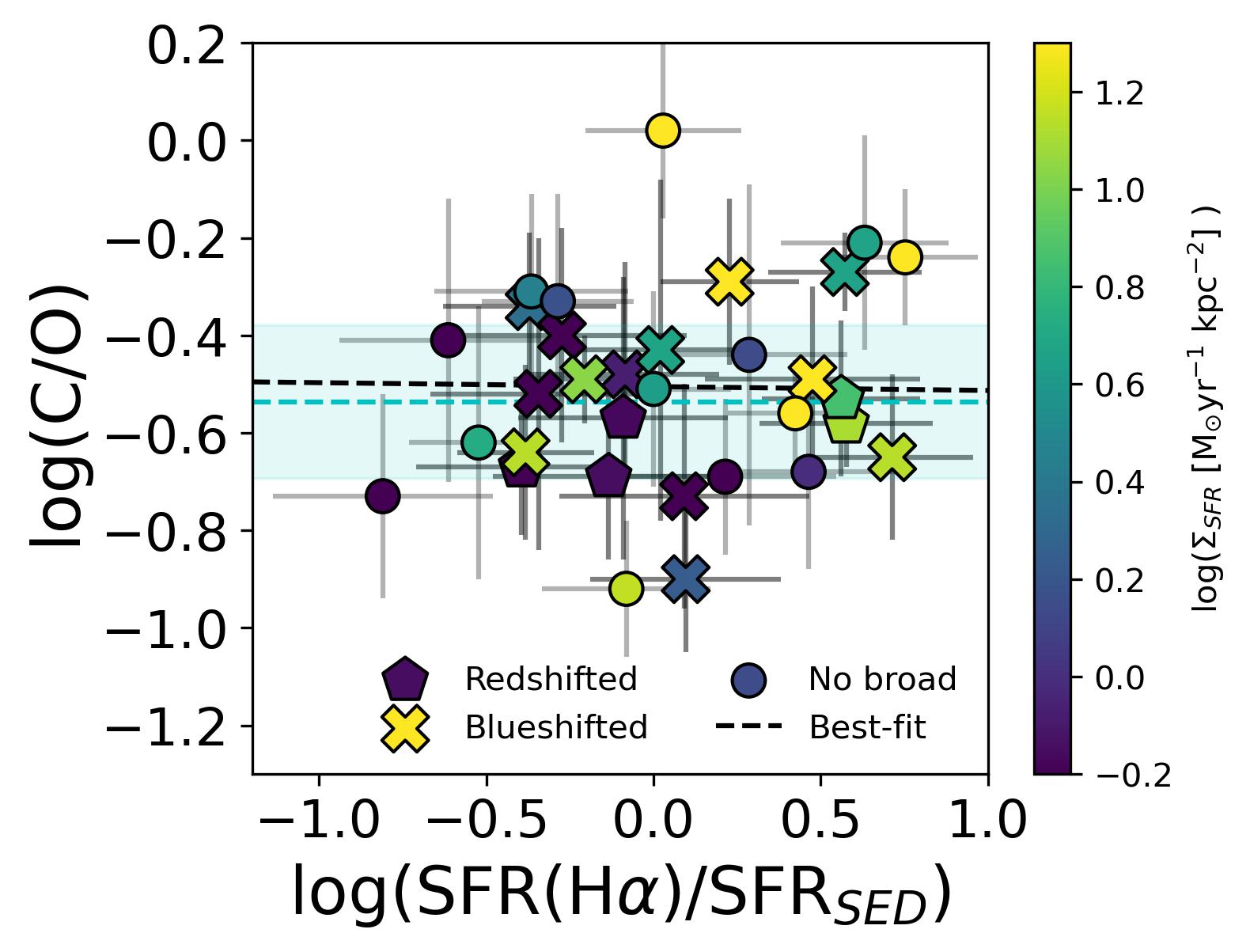}\\
    \caption{Relation of the chemical abundances (Oxygen abundance on top panels and C/O on bottom panels) with mass loading factor (left panels) and with burstiness (right panels). Our sample is divided into galaxies with broad component blueshifted (blue crosses) or redshifted (red pentagons) or without broad component detected (black squares). {The dashed cyan lines are the mean value, and the shaded region {is the mean observed scatter}. {Our sample is color-coded by $\Sigma_{\rm SFR}$.}}}
    \label{fig:loading-OH}
\end{figure*}

{Compared to recent observational studies using a similar two-Gaussian decomposition of strong emission lines, in Fig.~\ref{fig:loading} we typically find larger $\eta$ values than main sequence galaxies at cosmic noon \citep[e.g.][]{Concas2022} and nearby dwarf galaxies \citep[e.g.][]{Marasco2022}. The relative disagreement with simulation predictions found by these authors, i.e., a dropoff of $\eta$ at M$_{\star}\lesssim$10$^{10}$M$_{\odot}$ compared to the predicted increasing trend, was interpreted as evidence of possible inefficient feedback efficiency in low-mass galaxies}. 
{However, our results suggest that the compactness of the SF regions in low-mass galaxies, i.e., their $\Sigma_{\rm SFR}$, is key to determining the impact of outflows and stellar feedback in low-mass systems. Our results are in good agreement with recent findings by \citep{McQuinn2019}, who also show a clear increasing trend for the outflow mass-loading factors of local 
dwarf galaxies with centrally concentrated star formation (green squares in Fig.~\ref{fig:loading}). Indeed, the few blue compact dwarfs included in \citep{Marasco2022} also show comparably larger $\eta$ values. This is also evident from Fig.~\ref{fig:loading} (bottom panel), in which we find a clear trend with galaxies of higher $\Sigma_{\rm SFR}$ showing higher $\eta$ (best-fit in green line). The resulting trend consistently includes local dwarfs and results from stacking of EELGs at $z\sim 3-4$ \citep{Gupta2022} (red stars in Fig.\ref{fig:loading}).}

{The above results strongly suggest that galaxies of a given stellar mass may have different feedback effects according to their SF surface density, with low-mass galaxies of larger $\Sigma_{\rm SFR}$ experiencing more effective feedback}.
{These results are in tension with simulations where an opposite trend is found \cite[e.g.][]{Nelson2019TNG,Pahl2022}. As discussed in \cite{Nelson2019TNG}, this trend depends on the velocity threshold of outflow particles and the tension with observations might be explained by small-scale relationships that are not considered in simulations \citep[see also][for a detailed discussion]{McQuinn2019}}.  {Within this context, we note that our sample of galaxies is mainly strong [OIII] emitters with high EWs, which are somehow pushing the relation with $\eta$ towards higher $\Sigma_{\rm SFR}$, as shown in Fig.~\ref{fig:loading}. Thus, it appears possible that low-mass galaxies with moderate SFR and lower emission-line EWs are more consistent with literature data and values observed in simulated haloes at low-$z$ ($z<0.05$)}.

{Interestingly, recent works suggest a relation between the SFR surface density and the escape of ionizing photons (i.e. $\Sigma_{\rm SFR}\propto f_{\rm esc}$), which is intimately related to the ability of stellar feedback, i.e. outflows, to clear out young star-forming regions from dust and neutral, thus creating optically thin channels from which ionizing  photons may eventually escape \citep[e.g.][]{Naidu2020,Flury2022}. Within this context, our results suggest that young low-mass galaxies with strongly mass-loaded outflows, i.e. showing broad emission line components, could be clear candidates to have favorable conditions for Lyman photon escape}. 
 
{Recently, the spatially resolved study of the Sunburst arc, a lensed metal-poor galaxy at $z=2.37$ showing LyC escape, presented by \cite{Mainali2022}, revealed a strong blue-shifted broad emission component in [OIII]. Remarkably, the broad-to-narrow ratio in the leaking clump is  120\%, whereas for the non-leaker regions this ratio falls to 35\%. If we compare these results with the f$_{B}$ values obtained for our sample and other LyC indirect diagnostics \citep[e.g.][]{Flury2022}, we find that only a few  galaxies with the higher f$_{B}$ and $\Sigma_{\rm SFR}$, as well as high [OIII]/[OII], could be considered as our best candidates for LyC leakage.}

\subsection{{Effects of stellar feedback on chemical abundances}}\label{sec:outflows-MZR}

{We discuss the impact of  stellar feedback, traced by outflows, in the chemical abundances of the host galaxies. We use Fig.~\ref{fig:loading-OH}, in which we plot oxygen and carbon nebular abundances versus $\eta$ and $\beta = \log(\rm SFR_{H\alpha}/SFR_{\rm SED}$), the ratio between the instantaneous SFR, traced by H$\alpha$, and the SED-based SFR$_{\rm SED}$, tracing longer-timescales. $\beta$ is often considered a burstiness parameter  
\citep{Scalo1986,Guo2016}. }

{For our sample, we find no significant correlation ($\rho=-0.15$, $0.6\sigma$ significance) between metallicity and $\eta$ suggesting that metallicity is insensitive to the strength of the outflow at global spatial scales. We do not find compelling evidence that younger starbursts (high $\beta$) show lower metallicities ($\rho=-0.04$, 0.17$\sigma$) significance, suggesting also that metallicity is insensitive to SF timescale at global spatial scales.
No correlation is found with C/O neither ($\rho\sim 0$, 0.1$\sigma$ significance). 
These results are somehow consistent with the position of our sample in the mass-metallicity-SFR relations and the large scatter they show in the relation between $\eta$ and M$_{\star}$}. 

{However, galaxies with higher $\eta$ appear to have a weak increasing trend 
with the C/O abundance, which is slightly larger for galaxies with stronger outflows and denser star formation. This implies a certain level of selective enrichment due to outflows that could be in place in such galaxies. } 
{Our best fit is consistent  with a weak correlation  ($\rho$=0.42, 2$\sigma$ significance), not far from the  scatter we measure around the mean C/O value for the sample (cyan regions in Fig. \ref{fig:loading-OH}). For this  reason, a more complete analysis of the impact of stellar feedback on the chemical properties of the  galaxies will be done in future works.}  

{Finally, in Fig.~\ref{fig:loading-OH} we identify galaxies with blue- and red-shifted broad components to explore potential differences for in(out)flows, but we do not find any clear distinction among them and compared them to galaxies without broad emission. 
While this may suggest that we do not detect those components because of the geometry of the gas flow or the depth of our spectra,  rather than because of different nebular physical conditions, it may also indicate that a global parameter such as $\beta$ is not efficient to identify chemo-dynamical differences in these unresolved galaxies}.

Therefore, {we conclude that unresolved spectroscopy is likely insufficient to discern between these two possible scenarios in our sample. Spatially resolved spectroscopy is needed to consider the geometry of the ISM, compare with models and further explore the connection between gas flows, chemical abundances, and star formation of galaxies at $z\sim$3, which is now possible} using the NIRSpec IFU onboard the JWST.  

\section{Conclusions}\label{sec:conclusions}

In this work, we {present a detailed analysis of} the chemical abundances and kinematics of the ionized gas of low mass (10$^{7.9}$-10$^{10.3}$M$_{\odot}$) SFGs at $z\sim$3. We use new follow-up NIR spectroscopy for a sample of 35 SFGs selected on the basis of their rest-UV emission line properties (from Ly$\alpha$ to CIII]) from two previous works using ultra-deep optical spectra of the VANDELS \citep{Llerena2022} and VUDS \citep{Amorin2017} surveys. For VANDELS targets, our sample was assembled from  Keck/MOSFIRE spectra of the  NIRVANDELS survey \citep{Cullen2021}. For VUDS targets, our sample was assembled from MOSDEF spectra \citep{Kriek2015} and from new {spectra obtained} with VLT/X-shooter and Magellan/FIRE.

We focus our analysis of the NIR spectra on strong emission lines in the rest-optical, from [OII]$\lambda$3727 to H$\alpha$. We characterize the main properties of the sample based on the UV and optical datasets. We discuss scaling relations involving galaxies' gas metallicity and C/O abundances, which are derived using T$_{\rm e}$-consistent methods based on photoionization models and the observed UV and optical emission line ratios. In addition, using the available high-resolution spectra, we perform an analysis of the [OIII]$\lambda\lambda$4959,5007 emission line profiles with  a multi-Gaussian fitting technique to investigate the ionized gas kinematics of the galaxies and discuss the connection between stellar feedback and chemical enrichment in these young low-mass SFGs. We summarize our main results and conclusions as follows:

\begin{itemize}
    \item According to diagnostic diagrams based on both UV and optical emission line ratios, the dominant source of ionization in our sample of SFGs is massive stars. While 14\% of the sample show UV emission line ratios that are closer to those expected from AGN models, we find that their optical line ratios are instead consistent with pure stellar photoionization. Overall, our sample is characterized by high [OIII]/H$\beta>4$ ratios, which suggests high ionization conditions in the ISM.

    \item{We find rest-frame EW(CIII]) ranging from 1\r{A} to 15\r{A} and EW([OIII]) ranging from 102\r{A} to 1715\r{A}. We derive positive correlations between the EWs of bright UV and optical emission lines. About {15\%} of our sample show EW([OIII]$\lambda\lambda$4959,5007)$>1000$\r{A} that closely resemble those measured in $z>6$ EoR galaxies with photometric data \citep[e.g.][]{Endsley2021} and, more recently, with JWST spectroscopy \citep[e.g.][]{Matthee2022}.}
    
    \item {For galaxies with reliable measurements of the OIII]$\lambda$1666/[OIII]$\lambda$5007 ratio, we find mean electron temperatures T$_{\rm e}$=1.8$\times 10^4$K. Consequently, we use the code \hcmuv\, based on UV photoionization models to consistently derive low gas-phase metallicities and C/O abundances. We find a wide range of metallicity (12+log(O/H)$\sim$7.5-8.5) with a mean value of 12+log(O/H)=7.91 or 17\% solar. Using alternative methods, we find differences of up to $\sim 0.3$ dex toward higher (lower) metallicities when using pure optical (UV) strong-line calibrations, which are larger than typical uncertainties. We also derive a wide range of C/O abundance ratios ranging from log(C/O)$= -$0.9 to log(C/O)$=$-0.15 (23\% and 128\% solar, respectively) with a mean value of log(C/O)$=-0.52$ (54\% solar) that is consistent with previous results for SFGs at $z\sim3$ based on stacking spectra \citep{Shapley2003,Llerena2022}. Both oxygen and carbon abundances for the highest EW galaxies in our sample are in excellent agreement with values obtained for galaxies at $z>6$ with JWST spectra \citep{Arellano2022, Jones2023} }

     \item Our sample follows a mass-metallicity relation with a slope consistent with previous work at similar redshifts but showing an offset of about 0.3 dex to lower metallicities, which appears consistent with the low-metallicity envelope of the MZR scatter \citep{Curti2020,Sanders2021}. While these differences could be explained by the different methods used to estimate the metallicities, we conclude that the high ionization properties of our sample are most likely driving these offsets. Furthermore, we find that for a given stellar mass, {galaxies with 
     lower metallicities tend to show larger deviations from the FMR.} These results suggest that our SFGs are experiencing a rapid and active episode of massive star formation in which outflows from stellar feedback and accretion of fresh gas can be acting as significant regulators of their mass and metal content. 
     
     \item From the analysis of the CO-O/H relation, {we find no apparent increase of C/O with metallicity, as models predict \citep[e.g.][]{Mattsson2010,Nicholls2017}, but a large scatter of C/O values around metallicity $\sim10-20$\% solar. On the other hand, our galaxies appear consistent with an increase of C/O with stellar mass, suggesting that a fraction of their C/O may have a secondary origin. One possible interpretation for these trends is that a recent metal-poor inflow may dilute O/H while keeping the C/O as large as expected for their stellar mass. To explore further this and other possible interpretations larger representative samples over a wide range of   mass and metallicity are needed. }

    \item  {From a detailed multi-Gaussian component fitting of [OIII]$\lambda\lambda$4959,5007 line profiles, we find 65\% of our galaxies showing two distinct kinematic components: a narrow component with intrinsic velocity dispersion of $\sigma_N\sim$ 57 km s$^{-1}$ accounting for the core of the lines and a broader component with $\sigma_B\sim$121 km s$^{-1}$ that best fit the extended line wings. We find the broad component is typically blue- or red-shifted by $\sim 30-40$ km s$^{-1}$ with respect to the narrow one in most galaxies. Following the close similarities with local analogs, such as the Green Peas \citep{Amorin2012,Hogarth2020}, we interpret the narrow and broad kinematic components as gas tracing virial motions and {turbulent outflowing ionized gas driven by strong star formation}, respectively.  From our kinematic analysis, we find typical outflow velocities of $\sim 280$ km s$^{-1}$, which are found to correlate weakly with stellar mass but strongly with the instantaneous SFR traced by Balmer lines and $\Sigma_{\rm SFR}$}. 
     
     \item From our kinematic analysis, we find a mean mass-loading factor $\eta=0.54$ {(with a large range of  0.05-3.26 and a typical uncertainty of 0.3)} that is larger compared to the typical value observed in SFGs at similar redshift. 
     {We find galaxies with more compact star formation, i.e. $\Sigma_{\rm SFR} \gtrsim$10$M_{\odot}$yr$^{-1}$kpc$^{-2}$, showing larger $\eta$ for a given $M_{\star}$ at the stellar mass range covered by our sample (log(M$_{\star}/{\rm M}_{\odot})<10.2$). 
    This suggests that for a given stellar mass, denser starbursts in low-mass, low-metallicity galaxies produce stronger outflows.}
     This indicates that stellar mass alone, as concluded by some studies at lower redshift, does not necessarily determine how effectively gas is removed due to stellar feedback and that the star formation and ISM densities can regulate this process in low-mass galaxies, as some simulations predict. 
\end{itemize}

Overall, our results suggest a complex interplay between star formation, gas kinematics, and chemical enrichment in relatively young galaxies at $z\sim 3$. When observed during a young burst of SF, ionization properties are extreme and their chemical abundances are strongly regulated by their significant gas accretion and stellar feedback, {which make them outliers of key scaling relations}. In this phase, SFGs may show broad emission line components imprinting the turbulent ionized gas that is outflowing from the starbursting regions.  While outflows appear ubiquitous in the rapid star-forming episodes of low-mass galaxies at $z\sim 3$, their role as a regulator of the gas metallicity could be significantly stronger in galaxies developing denser starbursts. Outflows are in turn  suggested as an important mechanism to shape the ISM properties and to facilitate the escape of ionizing photons. Due to the close resemblance of EELGs at $z>6$ with a subsample of galaxies in this work, we conclude that the above results suggest that similar findings could be common in galaxies at higher redshifts observed with deep JWST high-resolution spectra.

\begin{acknowledgements}
    {We thank the anonymous referee for the detailed review and useful suggestions that help to improve this paper}. This work is based on data products from observations made with ESO Telescopes at La Silla Paranal Observatory under ESO program ID 194.A-2003 (PIs: Laura Pentericci and Ross McLure). {This paper includes data gathered with the 6.5 meter Magellan Telescopes located at Las Campanas Observatory, Chile.} {We thank Ross McLure, Fergus Cullen, and James Dunlop as part of the NIRVANDELS team for providing us with part of the data used in this paper.} MLl acknowledges support from the National Agency for Research and Development ANID/Scholarship Program/Doctorado Nacional/2019-21191036. RA acknowledges support from ANID FONDECYT Regular Grant 1202007.
    
    This work has made extensive use of Python packages astropy \citep{astropy:2018}, numpy \citep{harris2020}, and Matplotlib \citep{Hunter:2007}.

\end{acknowledgements}

\bibliographystyle{aa}
\bibliography{main}

\begin{appendix}
\section{Spectra of the sample}
\begin{figure*}[h]
    \centering
    \includegraphics[width=0.99\textwidth]{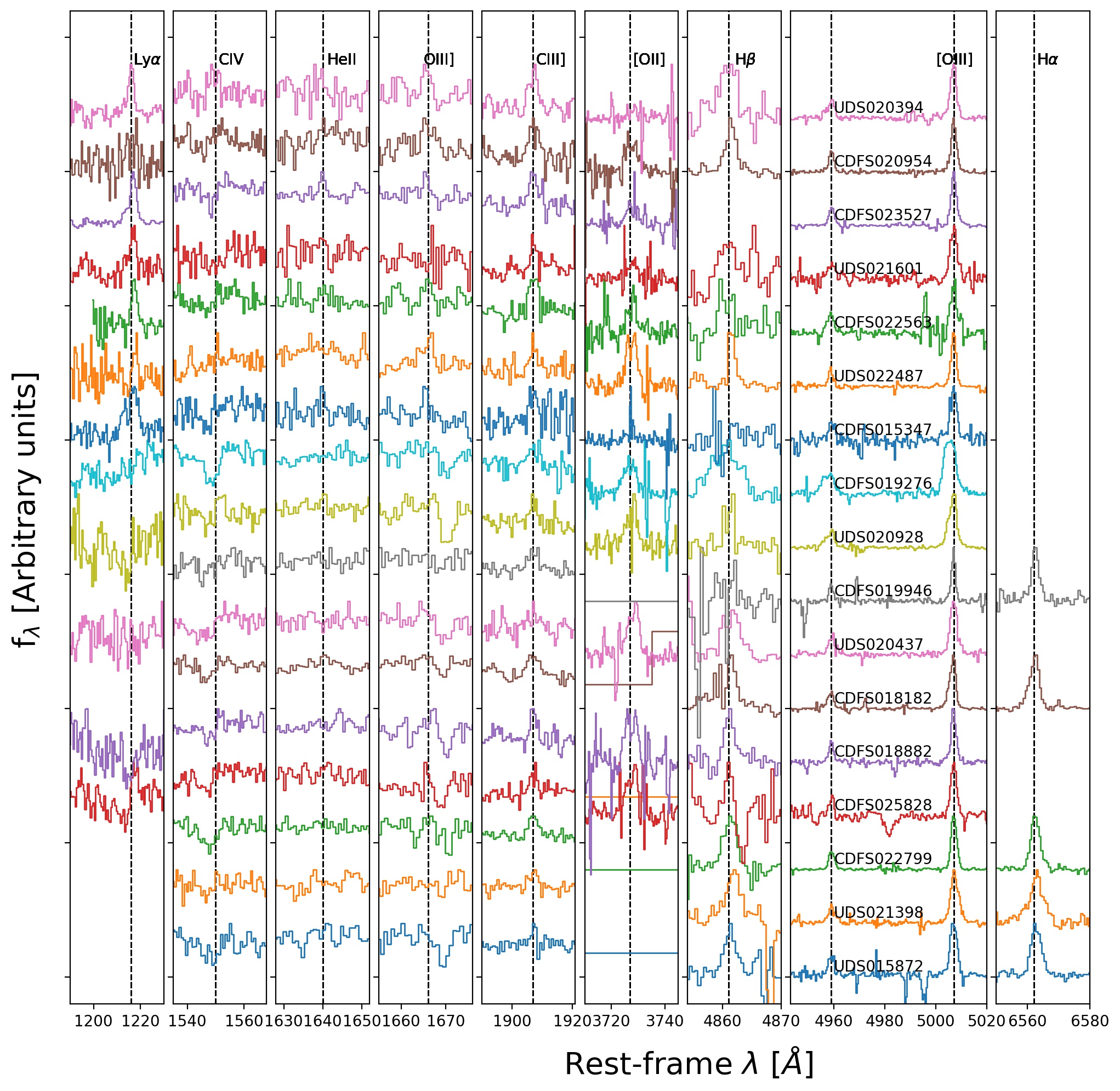}
    \caption{{Rest-frame spectra of the galaxies in the C3-VANDELS subsample. From left to right panels, we highlight the following emission-lines (which are marked with the black dashed line): Ly$\alpha$, CIV$\lambda\lambda$1548,51, HeII$\lambda$1640, OIII]$\lambda\lambda$1661,66, CIII]$\lambda\lambda$1907,09, [OII]$\lambda\lambda$3727,3729, H$\beta$, [OIII]$\lambda\lambda$4959,5007 and H$\alpha$. The flux density is in arbitrary units. Each galaxy is shown with the same color in each panel and its ID is in the [OIII]$\lambda\lambda$4959,5007 panel.}}
    \label{fig:spectraC3vandels}
\end{figure*}

\begin{figure*}[h]
    \centering
    \includegraphics[width=0.99\textwidth]{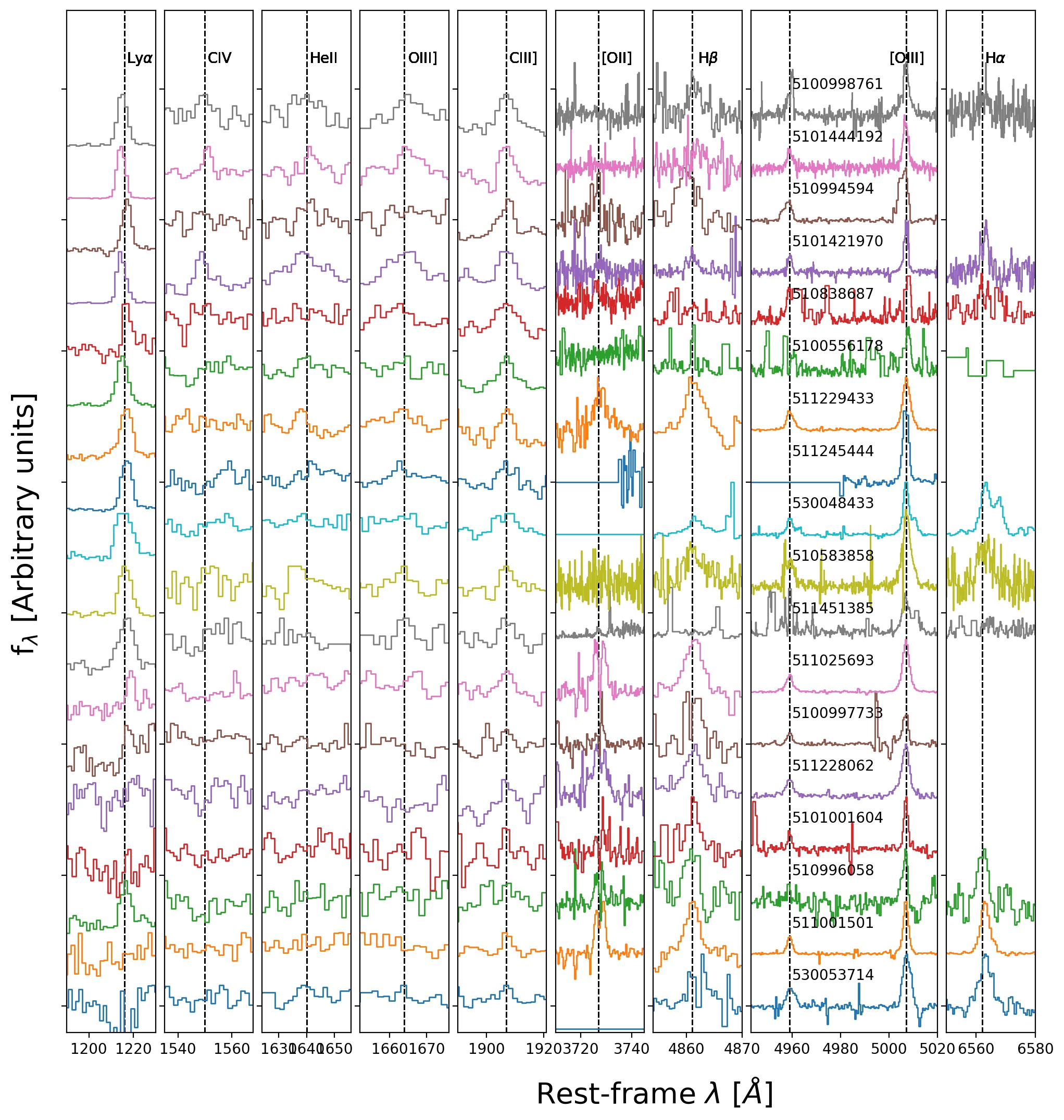}
    \caption{{Same as in Fig. \ref{fig:spectraC3vandels} but for the galaxies in the C3-VUDS sample.}}
    \label{fig:spectraC3vuds}
\end{figure*}

{In Fig. \ref{fig:spectraC3vandels} and \ref{fig:spectraC3vuds} we show the spectra of the galaxies used in this paper for both samples, C3-VANDELS and C3-VUDS, respectively. We highlight the spectral regions including the emission-lines used that includes Ly$\alpha$, CIV$\lambda\lambda$1548,51, HeII$\lambda$1640, OIII]$\lambda\lambda$1661,66, CIII]$\lambda\lambda$1907,09, [OII]$\lambda\lambda$3727,29, H$\beta$, [OIII]$\lambda\lambda$4959,5007 and H$\alpha$.}

\section{Plot of the SED fitting}
\begin{figure*}[ht]
    \centering
    \includegraphics[width=0.95\textwidth]{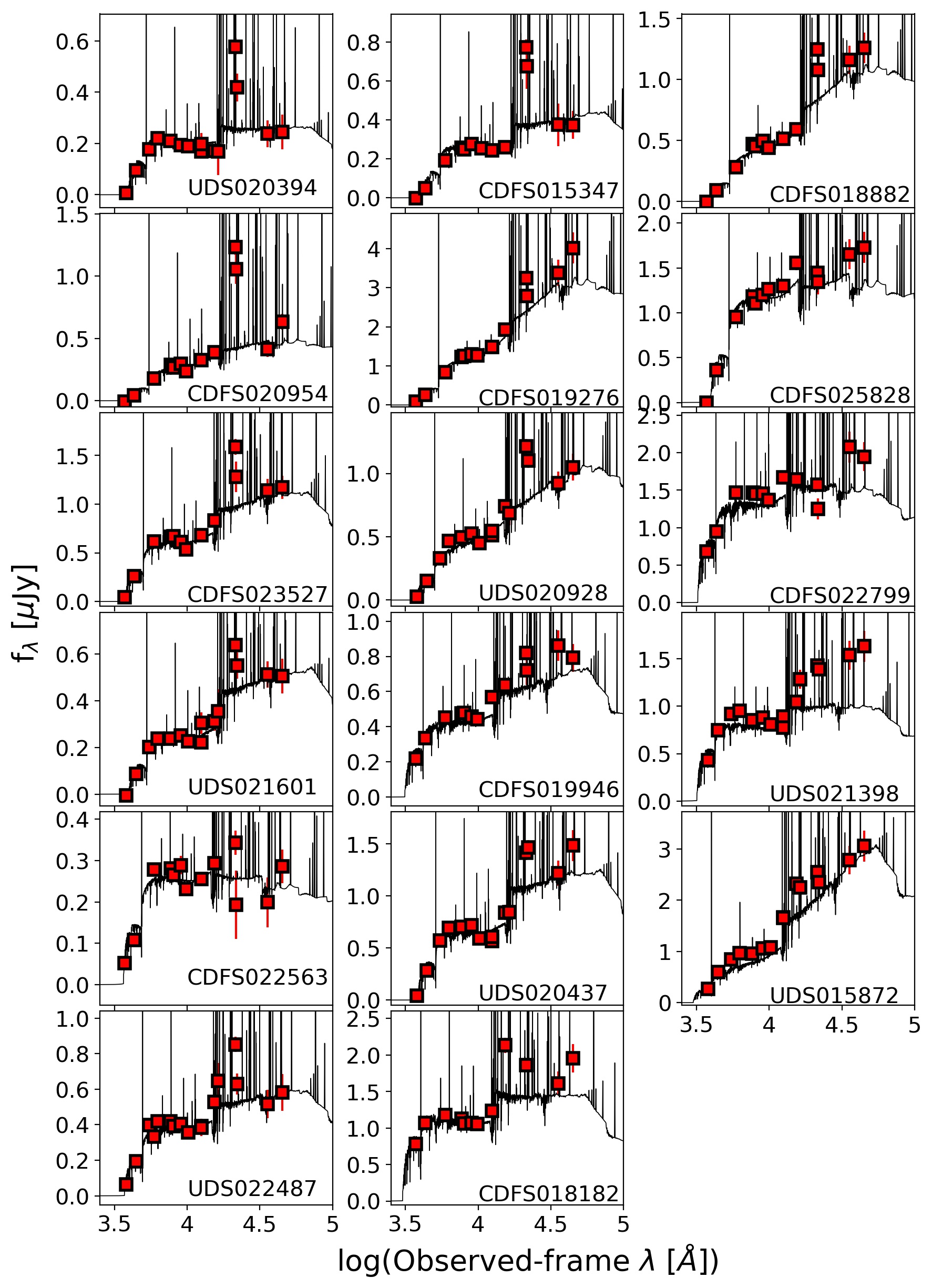}
    \caption{{SED model for the galaxies in the C3-VANDELS sample. The red squares are the photometric points, and the solid black line is the resulting spectrum from the SED fitting using BAGPIPES as described in Sec. \ref{sec:sed}.}}
    \label{fig:sedC3vandels}
\end{figure*}

\begin{figure*}[ht]
    \centering
    \includegraphics[width=0.95\textwidth]{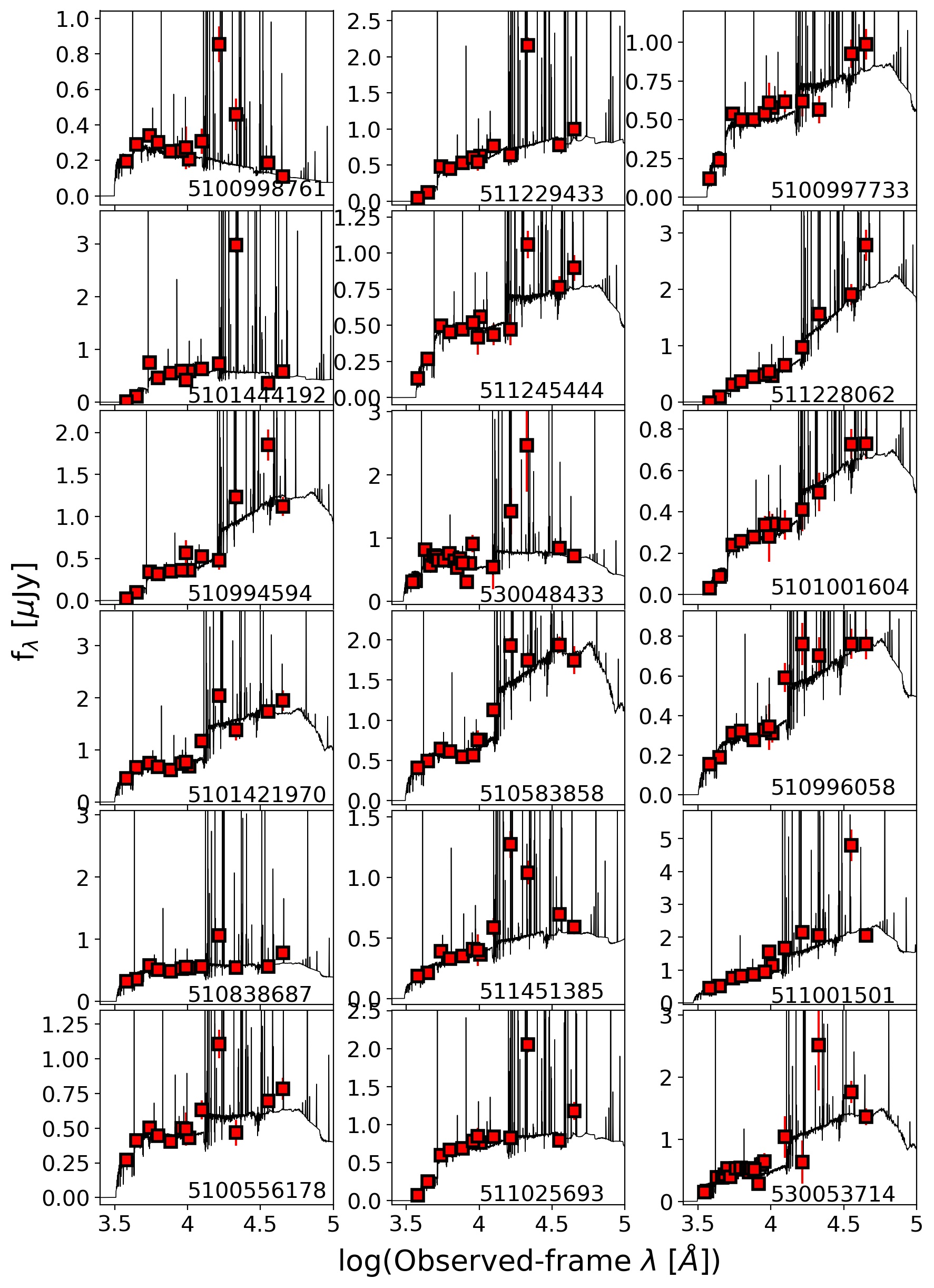}
    \caption{{Same as in Fig. \ref{fig:sedC3vandels} but for the galaxies in the C3-VUDS sample.}}
    \label{fig:sedC3vuds}
\end{figure*}

{In Fig. \ref{fig:sedC3vandels} and \ref{fig:sedC3vuds}, we display the photometry used and the SED model for each galaxy in both of the samples, C3-VANDELS and C3-VUDS, respectively.}

\section{HST imaging of the C3-VUDS sample}
\begin{figure}[ht]
    \centering
    \includegraphics[width=\columnwidth]{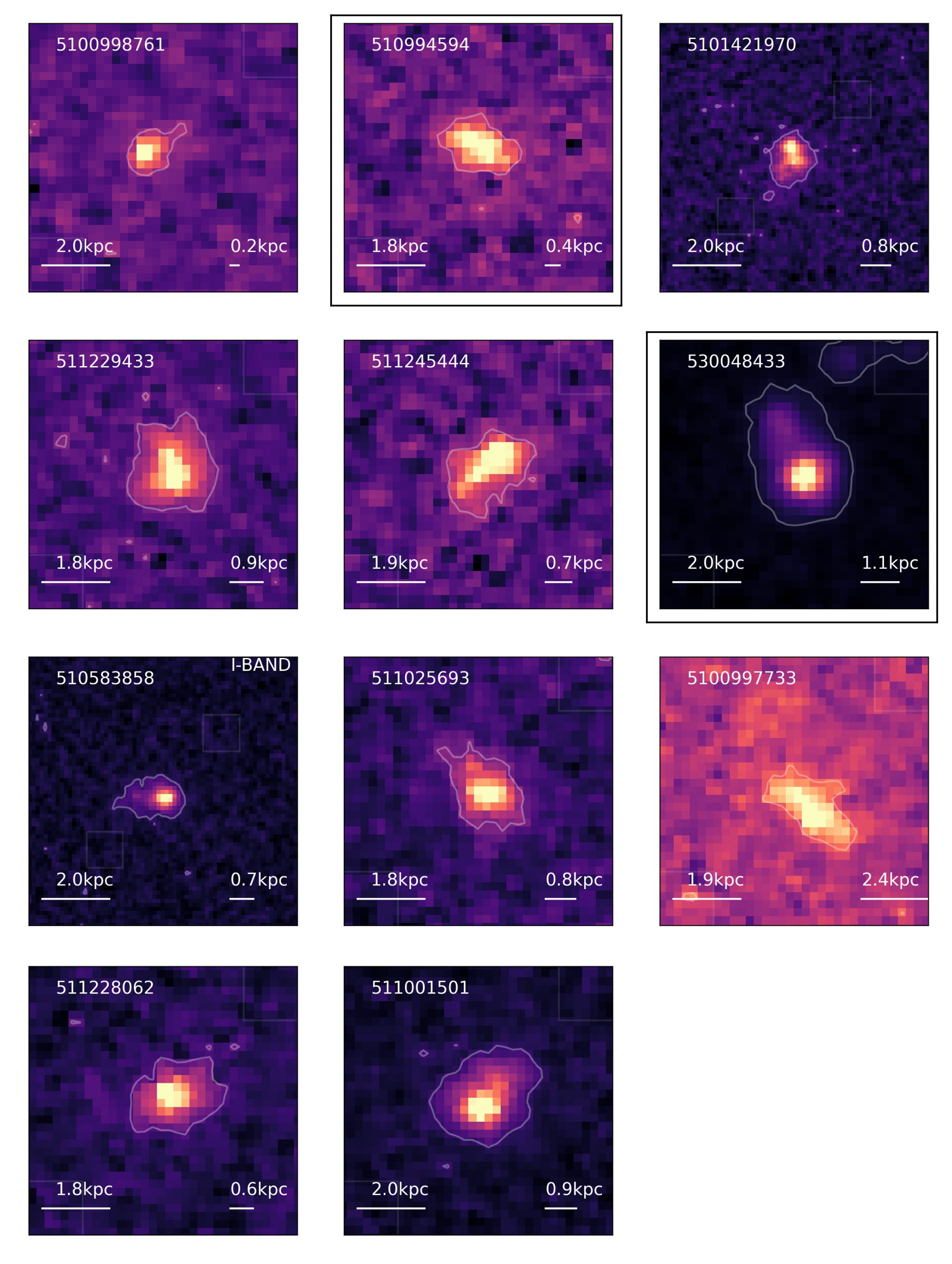}
    \caption{HST/F160W {images} \cite{Koekemoer2011} of the C3-VUDS sample with $\Delta$BIC$>2$, i.e., the subsample with a broad component in their [OIII] profile. The images are tracing the rest-optical. The white contour is the $3\sigma$ level. The physical scale of 0.5 arcsec at their redshift is shown on the left of each image, while on the right, the effective radius is shown. The galaxies with only i-band HST/F814W have a label. {The galaxies marked with a black square show two narrow components in their [OIII] profile.}}
    \label{fig:hstdeltabic>2vuds}
\end{figure}

\begin{figure}[ht]
    \centering
    \includegraphics[width=\columnwidth]{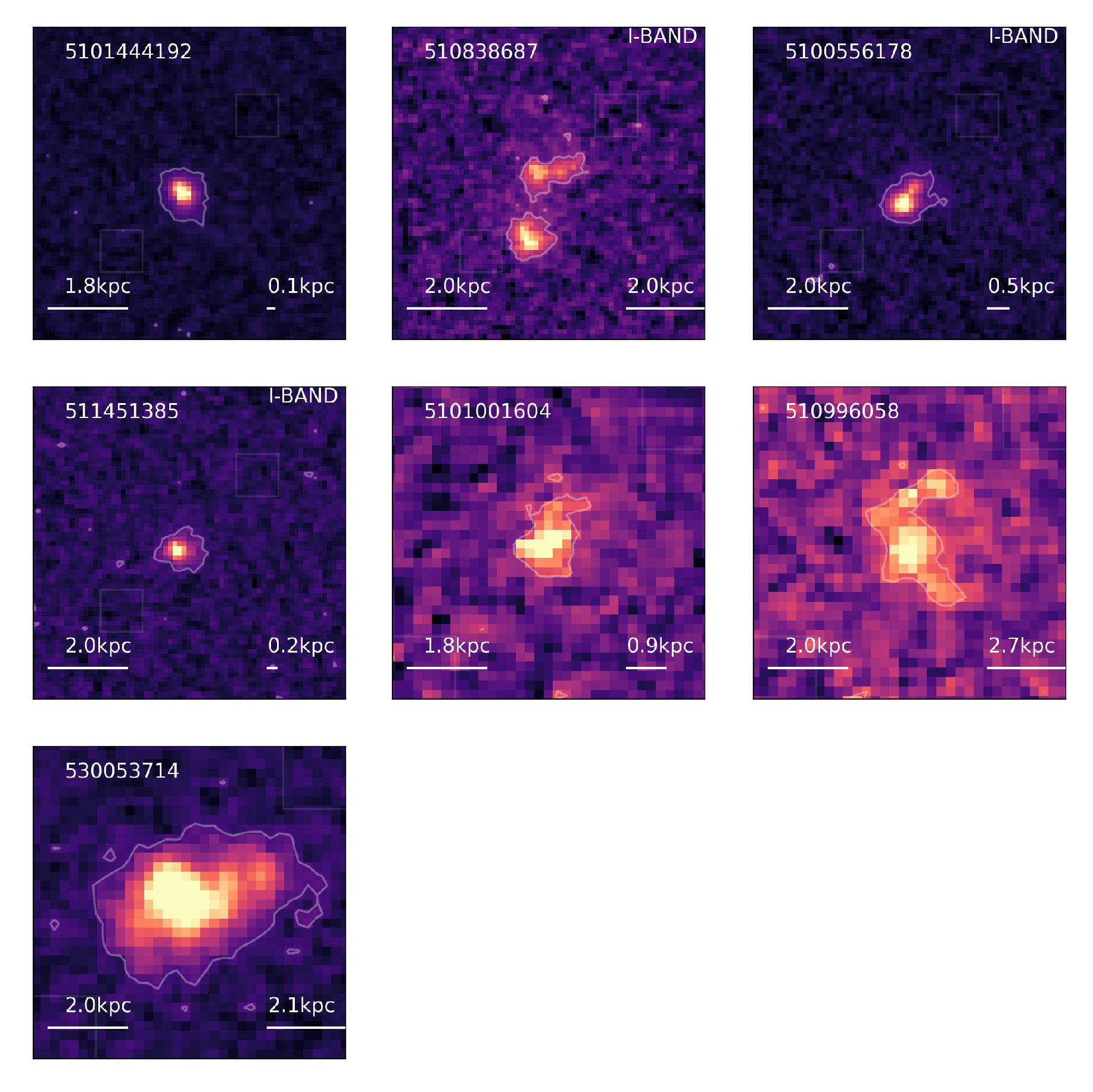}
    \caption{The same as in Fig. \ref{fig:hstdeltabic>2vuds} but for the C3-VUDS sample without features of a broad component in their [OIII] profile.}
    \label{fig:hstdeltabic<2vuds}
\end{figure}

{In Fig. \ref{fig:hstdeltabic>2vuds} and \ref{fig:hstdeltabic<2vuds} we show the HST images of the C3-VUDS sample for the galaxies with with $\Delta$BIC$>2$ and with $\Delta$BIC$<2$, respectively.}

\end{appendix}
\end{document}